\documentclass[11pt]{article}
\usepackage[utf8]{inputenc}
\usepackage{caption}
\usepackage{braket,slashed,bm}
\usepackage{jheppub}
\usepackage{simplewick}
\usepackage{array,multirow}
\usepackage{amsmath} \DeclareMathAlphabet{\mathpzc}{OT1}{pzc}{m}{it}
\usepackage[normalem]{ulem}
\usepackage{calligra}
\usepackage{floatrow}
\DeclareMathAlphabet{\mathpzc}{OT1}{pzc}{m}{it}
\usepackage{mathrsfs}
\usepackage{tikz}

\usepackage{subcaption}
\usepackage{lscape}
\usepackage{hyperref}
\usepackage{graphicx}
\usepackage{booktabs}
\newfloatcommand{capbtabbox}{table}[][\FBwidth]

\graphicspath{{./Figures/}}

\def\beq{\begin{equation}}
\def\eeq{\end{equation}}
\def\bea{\begin{eqnarray}}
\def\eea{\end{eqnarray}}
\def\nn{\nonumber \\}
\def\hyp{\mathsf{y}}

\def\gcb{{\overline g_{1}}}
\def\gcw{{\overline g_{2}}}

\newcommand{\Lagr}{\mathcal{L}}
\renewcommand{\to}{\rightarrow}

\begin{document}
\title{Consistent higher order $\sigma(\mathcal{G} \,\mathcal{G}\rightarrow h)$, $\Gamma(h \rightarrow \mathcal{G} \,\mathcal{G})$
and $\Gamma(h \rightarrow \gamma \gamma)$ in geoSMEFT}

\author[a]{Tyler Corbett,}

\author[b]{Adam Martin,}

\author[a,b]{and Michael Trott}

\affiliation[a]{Niels Bohr Institute, University of Copenhagen,
Blegdamsvej 17, DK-2100, Copenhagen, Denmark}

\affiliation[b]{Department of Physics, University of Notre Dame, Notre Dame, IN, 46556, USA}

\abstract{We report consistent results for $\Gamma(h \rightarrow \gamma \gamma)$, $\sigma(\mathcal{G} \,\mathcal{G}\rightarrow h)$ and $\Gamma(h \rightarrow \mathcal{G} \,\mathcal{G})$
in the Standard Model Effective Field Theory (SMEFT) perturbing the SM by corrections $\mathcal{O}(\bar{v}_T^2/16 \pi^2 \Lambda^2)$
in the Background Field Method (BFM) approach to gauge fixing, and to $\mathcal{O}(\bar{v}_T^4/\Lambda^4)$
using the geometric formulation of the SMEFT. We combine and modify
recent results in the literature into a complete set of consistent results, uniforming conventions, and
simultaneously complete the one loop results
for these processes in the BFM. We emphasize
calculational scheme dependence present across these processes, and how the operator and loop
expansions are not independent beyond leading order.
We illustrate several cross checks of consistency in the results.}
\maketitle
\setcounter{page}{3}
\flushbottom

\section{Introduction}
The Standard Model (SM) is not a complete description of observed phenomena in nature.
Neutrino masses clearly indicate that the SM must be extended with interactions that
couple to the SM states. However, there are no explicit collider results that directly evidence new
long-distance propagating states to add to those in the SM.
As a result, the SM is usefully thought of as an Effective Field Theory (EFT)
for measurements and current data analysis. In this approach, the effects of new states,
coupling to the SM through new interactions,
lead to Wilson coefficients of local contact operators at experimentally accessible energies.
This occurs under the assumption that
new physics states/dynamics are residing at scales $\Lambda$ larger than the Electroweak scale
($\sqrt{2 \, \langle H^\dagger H} \rangle \equiv \bar{v}_T$) enabling a Taylor expansion in $\bar{v}_T/\Lambda <1$.

The Standard Model Effective Field Theory (SMEFT) is defined by this Taylor expansion, for a review see Ref.~\cite{Brivio:2017vri}.
The SMEFT is also defined by several low energy assumptions;
that the spectrum at low energies is that in the SM, i.e. that there are no light hidden states in the spectrum with couplings
to the SM, and a $\rm SU(2)_L$ scalar doublet ($H$) with hypercharge
$\hyp_h = 1/2$ is present in the EFT in a manner that is respected by the power counting of the theory.

Studying experimental data using the SMEFT is a rich, and rapidly advancing, physics program. To robustly characterize
the projection of experimental results onto the leading (LO or dimension six) SMEFT perturbations of SM predictions,
it is important to calculate to sub-leading order in the loop expansion and also the
higher dimensional (dimension eight etc.) terms in the operator expansion in processes that have
experimental sensitivity to SMEFT perturbations. Knowledge of the sub-leading terms informs and defines
the theoretical error present in the projection of experimental results
onto leading order SMEFT results, when such sub-leading terms are neglected.
Such results also allow experimental fits to advance in the future to incorporate such
corrections as the data improves.

A key point of this paper is
the observation that the loop and operator expansions are not independent at sub-leading order in the SMEFT.
This fact is due to calculational scheme dependence
in fixing the SMEFT at leading order.
To illustrate this point, consider the perturbation due to a SMEFT operator to a dimensionless SM amplitude at dimension six
in an on shell process
\bea
\mathcal{A} = \mathcal{A}_{SM} + C_i \, N_i \, \frac{\bar{v}_T^2}{\Lambda^2} + \cdots
\eea
here $N_i$ is a numerical coefficient that is process dependent.
One can choose to normalize the Wilson coefficients by SM parameters $C_i \rightarrow g^j_{SM} \, C_i'$.
At dimension six such a choice is simply a rescaling. Developing a calculation to
$\mathcal{O}(\bar{v}_T^2/16 \pi^2 \Lambda^2)$, one loop finite terms are dictated by this choice.
This is due to finite renormalization of the parameters $g^j_{SM}$ and finite terms relating the
parameters $g^j_{SM}$ to input measurements at one loop. Similarly, when developing a prediction of
$\mathcal{A}$ the choice to rescale $C_i \rightarrow g^j_{SM} \, C_i'$ leads to
a specific scheme dependent set of dimension eight contributions relating $g^j_{SM}$ to input parameters including $\mathcal{O}(1/\Lambda^2)$
corrections. The operator expansion and the loop expansion are
not independent in the SMEFT. These expansions are tied together by scheme dependence
in fixing theory conventions.
Consequently, it is important that the loop corrections and dimension eight
corrections are formulated in a manner that is self consistent, so they can be meaningfully combined
when studying experimental data.
In this paper, we define consistent sub-leading
results, in both of these expansions, for
$\Gamma(h \rightarrow \gamma \gamma)$, $\sigma(\mathcal{G} \,\mathcal{G}\rightarrow h)$ and $\Gamma(h \rightarrow \mathcal{G} \,\mathcal{G})$.

The main results of this paper are given in Section \ref{numerics}. We report a set of numerical
results in both the $m_W$ and $\alpha$ input parameter schemes for
$\Gamma(h \rightarrow \gamma \gamma)$, $\sigma(\mathcal{G} \,\mathcal{G}\rightarrow h)$ and $\Gamma(h \rightarrow \mathcal{G} \,\mathcal{G})$
with a consistent set of corrections in both expansions defined to $\mathcal{O}(\bar{v}_T^2/16 \pi^2 \Lambda^2)$ and$\mathcal{O}(\bar{v}_T^4/\Lambda^4)$
as perturbations to the SM one loop amplitudes. These formulae can be used to generate
a consistent set of sub-leading terms in the SMEFT for these processes from inclusive LO simulation results
used in a global SMEFT fit.
\section{SMEFT notation}\label{setup}

The SM Lagrangian \cite{Glashow:1961tr,Weinberg:1967tq,Salam:1968rm} notation is fixed to be
\bea\label{sm1}
\mathcal{L} _{\rm SM} &=& -\frac14 G_{\mu \nu}^A G^{A\mu \nu}-\frac14 W_{\mu \nu}^I W^{I \mu \nu} -\frac14 B_{\mu \nu} B^{\mu \nu}  + \sum_{\psi} \overline \psi\, i \slashed{D} \, \psi, \\
&\,&\hspace{-0.75cm} + (D_\mu H)^\dagger(D^\mu H) -  \lambda \left(H^\dagger H -\frac12 v^2\right)^2 -  \left[H^{\dagger j} \, \overline d\, Y_d\, q_{j}
+ \widetilde H^{\dagger j} \overline u\, Y_u\, q_{j} + H^{\dagger j} \overline e\, Y_e\,  \ell_{j} + \hbox{h.c.}\right]. \nonumber
\eea
The chiral projectors have the convention $\psi_{L/R} = P_{L/R} \, \psi$ where
$P_{R} = \left(1 + \gamma_5 \right)/2$, and
the gauge covariant derivative is defined with a positive sign convention
\bea
D_\mu = \partial_\mu + i g_3 T^A A^A_\mu + i g_2  \sigma^I W^I_\mu/2 + i g_1 {\bf \hyp_i} B_\mu,
\eea
with $I=\{1,2,3\}$, $A=\{1\dots 8\}$ , $\sigma^I$ denotes the Pauli matrices and
${\bf \hyp_i}$ the $\rm U_Y(1)$ hypercharge generator with charge normalization ${\bf \hyp_i}= \{1/6,2/3,-1/3,-1/2,-1,1/2\}$ for $i =\{q,u,d,\ell,e,H \}$.
The SMEFT Lagrangian is
\begin{align}
 \Lagr_{\textrm{SMEFT}} &= \Lagr_{\textrm{SM}} + \Lagr^{(d)}, &   \Lagr^{(d)} &= \sum_i \frac{C_i^{(d)}}{\Lambda^{d-4}}\mathcal{Q}_i^{(d)}
 \quad \textrm{ for } d>4.
\end{align}
The SM Lagrangian notation and conventions are consistent with Refs.~\cite{Grzadkowski:2010es,Alonso:2013hga,Brivio:2017vri,Brivio:2017btx,Helset:2020yio,Brivio:2020onw}.
The operators $\mathcal{Q}_i^{(d)}$ are labelled with a mass dimension $d$ superscript
and multiply unknown Wilson coefficients $C_i^{(d)}$.  For compact dimensionless notation we
define $\tilde{C}^{(d)}_i \equiv C^{(d)}_i \bar{v}_T^{d-4}/\Lambda^{d-4}$. Here we have introduced
$\bar{v}_T$ which is the minimum of the potential in the SMEFT taking into account the presence of
higher dimensional operators \cite{Alonso:2013hga}. In this paper we generally use $\delta$
as notation to indicate a perturbation in the $1/\Lambda$ expansion in the SMEFT, while we use $\Delta$ to indicate
a perturbation in the loop expansion.

 We use the Warsaw basis \cite{Grzadkowski:2010es} for
$\mathcal{L}^{(6)}$ and otherwise Ref.~\cite{Helset:2020yio} for operator conventions. Due to strong constraints from low energy CP violating observables~\cite{Cirigliano_2016}, will restrict our study to CP even operators only.
Our approach to next to leading order interactions in the SMEFT, in both the loop expansion and the
operator expansion is fundamentally organized in the Geometric approach to the SMEFT (geoSMEFT).
\subsection{geoSMEFT}
The geoSMEFT is a organization of the physics of the SMEFT, focused on the fact that
this theory is an EFT with towers
of higher dimensional operators, including scalar fields that can take on vacuum expectation values.
Subsets of these operators can dress composite operator forms
with towers of scalar fields contracted with one another, and with symmetry generators.
Physical predictions in the SMEFT cannot depend on the coordinate choice of the scalar fields.
As such the effect of the scalar dressings of interactions terms are required to be encoded
in experimental measurements through scalar coordinate independent quantities. These quantities
define a series of scalar geometries that dress the composite operators. As the SMEFT contains an H field that can take on
a vacuum expectation value, geometric effects are associated with the $\bar{v}_T/\Lambda$ expansion in this theory.
The geoSMEFT \cite{Corbett:2019cwl,Helset:2020yio,Hays:2020scx} makes this physics manifest
by organizing the theory in terms of field-space connections $G_i$
multiplying composite operator forms $f_i$, represented schematically by
\bea\label{basicdecomposition}
\Lagr_{\textrm{SMEFT}} = \sum_i G_i(I,A,\phi \dots) \, f_i ,
\eea
where $G_i$ depend on the group indices $I,A$ of the (non-spacetime) symmetry groups,
and the scalar field coordinates of the composite operators. Powers of $D^\mu H$ are included in $f_i$.
The field-space connections depend on the coordinates of the Higgs scalar doublet expressed in terms of
real scalar field coordinates, $\phi_I = \{\phi_1,\phi_2,\phi_3,\phi_4\}$.
Of particular importance are the field space connections $h_{IJ},g_{AB}$ dressing
the Higgs kinetic term and the generalized Yang Mills term built out of $\mathcal{W}^A$
with $\mathcal{W}^A = \{W^1,W^2,W^3,B\}$.
These field space connections are
\begin{align}
	\mathcal{L}_{SMEFT} &= \frac{1}{2} h_{IJ}(\phi) (D_{\mu} \phi)^{I} (D^{\mu} \phi)^{J}
	- \frac{1}{4} g_{AB}(\phi) {\mathcal{W}}^{A}_{\mu\nu} {\mathcal{W}}^{B\mu\nu} + \cdots,
\end{align}
with $A =\{1,2,3,4\}$ and couplings $\alpha_A = \{g_2, g_2, g_2, g_1\}$.
The mass eigenstate
field coordinates are $\mathcal{A}^A = \{\mathcal{W}^+,\mathcal{W}^-,\mathcal{Z},\mathcal{A}\}$.
Similarly the mass eigenstate ghost field is defined as ${c^A = \{c_{{W}^+},c_{{W}^-},c_{Z},c_{A}\}}$.
Our notation is such that the covariant derivative acting on the bosonic fields of the SM in the doublet,
using real scalar field coordinates, is given by \cite{Helset:2018fgq}
\bea
(D^{\mu}\phi)^I &=& (\partial^{\mu}\delta_J^I - \frac{1}{2}\mathcal{W}^{A,\mu}\tilde\gamma_{A,J}^I)\phi^J,
\eea
with symmetry generators/structure constants ($\tilde{\epsilon}^{A}_{\, \,BC},\tilde{\gamma}_{A,J}^{I}$).
See Refs.~\cite{Helset:2018fgq,Helset:2020yio} for the generators/structure constants for the real scalar representation.

The transformation of the scalar fields to the mass eigenstates at all orders in the
$\bar{v}_T/\Lambda$ expansion is given by
\bea
\phi^{J} = \sqrt{h}^{JK} V_{KL} \Phi^{L}.
\eea

The transformation of the gauge/ghost fields, gauge parameters into mass eigenstates in the SMEFT
are given by
\begin{align}\label{basicrotations}
\mathcal{W}^{A,\nu} &=  \sqrt{g}^{AB} U_{BC} \mathcal{A}^{C,\nu}, \\
u^A &= \sqrt{g}^{AB} U_{BC} \, c^C, \\
\alpha^{A} &= \sqrt{g}^{AB} U_{BC} \mathcal{\beta}^{C},
\end{align}
with $\Phi^L = \{\Phi^+, \Phi^-, \chi, H \}$. $\mathcal{\beta}^{C}$ is obtained directly
from $\alpha^{A}$ and $U_{BC}$. Note that $\alpha_A \mathcal{W}^{A,\nu}$ linear terms in the
covariant derivative are unchanged by these transformations at all orders in
the $\bar{v}_T/\Lambda$ expansion.\footnote{Here the inverses are defined via
$\sqrt{g}^{AB} \sqrt{g}_{BC} \equiv \delta^A_C$ and $\sqrt{h}^{IJ} \sqrt{h}_{JK} \equiv \delta^I_K$.
The matrix square roots of these field space connections are $\sqrt{g}_{AB} = \langle g_{AB} \rangle^{1/2}$,
and $\sqrt{h}_{IJ} = \langle h_{IJ} \rangle^{1/2}$ and $\langle \rangle$ indicates
a background field expectation value. The SMEFT perturbations are small corrections to the SM,
so the field-space connections are positive semi-definite matrices,
with unique positive semi-definite square roots.
We also use the notation $\hat{g}_{AB} = \langle g_{AB} \rangle$ and $\hat{h}_{IJ} = \langle h_{IJ} \rangle$
for the background field expectation values of the metrics at times.} The matrices $U,V$ are unitary rotations; i.e. orthogonal matricies whose transpose is equal to the matrix inverse,
and given by
\begin{align*}
 U_{BC} &= \begin{bmatrix}
   \frac{1}{\sqrt{2}} & \frac{1}{\sqrt{2}} & 0 & 0 \\
   \frac{i}{\sqrt{2}} & \frac{-i}{\sqrt{2}} & 0 & 0 \\
   0 & 0 & c_{\overline{\theta}} & s_{\overline{\theta}} \\
   0 & 0 & -s_{\overline{\theta}} & c_{\overline{\theta}}
 \end{bmatrix},& \quad
 V_{JK} &= \begin{bmatrix}
   \frac{-i}{\sqrt{2}} & \frac{i}{\sqrt{2}} & 0 & 0 \\
   \frac{1}{\sqrt{2}} & \frac{1}{\sqrt{2}} & 0 & 0 \\
 0 & 0 & -1 & 0 \\
 0 & 0 & 0 & 1
 \end{bmatrix}.
\end{align*}
Here the angle is defined via the generalized Yang Mills field space metric
\bea
s_{\bar{\theta}}^2 &=& \frac{(g_1 \sqrt{g}^{44} - g_2 \sqrt{g}^{34})^2}{g_1^2 [(\sqrt{g}^{34})^2+ (\sqrt{g}^{44})^2]+ g_2^2 [(\sqrt{g}^{33})^2+ (\sqrt{g}^{34})^2]
- 2 g_1 g_2 \sqrt{g}^{34} (\sqrt{g}^{33}+ \sqrt{g}^{44})}.
\eea

The masses and couplings that follow in the geoSMEFT are consistent with those previously defined
to $\mathcal{L}^{(6)}$ in Ref.~\cite{Alonso:2013hga} and used in SMEFTsim, see Refs.~\cite{Brivio:2017btx,Brivio:2019myy}.
We explicitly list these results here for completeness, the masses are
\begin{align}
\bar{M}_{W}^2 &= \frac{\bar g_2^2 \bar{v}_T^2}{4}, \\
\bar{M}_{Z}^2 &= \frac{\bar{v}_T^2}{4}(\bar g_1^2+\bar g_2^2)+\frac{1}{8} \, \bar{v}_T^2 \, (\bar g_1^2+\bar g_2^2) \, \tilde{C}_{HD} +\frac{1}{2} \bar{v}_T^2 \, \gcb \, \gcw \, \tilde{C}_{HWB},\\
\bar{m}_{h}^2 &= 2 \lambda \bar{v}_T^2 \left[1- 3 \frac{\tilde{C}_H}{2 \, \lambda} + 2 \left(\tilde{C}_{H \Box} - \frac{\tilde{C}_{HD}}{4}\right)\right].
\end{align}
The geometric SMEFT couplings with $\mathcal{L}^{(6)}$ corrections are
\begin{align}
\bar{e} &= \frac{\gcb \, \gcw}{\sqrt{\bar g_1^2+\bar g_2^2}} \left[1- \frac{\gcb \, \gcw}{\bar g_1^2+\bar g_2^2} \, \tilde{C}_{HWB} \right],
& \quad
\bar{g}_Z &= \sqrt{\bar g_1^2+\bar g_2^2}+ \frac{\gcb \, \gcw}{\sqrt{\bar g_1^2+\bar g_2^2}} \, \tilde{C}_{HWB}, \\
\gcb &= g_1(1+ \tilde{C}_{HB}),& \quad
\gcw &= g_2(1+ \tilde{C}_{HW}).
\end{align}
\subsubsection{Higgs gluon connection in geoSMEFT}
The Higgs-gluon field space metric is defined as
\bea
\Lagr_{\textrm{SMEFT}} \supset - \frac{1}{4} k_{\mathpzc{AB}}(\phi) G^{\mathpzc{A},\mu \nu} G_{\mathpzc{B},\mu \nu},
\eea
with $\mathpzc{A,B,C \dots}$ running over $\{1 \cdots 8 \}$ and
\bea
k_{\mathpzc{AB}}(\phi)= \left(1- 4 \, \sum_{n=0}^\infty C_{HG}^{(6+ 2n)} \, \left(\frac{\phi^2}{2}\right)^{n+1} \right) \delta_{\mathpzc{AB}}.
\eea
The field space connection $k_{\mathpzc{AB}}(\phi)$ is trivial in colour space,
as the Higgs carries no $\rm SU(3)$ charge. It is convenient to explicitly pull the trivial colour structure out of the
field space connection $k_{\mathpzc{AB}}(\phi) \rightarrow \kappa(\phi) \, \delta_{AB}$.
For the gluon field and coupling the transformations to canonically normalized fields are given by
\begin{align}
G^{A,\nu} &=  \sqrt{\kappa} \, \mathcal{G}^{A,\nu}, \\
\bar{g}_3 &= g_3 \, \sqrt{\kappa}.
\end{align}

We refer to the two index field space connections $k_{\mathpzc{AB}}$, $h_{IJ}$, $g_{AB}$ as field space metrics in this work.
These metrics are defined at all orders in the geoSMEFT organization of the SMEFT operator expansion in Ref.~\cite{Helset:2020yio}.

\subsection{...and the Background Field Method}
The geoSMEFT is a formulation of the SMEFT where the background scalar field expansions
of interaction terms are key to the organization of the theory.
The scalar expansion is associated with the scale $\bar{v}_T$.
The geoSMEFT approach, which organizes the operator expansion, was actually discovered by formulating the background field
method (BFM) \cite{DeWitt:1967ub,tHooft:1973bhk,Abbott:1981ke} in the SMEFT, with the aim of gauge fixing the theory in a manner
that is invariant under background field transformations for the purpose of loop corrections, a result reported in
Ref.~\cite{Helset:2018fgq}.\footnote{The issue of gauge fixing the SMEFT in the BFM was first discussed as a novel challenge in Ref.~\cite{Hartmann:2015oia}.
See Refs.~\cite{Ghezzi:2015vva,Dedes:2017zog,Misiak:2018gvl} for other approaches to gauge fixing the SMEFT.}

The BFM gauge fixes in a manner that leaves the effective action invariant under background field gauge transformations
($\Delta' F$) in conjunction with a linear change of variables on the quantum fields, see Ref.~\cite{Abbott:1981ke,Corbett:2019cwl}.
The fields are split into quantum (un-hated) and classical (hatted) background
fields: $F \rightarrow F+ \hat{F}$.\footnote{We also use a hat superscript for Lagrangian parameters
that are fixed to numerical values by relating to input parameter measurements.}
The classical fields are associated with the external states
of the $S$-matrix in an LSZ procedure \cite{Lehmann:1954rq}. The scalar field expectation value
key to the geoSMEFT organization of the theory is associated with an external background Higgs field, so that
\begin{align}
\hat{H}(\hat{\phi}_I) &= \frac{1}{\sqrt{2}}\begin{bmatrix} \hat{\phi}_2+i\hat{\phi}_1 \\ \hat{\phi}_4 + \bar{v}_T - i \hat{\phi}_3\end{bmatrix},
& \quad  H(\phi_I) &= \frac{1}{\sqrt{2}}\begin{bmatrix} \phi_2+i\phi_1 \\ \phi_4 - i\phi_3\end{bmatrix}.
\end{align}
Associating the vev with the external background Higgs field should not be over interpretted, as a tadpole scheme
counter term condition applies to all tadpole topologies.

The BFM generating functional of the geoSMEFT is given by
\begin{align}
Z[\hat{F},J]= \int \mathcal{D} F \,{\rm det}\left[\frac{\Delta' \mathcal{G}^A}{\Delta' \alpha^B}\right]e^{i \int d x^4 \left(S[F + \hat{F}] + \Lagr_{\textrm{GF}} +
{\rm source \, terms} \right)} \nonumber.
\end{align}
The generating functional is integrated over the quantum field configurations via $\mathcal{D} F$,
with $F$ field coordinates describing all long-distance propagating states. The sources $J$
couple only to the quantum fields \cite{'tHooft:1975vy}.
In the BFM, relationships between Lagrangian parameters (including renormalization constants) due to unbroken
background  $\rm SU(2)_L \times U(1)_Y$ symmetry then follow a ``naive" (classical) expectation when
quantizing the theory. This pattern is naturally related to the geoSMEFT generalization of the SM Lagrangian terms.
By which we mean the geometric masses and couplings appear in the Ward identities of the theory in the BFM \cite{Corbett:2019cwl}
in a natural generalization of the mass and coupling dependence in the SM into that of the geoSMEFT.

Ref.~\cite{Helset:2018fgq} reported that a minimal gauge fixing term in the BFM for the SMEFT, for the EW sector is
\begin{align}\label{gaugefixing1}
	\Lagr^{EW}_{\textrm{GF}} &= -\frac{\hat{g}_{AB}}{2 \, \xi} \mathcal{G}^A \, \mathcal{G}^B, &
\mathcal{G}^X &\equiv \partial_{\mu} \mathcal{W}^{X,\mu} -
		\tilde\epsilon^{X}_{ \, \,CD}\hat{\mathcal{W}}_{\mu}^C \mathcal{W}^{D,\mu}
    + \frac{\xi}{2}\hat{g}^{XC}
		\phi^{I} \, \hat{h}_{IK} \, \tilde\gamma^{K}_{C,J} \hat{\phi}^J,
\end{align}
for the QCD coupling we have analogously the BFM gauge fixing term \cite{Corbett:2020bqv}
\begin{align}\label{gaugefixing2}
	\Lagr^{QCD}_{\textrm{GF}} &= -\frac{\hat{\kappa}}{2 \, \xi_G} \, \mathcal{G}^{\mathpzc{A}} \, \mathcal{G}_{\mathpzc{A}},
\end{align}
where
\bea
\mathcal{G}^{\mathpzc{A}} &\equiv \partial_{\mu} \mathcal{G}^{\mathpzc{A},\mu} - \frac{\bar{g}_3}{\sqrt{\kappa}} \, f^{\mathpzc{ABC}} \,
\hat{\mathcal{G}}_{\mu,\mathpzc{B}} \, \mathcal{G}_{\mu, \mathpzc{C}}.
\eea

This approach to BFM gauge fixing in the SMEFT has a very intuitive interpretation.
The $n$-point interactions are dressed by scalar fields defining field space manifolds in the geoSMEFT.
These scalar dressings are curved spaces, whose curvature is associated with the
power counting expansion in $\bar{v}_T/\Lambda$.
The SM gauge fixing is promoted to a gauge fixing on these curved spaces
by upgrading the naive squares of fields in the gauge fixing term, to less-naive
contractions of fields through the Higgs background field space metrics $\hat{g}_{AB}, \hat{h}_{IK}, \hat{\kappa}$.\footnote{
When the field space metrics are trivialized to their values in the SM: $\hat{h}_{IJ} = \delta_{IJ}$,
$\hat{g}_{AB} = \delta_{AB}$ and $k_{\mathpzc{AB}}=\delta_{\mathpzc{AB}}$. The field space manifolds are no longer curved due to SMEFT corrections
in this $\bar{v}_T/\Lambda \rightarrow 0$ limit. The gauge fixing term in the BFM
then simplifies to that of the SM, as given in Ref.~\cite{Shore:1981mj,Einhorn:1988tc,Denner:1994xt}.}

Both the BFM and the geoSMEFT are approaches
to the SMEFT that are focused on the background fields
present in the theory, defining the background field manifolds (in scalar space)
that effect scattering experiments.
The geoSMEFT $\Lagr^{(8)}$ corrections are naturally
associated in a consistent manner with a loop expansion of the theory defined using the BFM approach to gauge fixing.
Again, we stress that the normalization of the operators at $\mathcal{L}^{(6)}$ effects both the perturbative expansion
through the gauge fixing term and finite renormalization effects, and also the $\mathcal{L}^{(8)}$ corrections.
This point is a key organizing principle of this paper.

\section{One loop formulation in the BFM}\label{oneloopBFM}
The formulation of the processes $\Gamma(h \rightarrow \gamma \gamma)$, $\sigma(\mathcal{G} \,\mathcal{G}\rightarrow h)$ and $\Gamma(h \rightarrow \mathcal{G} \,\mathcal{G})$
in the SMEFT to $\mathcal{O}(1/\Lambda^2 16 \pi^2)$ requires the definition of input parameters to one loop
in the SMEFT, and the one loop renormalization of the theory, including finite terms, must be fixed with renormalization
conditions. Note that we use $\phi_4 \equiv h$ interchangeably in the results below.

Some initial steps in this direction were made in the BFM in Refs.~\cite{Hartmann:2015oia,Hartmann:2015aia,Hartmann:2016pil}.
These efforts were hampered by the lack of a precise understanding of gauge fixing in the BFM in the SMEFT at the time of this
initial work. Further essential developments leading to the results collected and completed here
were developed in Refs.~\cite{Jenkins:2017jig,Jenkins:2017dyc,Dekens:2019ept}, which
matched the SMEFT (in the BFM) to the low energy effective field theory (LEFT).
The importance of these matching results is that
input parameters measured at low scales, such as $\alpha_{ew}, G_F$ are now consistently characterized
to one loop in the BFM. We use these results to complete the characterization of the
three essential processes of interest at LHC in the BFM to $\mathcal{O}(1/\Lambda^2 16 \pi^2)$
and to perform several consistency cross checks in the results.

\subsection{Wavefunction/Mass Renormalization}\label{wavefcn}

For the processes of interest in this work, we require the wavefunction renormalization of the Higgs,
photon and gluon fields. Renormalization constants are introduced for the background fields
and the couplings (here a $0/r$ superscript means a bare/renormalized parameter) via
\bea
\hat{\phi}_4^0 &=& Z^{1/2}_{\hat{\phi}_4} \, \hat{\phi}_4^{(r)}, \\
\hat{\mathcal{A}}_\mu^0 &=& Z^{1/2}_{\hat{\mathcal{A}}} \, \hat{\mathcal{A}}_\mu^{(r)}, \\
\hat{\mathcal{G}}_\mu^0 &=& Z^{1/2}_{\hat{\mathcal{G}}} \, \hat{\mathcal{G}}_\mu^{(r)}, \\
\bar{e}^0 &=& Z_{e} \, \bar{e}^{(r)} \, \mu^{\epsilon}, \\
\bar{g}_3^0 &=& Z_{g} \, \bar{g}_3^{(r)} \, \mu^{\epsilon}, \\
\bar{v}_T^{0} &=& Z_v^{1/2}  \, \bar{v}_T^{(r)},
\eea
and the divergent contributions to the tadpole contribution to the Higgs vev $(\Delta v/\bar{v}^0_T)_{\it div}$.
We frequently suppress the explicit $0/r$ superscripts below.
We use dimensional regularization and work in $d= 4- 2 \epsilon$ dimensions.
The factor of $\mu$ is included in the coupling renormalization to render the renormalized coupling
dimensionless \cite{Manohar:2000dt}.
Each of the renormalization constants is expanded as $ Z_i = 1 + \Delta Z_i + \cdots$.
Our notation is we use $\Delta Z_i$ for the divergence
chosen to cancel in a $\rm \overline{MS}$ subtraction.\footnote{
We use a $0$ superscript for both bare parameters and matrix elements at tree level at times.
These notational choices are made due to historical conventions.} The notation $\Delta R_i$ is reserved
for the finite renormalization factors. We generally use $\Delta$ to indicate a loop correction to a Lagrangian parameter.

To define the mass and wavefunction counter-terms we define the decomposition of the two point functions
for vector fields $\hat{V},\hat{V}'$
\bea
-i \Gamma^{\hat{V},\hat{V}'}_{\mu \nu}(k)&=&
\left(-g_{\mu \nu} k^2 + k_\mu k_\nu + g_{\mu \nu} \bar{M}_{\hat{V}}^2\right)\delta^{\hat{V} \hat{V}'}
+\left(-g_{\mu \nu} +\frac{k_\mu k_\nu}{k^2}  \right) \Sigma_{T}^{\hat{V},\hat{V}'}- \frac{k_\mu k_\nu}{k^2}
\Sigma_{L}^{\hat{V},\hat{V}'}.
\eea

The photon and Higgs wavefunction finite terms are fixed by the on-shell conditions
\bea
\Delta R_{\hat{A}} &=& - \left.\frac{\partial \Sigma^{\hat{A}\hat{A}}_T}{\partial k^2}\right|_{k^2=0}, \\
\Delta R_{\hat{\phi}_4} &=&- \left.\frac{\partial\Pi_{\hat{\phi}_4 \phi_4}(p^2)}{\partial p^2}\right|_{p^2=\bar{m}_h^2}.
\eea
the divergent terms have the same definition with $\Delta R_{i} \rightarrow \Delta Z_{i}$.
The masses counter-terms, and finite contributions to the masses to one loop, are defined via the transverse
two point functions. In the case of the divergences we define
\bea
\Delta Z_{m_{W}^2} &= {\rm Re} \left(\Sigma_T^{\hat{\mathcal{W}}\hat{\mathcal{W}}}(\bar{m}_{W}^2) \right)_{div}, \\
\Delta Z_{m_{Z}^2} &= {\rm Re} \left(\Sigma_T^{\hat{\mathcal{Z}}\hat{\mathcal{Z}}}(\bar{m}_{Z}^2) \right)_{div},
\eea
so that to one loop bare masses are related to the renormalized masses as
\bea
(\bar{m}_{W}^{(0)})^2 &=  (\bar{m}_{W}^{(r)})^2 + \Delta Z_{m_{W}^2},  \quad (\bar{m}_{Z}^{(0)})^2 &= (\bar{m}_{Z}^{(r)})^2+ \Delta Z_{m_{Z}^2}.
\eea
Note that when a FJ tadpole scheme \cite{Fleischer:1980ub} is used, the tadpole contributions to $\Sigma_T$ are zero by
definition and the one loop shift in the minimum of $\bar{v}_T$ (denoted $\Delta \bar{v}_T$) is required to be included
in predictions;
see Section \ref{vev} for more details.
\subsubsection{Divergent counter-terms}
The counter-terms in the
SM in the BFM are
\begin{align}
\Delta Z_{\hat{\phi}_4} &= \frac{(3+ \xi)(\bar{g}_1^2 + 3 \,\bar{g}_2^2)}{64 \pi^2 \epsilon} - \frac{Y}{16 \pi^2 \epsilon}, \\
\Delta Z_{\hat{A}} &= - \frac{\bar{g}_1^2 \, \bar{g}_2^2}{\bar{g}_1^2 + \bar{g}_2^2} \left[\frac{32}{9} \, n - 7 \right] \, \frac{1}{16 \pi^2 \, \epsilon}, \\
\Delta Z_{\hat{G}} &= \left[ \frac{11 \, N_c}{3} - \frac{2 \, n}{3} \right] \, \frac{\bar{g}_3^2}{16 \, \pi^2 \, \epsilon},
\end{align}
where  $n$ sums over the fermion generations and here we have introduced the notation
\begin{align}
Y &= \text{Tr}\left[N_c Y_u^\dagger Y_u + N_c Y_d^\dagger Y_d + Y_e^\dagger Y_e\right].
\end{align}
The use of the background field method leads to relationships between the finite and divergent counter-terms
in the theory, due to the unbroken background field symmetries. Some of these relationships are
\begin{align}
(\sqrt{Z_v}+ \frac{\Delta v}{\bar{v}^0_T} )_{div} &= \frac{\Delta Z_{\hat{\phi}_4}}{2}, \\
\Delta Z_{\hat{\mathcal{Z}} \,\hat{\mathcal{A}}} &= 0, \\
\Delta Z_{\hat{\mathcal{A}} \,\hat{\mathcal{A}}} &= - 2 \Delta Z_{e},\\
\Delta Z_{\hat{\mathcal{G}} \,\hat{\mathcal{G}}} &= - 2 \Delta Z_{g_3}.
\end{align}

For the finite terms we adopt
on-shell conditions fixing the mass and wavefunction finite terms,
as in the case of the BFM SM calculations in Ref.~\cite{Denner:1994xt}.
The mass counter-terms\footnote{Note our convention of adding the shift in the vev in this expression
functionally differs from Ref.~\cite{Denner:1994xt} which did not impose a FJ tadpole scheme and separated out the tadpole contributions
as a distinct (equivalent) contribution the the physical masses.}, are
\cite{Denner:1994xt,Corbett:2020ymv}
\begin{align}
\Delta Z_{m_{W}^2} &= {\rm Re} \left(\Sigma_T^{\hat{\mathcal{W}}\hat{\mathcal{W}}}(\bar{m}_{W}^2) \right)_{div}, \nn
&= \frac{\bar{g}_2^2 \,  (\xi + 3) \, (\bar{g}_1^2 + 3 \bar{g}_2^2) \bar{v}^2_T}{256 \pi^2 \epsilon}
+ \frac{4 \, \bar{g}_2^2}{3} \, \frac{\bar{m}_{W}^2}{16 \, \pi^2 \, \epsilon} \, n - \frac{43 \, \bar{g}_2^2 \,\bar{m}_{W}^2}{96 \, \pi^2 \, \epsilon}
-\sum_\psi \frac{N_C^\psi \, m_\psi^2 \, \bar{g}_2^2}{32 \pi^2 \epsilon}, \\
\Delta Z_{m_{Z}^2} &= {\rm Re} \left(\Sigma_T^{\hat{\mathcal{Z}}\hat{\mathcal{Z}}}(\bar{m}_{Z}^2) \right)_{div}, \nn
&= \frac{(\bar{g}_1^2+ \bar{g}_2^2) (\xi + 3) \, (\bar{g}_1+ 3 \, \bar{g}_2^2)\bar{v}_T^2}{256 \, \pi^2 \, \epsilon}
+ \frac{5 \bar{g}_1^4+3 \bar{g}_2^4}{\bar{g}_1^2+\bar{g}_2^2}\frac{\bar{m}_{Z}^2}{36\pi^2\epsilon}n
+ \frac{\bar{g}_1^4 - 43 \, \bar{g}_2^4}{\bar{g}_1^2 + \bar{g}_2^2}
\frac{\bar{m}_{Z}^2}{96 \, \pi^2 \, \epsilon}, \\
&-\sum_\psi N_C^\psi \, \frac{\hat{m}_\psi^2 \, (\bar{g}_1^2+\bar{g}_2^2)}{32 \pi^2 \epsilon}. \nonumber
\end{align}
The gauge dependent part of the mass counter terms are
\begin{align}
\Delta Z_{m_{W}^2}(\xi) &= \frac{\bar{g}_2^2 \, \bar{v}_T^2}{2}  \, \frac{(\bar{g}_1^2 + 3 \, \bar{g}_2^2)}{128 \pi^2 \epsilon} \, \xi, \\
\Delta Z_{m_{Z}^2}(\xi)  &= \frac{(\bar{g}_1^2+\bar{g}_2^2) \, \bar{v}_T^2}{2}  \, \frac{(\bar{g}_1^2 + 3 \, \bar{g}_2^2)}{128 \, \pi^2\, \epsilon} \, \xi,
\end{align}
which exactly cancels against
the gauge dependence in the net one point function contribution to the pole position of the bare propagator mass via
vev renormalization \cite{Denner:1994xt}
\bea
\langle \hat{\mathcal{V}} \hat{\mathcal{V}} \hat{\phi}_4  \rangle \, \bar{v}_T \, (\sqrt{Z_v}+ \frac{\Delta v}{\bar{v}_T} )_{div} = \langle \hat{\mathcal{V}} \hat{\mathcal{V}} \hat{\phi}_4  \rangle
 \, \bar{v}_T \, \frac{\bar{g}_1^2 + 3 \,\bar{g}_2^2}{128 \pi^2 \epsilon} \, \xi,
\eea
with $\langle \hat{\mathcal{W}} \hat{\mathcal{W}} \hat{\phi}_4 \rangle  = \bar{g}_2^2 \bar{v}_T/2$ and
$\langle \hat{\mathcal{Z}} \hat{\mathcal{Z}} \hat{\phi}_4  \rangle  = (\bar{g}_1^2 + \bar{g}_2^2) \bar{v}_T/2$.
The gauge dependence canceling also holds for the gauge dependent finite terms including the tadpole contributions
leading to a shift in the vev ($\Delta v$) effectively included in $\Gamma^{\hat{V},\hat{V}'}_{\mu \nu}(k)$.

\subsubsection{Renormalized masses, finite terms}

The sum of the radiative corrections to the $W$ and $Z$--bosons include a series of terms
that individually depend on the gauge parameter, but the
masses themselves, summing all such contributions (include tadpole contributions in the BFM), are gauge independent.
We have explicitly verified this is the case. We give the results for
$\xi=1$ for the one loop renormalized masses, purely for brevity of presentation.
The masses to one loop after $\rm \overline{MS}$ counter-term subtractions are
\begin{align*}
\frac{(\bar{m}_{Z}^{(r)})^2}{\bar{m}_{Z}^2} &\equiv 1 + \Delta  R_{m_{Z}^2}, &
\frac{(\bar{m}_{W}^{(r)})^2}{\bar{m}_{W}^2} &\equiv 1 + \Delta  R_{m_{W}^2},\nonumber\\
&=1+ 2 \, \frac{\Delta v}{\bar v_T}-\frac{\Delta M_Z^2}{\bar m_Z^2}, & &=1+ 2 \frac{\Delta v}{\bar v_T}-\frac{\Delta M_W^2}{\bar m_W^2},
\end{align*}
where $\Delta M_Z^2,\Delta M_W^2$ are reported in Appendix \ref{appendix}.
\subsubsection{Photon wavefunction and electric charge finite terms}
Using the BFM leads to several cross checks in calculations and relationships between
the divergent and finite parts of one loop counter-terms.
The Ward identities in the SMEFT in the BFM \cite{Corbett:2019cwl}
give the relation between the renormalization constants at one loop:
\bea\label{BFMphoton}
\Delta Z_e  &= - \frac{1}{2} \Delta Z_{\hat{\mathcal{A}}}, \nonumber \\
\Delta R_e  &= - \frac{1}{2} \Delta R_{\hat{\mathcal{A}}}.
\eea
Directly calculating the finite contributions to the photon wavefunction renormalization in the BFM
using the code package reported in Ref.~\cite{Corbett:2020bqv}
we find:
\bea
\Delta R_{\hat{\mathcal{A}}} = \frac{\bar{g}_1^2 \bar{g}_2^2}{(\bar{g}_1^2 + \bar{g}_2^2)} \,
\left[-\frac{7}{16 \pi^2} \log \left(\frac{\mu^2}{\bar{m}_W^2} \right)
+ \sum_{\psi} \frac{N^\psi_c Q^2_\psi}{12 \pi^2}
\,  \log \left(\frac{\mu^2}{\bar{m}_\psi^2} \right) - \frac{1}{24 \pi^2}\right].
\eea
This result can be directly compared to the matching onto $e$ from the SMEFT to the LEFT at one loop
using the BFM reported in Ref.~\cite{Dekens:2019ept}. This matching result is restricted to the
case of the top loop integrated out and is given as
\bea
\Delta R_e = \frac{\bar{g}_1^2 \bar{g}_2^2}{(\bar{g}_1^2 + \bar{g}_2^2)} \,
\left[\frac{7}{32 \pi^2} \log \left(\frac{\mu^2}{m_W^2} \right)
- \frac{N^{t}_c Q^2_{t}}{24 \pi^2}
\,  \log \left(\frac{\mu^2}{\bar{m}_t^2} \right) + \frac{1}{48 \pi^2}\right],
\eea
which satisfies the expected relationship given in Eqn.~\eqref{BFMphoton}.
Note that this result is also consistent with the onshell renormalization result reported
for these finite terms in Ref.~\cite{Sirlin:1980nh}.

As the electric coupling is extracted at scales $q^2 \rightarrow 0$, far below $\Lambda_{QCD}$
where the quarks have hadronized, it is useful to rearrange the finite terms into the
form
\bea
- i \, \left[\frac{4 \, \pi \, \hat{\alpha}(q^2)}{q^2}\right]_{q^2 \rightarrow 0}
\equiv \frac{- i \, (\bar{e}_0 + \Delta R_e)^2}{q^2} \left[1 +  {\rm Re} \frac{\Sigma^{AA}(\bar{m}_Z^2)}{\bar{m}_Z^2}
-\nabla \alpha \right]
\eea
where
\bea
\nabla \alpha = \left[\frac{{\rm Re} \Sigma^{AA}(\bar{m}_Z^2)}{\bar{m}_Z^2}  - \left[\frac{\Sigma^{AA}(q^2)}{q^2} \right]_{q^2 \rightarrow 0}\right].
\eea
The known SM results for  $\nabla \alpha$ can then be used as
$\nabla \alpha = \nabla \alpha_\ell + \nabla \alpha_t + \nabla \alpha_{had} + \nabla \alpha_{\rm \overline{MS} -os}$,
where $\nabla \alpha_\ell  = 0.03150$,  $\nabla \alpha_t = -0.0007$, $\nabla \alpha_{had} \approx 0.02764$ and $\nabla \alpha_{\rm \overline{MS} -os} \approx 0.0072$
\cite{Wells:2005vk,Agashe:2014kda,Baikov:2012zm,PhysRevD.22.971}.
The result for $\Sigma^{AA}(\bar{m}_Z^2)$ is purely transverse and is also required, explicitly
\begin{equation}
\begin{array}{rl}
\Sigma^{AA}_T&=-\frac{\bar g_1^2\bar g_2^2}{16\pi^2(\bar g_1^2+\bar g_2^2)}\left[7k^2\left(2+\log\left[\frac{\mu^2}{\bar m_W^2}\right]+{\rm Disc}[\sqrt{k^2},\bar m_W,\bar m_W]\right)+4\bar m_W^2(2+{\rm Disc}[\sqrt{k^2},\bar m_W,\bar m_W])\right]\\
&+\sum_\psi\frac{g_1^2g_2^2Q_\psi^2 N_c}{36(g_1^2+g_2^2)\pi^2k^2}\left[12\bar m_\psi^2+3{\rm Disc}[\sqrt{k^2},\bar m_\psi,\bar m_\psi](2\bar m_\psi^2+k^2)+k^2\left(5+3\log\left(\frac{\mu^2}{\bar m_\psi^2}\right)\right)\right].
\end{array}
\end{equation}
The ${\rm Disc}$ function is reported in Appendix \ref{appendix}.
\subsubsection{Higgs wavefunction finite terms}
Unlike the case of the photon, the Higgs wavefunction renormalization is not gauge invariant on its own in the BFM.
The combination of the Higgs wavefunction renormalization with a subclass of one one loop diagrams in each calculation,
and the one loop corrections to the vev, is however gauge independent, see Section \ref{gaugeparameter}. We present results for $\xi = 1$,
for the sake of brevity. The finite results for the Higgs wavefunction renormalization
in the BFM are \cite{Hartmann:2015oia}
\begin{align}
16 \, \pi^2 \,  \Delta R_{\phi_4} &=   2 \,  \lambda \, \Big(6  - \sqrt{3} \pi - \mathcal{J}_x[\bar{m}_Z^2] - 2 \, \mathcal{J}_x[\bar{m}_W^2] \Big)
+2 \, \bar{g}_2^2 \Bigg(
  \Big( \mathcal{J}_x[\bar{m}_W^2]- \frac{1}{2} \Big)  \,  \left(1 -\frac{3 \bar{m}_W^2}{\bar{m}_h^2} \right)  - \mathcal{I}[\bar{m}_W^2]\Bigg),
\nn
 & +\left(\sum_\psi \, y_\psi^2 \, N_c - \bar{g}_1^2-3  \bar{g}_2^2\right) \log
    \left(\frac{\bar{m}_h^2}{\mu ^2}\right)  +\left( \bar{g}_1^2+ \bar{g}_2^2\right) \, \Bigg( \Big( \mathcal{J}_x[\bar{m}_Z^2] - \frac{1}{2}\Big)\left(1 - \frac{3
   \bar{m}_Z^2}{\bar{m}_h^2}\right) -\mathcal{I}[\bar{m}_Z^2] \Bigg),
   \nn
   & +\sum_\psi \, y_\psi^2 \, N_c \Bigg(1+
   \left(1+\frac{2 m_\psi^2}{\bar{m}_h^2}\right)\, \mathcal{I}[m_\psi^2] -\frac{2 m_\psi^2}{\bar{m}_h^2} \,\log
   \left(\frac{m_\psi^2}{\bar{m}_h^2}\right) \Bigg).
\end{align}
The one loop functions $\mathcal{I},\mathcal{J}_x$ are defined in Appendix \ref{appendix}.

\subsection{vev dependence, tadpoles and $G_F$ extractions}\label{vev}

\subsubsection{One loop vev definition and tadpole scheme}
The tree level vacuum expectation value of the Higgs is defined by the potential with convention
\bea
\mathcal{L}_{SM}^V = - \lambda \left(H^\dagger \, H - \frac{v^2}{2} \right)^2.
\eea
In the SMEFT the tree level vacuum expectation value of the Higgs is augmented to include corrections due to
the tower of higher dimensional operators, introducing $\bar{v}_T$ as the tree level vacuum expectation value \cite{Alonso:2013hga}.
An all orders solution of $\bar{v}_T$ in the SMEFT in terms of Wilson coefficients is given in Ref.~\cite{Hays:2020scx}.

The parameter $\bar{v}_T \equiv \sqrt{\langle 2 H^\dagger H \rangle}$ in the SMEFT is defined as the minimum of the potential, including
corrections due to higher-dimensional operators.
Our approach to tadpole corrections is to directly calculate the tadpoles and then use the
FJ tadpole scheme \cite{Fleischer:1980ub} as a conventional choice
for defining the contributions to $S$ matrix elements.\footnote{For further discussion on this scheme see Ref.~\cite{Denner:1994xt,Denner:2018opp,Denner:2019vbn}.}

The one loop correction ($\Delta v$) to the vacuum expectation value is fixed
by the condition that the one point function of the Higgs field vanishes, including the factor of $\Delta v$.
$\Delta v$ is defined by the condition $T=0$ on \cite{Hartmann:2015oia}
\bea
T &=&   \, \bar{m}_{h}^2 \, h \,\bar{v}_T \, \frac{1}{16\pi^2}  \left[- 16 \pi^2 \, \frac{\Delta v}{\bar{v}_T}   + 3  \, \lambda \left(1+ \log \left[\frac{\mu^2}{\bar{m}_{h}^2} \right] \right) + \frac{1}{4} \,  \bar{g}_2^2 \,
\xi   \left(1+ \log \left[\frac{\mu^2}{\xi \, \bar{m}_{W}^2} \right] \right)  \right.,  \\
&\,& \hspace{2.3cm} \left. + \frac{1}{8}(\bar{g}_1^2+ \bar{g}_2^2) \, \xi \, \left(1+ \log \left[\frac{\mu^2}{\xi \, \bar{m}_{Z}^2} \right] \right)
- 2 \sum_\psi y_\psi^2  \, N_c \frac{m_\psi^2}{\bar{m}_{h}^2}\left(1+ \log \left[\frac{\mu^2}{m_\psi^2} \right] \right)  \right., \nn
&\,&  \hspace{2.3cm} \left.  + \frac{\bar{g}_2^2}{2} \frac{\bar{m}_{W}^2}{\bar{m}_{h}^2} \left(1+ 3\log \left[\frac{\mu^2}{\bar{m}_{W}^2} \right] \right)
 +\frac{1}{4} (\bar{g}_1^2 + \bar{g}_2^2) \frac{\bar{m}_{Z}^2}{\bar{m}_{h}^2} \left(1+ 3\log \left[\frac{\mu^2}{\bar{m}_{Z}^2} \right] \right) \right]. \nonumber
\eea
Due to this choice we can neglect the further effects of tadpole diagrams, following Ref.\cite{Fleischer:1980ub,Denner:1994xt}.
The cost of this approach is including the explicit one loop result for $\Delta v$ in predictions.
Note the dependence of $\xi$ in the vev at one loop, indicating that it must be combined with the one loop gauge dependent
terms in the Higgs wavefunction renormalization factor, or the gauge dependent two point functions, to
obtain a gauge independent result.\footnote{Recently the one loop tadpole in this calculational scheme was reported at all orders
in the $\bar{v}_T/\Lambda$ expansion in Ref.~\cite{Corbett:2021jox}.}

\subsubsection{$G_F$ extraction one loop results}
The numerical extraction of the vev is through $\mu^- \rightarrow e^- + \bar{\nu}_e + \nu_\mu$,
which also has one loop corrections.
The proper approach to using such a measurement for SMEFT applications is to match to the
effective theory when the top, Higgs, $W,Z$ are integrated out in sequence, defining the low
energy effective field theory (LEFT). Subsequently the
one loop improvement of the extraction of $G_F$ in the LEFT is done, with a chosen tadpole scheme and gauge
fixing.
This calculation in defining and matching to the LEFT at one loop order was recently
reported in Refs.~\cite{Jenkins:2017jig,Jenkins:2017dyc,Dekens:2019ept}. We use these results in this work.
Using the operator basis defined in \cite{Jenkins:2017jig}, the relevant one loop matching is onto
the operator

\bea
\mathcal{L}_{LEFT} \supset L^{V,LL} (\bar{\nu}_{L, \mu} \gamma^\mu \nu_{L, e})(\bar{e}_L \gamma_\mu \mu_L).
\eea
The one loop matching to $\Delta L^{V,LL}$ in the BFM was
reported in Ref.~\cite{Dekens:2019ept} as
\bea
\bar{v}_T^2 \, \Delta L^{V,LL} &=&
\frac{\left(7 \bar{m}_{h}^4+\bar{m}_{h}^2
   \left(2 m_t^2 \, N_c -5 \left(2 \bar{m}_{W}^2+\bar{m}_{Z}^2\right)\right)+4
   \left(-4 \, m_t^4 \, N_c +2 \,\bar{m}_{W}^4+ \bar{m}_{Z}^4 \right)\right)}{16 \pi ^2 \,
  \bar{m}_{h}^2 \, \bar{v}_T^2}, \nn
&+&\frac{3 (\bar{m}_{h}^4 - 2 \bar{m}_{h}^2 \, \bar{m}_{W}^2)}{8 \pi^2 \, \bar{v}_T^2 (\bar{m}_{h}^2 -\bar{m}_{W}^2)}
  \log \left(\frac{\mu^2}{\bar{m}_{h}^2} \right)
+ \frac{m_t^2 \,N_c \,(\bar{m}_{h}^2 - 4 m_t^2)}{4 \pi^2 \bar{m}_{h}^2 \, \bar{v}_T^2}
\log \left(\frac{\mu^2}{m_t^2} \right), \nn
&+&\frac{3((\bar{m}_{h}^2(\bar{m}_{Z}^4-2 \bar{m}_{W}^2 \bar{m}_{Z}^2) + 2 \bar{m}_{Z}^4 (\bar{m}_{W}^2-\bar{m}_{Z}^2))}
{8 \pi^2 \bar{m}_{h}^2  \bar{v}_T^2 (\bar{m}_{W}^2-\bar{m}_{Z}^2)}
\log \left(\frac{\mu^2}{\bar{m}_{Z}^2} \right), \\
&-& \frac{3 \, \bar{m}_{W}^2 \left(\bar{m}_{h}^4 \left(\bar{m}_{W}^2 -2 \bar{m}_{Z}^2\right)+\bar{m}_{h}^2
   \left(7 \,\bar{m}_{W}^2  \bar{m}_{Z}^2-6 \bar{m}_{W}^4\right)+4 \bar{m}_{W}^4
   ( \bar{m}_{W}^2 - \bar{m}_{Z}^2 )\right)}{8 \pi ^2\bar{m}_{h}^2 \bar{v}_T^2
   (\bar{m}_{h}^2-\bar{m}_{W}^2)(\bar{m}_{W}^2-\bar{m}_{Z}^2)}
\log \left(\frac{\mu^2}{\bar{m}_{W}^2} \right). \nonumber
\eea
Here we have set the evanescent scheme parameter in this result
(bEvan$=1$) to be consistent with naive tree level Fierz identities used in the matching.
The matching result from the SMEFT to the LEFT onto this operator is gauge independent in the BFM \cite{Dekens:2019ept}
with the chosen tadpole scheme. We have suppressed the flavour rotation matrices.
\begin{figure}[ht!]
\includegraphics[width=0.25\textwidth]{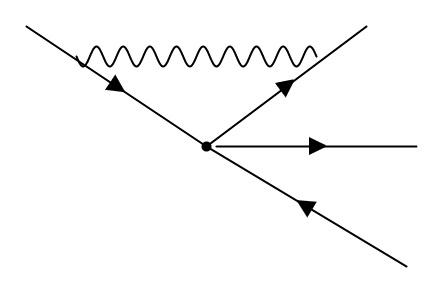}
\caption{One loop QED contribution to $\mu^- \rightarrow e^- + \bar{\nu}_e + \nu_\mu$.}
\label{fig:1}
\end{figure}
Both the full and effective theory contains a photon loop that cancels in the matching.
The matrix element to one loop for the parameter extraction
includes a photon loop in the LEFT ($\Delta L_{ew}^{V,LL}$),
see Fig.~\ref{fig:1}, that is consistent with Refs.~\cite{Kallen:1968gaa,Green:1980bd}
giving
\bea
\Delta L_{ew}^{V,LL} &=& -\frac{\alpha}{4 \pi}\left(\pi^2 - \frac{25}{4}\right).
\eea
 Adding all
of these results together, and including the leading matching in the SMEFT onto the LEFT,
the net result is
\bea
L^{V,LL} &\equiv& - \frac{4 \, \hat{G}_F}{\sqrt{2}}, \\
 &=& -\frac{2}{\bar{v}_T^2}\left(1 +\Delta L_{ew}^{V,LL}\right)
+ \Delta L^{V,LL} - 2 \sqrt{2} \, \frac{\delta G_F}{\bar{v}_T^2}. \nonumber
\eea
Here $\hat{G}_F$ is the measured value of the Fermi constant, and to $\mathcal{L}^{(6)}$
\bea
\delta G_F^{(6)} \equiv  \frac{1}{\sqrt{2}}
\left(\tilde{C}^{(3)}_{\substack{Hl \\ ee }} +  \tilde{C}^{(3)}_{\substack{Hl \\ \mu\mu }}
- \frac{1}{2}\left(\tilde{C}'_{\substack{ll \\ \mu ee \mu}} +  \tilde{C}'_{\substack{ll \\ e \mu\mu e}}\right)\right).
\eea
The notation is consistent with Refs.~\cite{Brivio:2017btx,Brivio:2019myy}.
Considering both loop corrections and SMEFT perturbations replacing the vev
in terms of the input parameter $\hat{G}_F$ is given by
\bea
\bar{v}_T = \hat{v}_T \left(1 + \Delta G_F + \frac{\delta G_F^{(6)}}{\sqrt{2}}  \right),
\label{eq:deltaGFdefn}
\eea
where $\hat{v}_T^2 \equiv 1/(\sqrt{2} \, \hat{G}_F)$ and we define
\bea
\Delta G_F = - \frac{\bar{v}_T^2}{4} \Delta L^{V,LL} + \frac{\Delta L_{ew}^{V,LL}}{2}.
\eea
This result affords a cross check in an observable to observable relationship.
Utilizing a FJ tadpole scheme, as in the calculations reported above, one expects
large perturbative corrections $\propto m_t^4$ desending
from finite terms in the tadpoles shifting the vev to cancel between observables \cite{Hartmann:2015oia}.
We find that the large $m_t^4$ contribution exactly cancels as expected via the appearance in $M_1$
(defined below) of a term
\bea
\frac{\Delta v}{\bar{v}_T} \propto - \frac{2 y_t^2}{16 \pi^2} N_C \frac{m_f^2}{\bar{m}_h^2} \left[1+ \log \left(\frac{\mu^2}{m_f^2}
\right) \right].
\eea
Further, as the observables we consider are $\propto \bar{v}_T$, it follows that the observables are proportional to
\bea
\bar{v}_T = \hat{v}_T \left[1 +\frac{2 y_t^2}{16 \pi^2} N_C \frac{m_f^2}{\bar{m}_h^2} \left[1+ \log \left(\frac{\mu^2}{m_f^2}\right)\right] + \cdots \right].
\eea
This cancelation also extends to several other gauge independent terms in $\Delta v/v$ and $\Delta L^{V,LL}$,
affording a non-trivial cross check of the results.
\subsubsection{Gauge parameter independence}\label{gaugeparameter}
All of the observables we consider are linear in the Higgs field and the vev.
As such, a common cancelation of gauge dependence occurs in the one loop matrix element in each case.
Consider the relevant terms in the geoSMEFT organization of the SMEFT
\bea
\mathcal{L}_{SMEFT} \supset -\frac{1}{4} \, k_{\mathpzc{AB}} \, \mathcal{G}_{\mu \nu}^{\mathpzc{A}} \, \mathcal{G}^{\mu \nu,\mathpzc{B}}
-\frac{1}{4} \, g_{AB} \, \mathcal{W}_{\mu \nu}^A \, \mathcal{W}^{\mu \nu,B}
\eea
For the observables considered  the tree level three-point amplitudes are
\begin{align}\label{threepointamps}
\langle \mathcal{G}(p_1) \mathcal{G}(p_2)| \phi_4 \rangle  &= - \langle \, \mathcal{G}^{\mu\nu} \mathcal{G}_{\mu \nu} \phi_4\rangle \frac{\sqrt{h}^{44}}{4} \,
\langle \frac{\delta \kappa(\phi)}{\delta \phi_4} \rangle  \sqrt \kappa^2, \\
\label{eq:hggvertex}
\langle \phi_4 |\mathcal{A}(p_1) \mathcal{A}(p_2) \rangle &= -\langle \phi_4 \mathcal{A}^{\mu\nu} \mathcal{A}_{\mu \nu} \rangle \frac{\sqrt{h}^{44}}{4} \left[
	\langle \frac{\delta g_{33}(\phi)}{\delta \phi_4}\rangle \frac{\overline{e}^2}{g_2^2}  +
 2\langle \frac{\delta g_{34}(\phi)}{\delta \phi_4}\rangle \frac{\overline{e}^2}{g_1 g_2}  +
  \langle \frac{\delta g_{44}(\phi)}{\delta \phi_4}\rangle \frac{\overline{e}^2}{g_1^2}
	\right]. \nn
\end{align}
The photon field is canonically normalized in the process of rotating to the mass eigenstate field in the geoSMEFT,
introducing coupling dependence, while the gluon field is directly canonically normalized by $\kappa^2$.
The cancellation in gauge dependence at one loop occurs in a common fashion. Denoting a general gauge field with $F$,
and a general variation of a field space metric with respect to the Higgs field with $\delta M_{AB}/\delta \phi_4$,
one has the tree level dependence
\bea
\langle \phi_4 | F(p_1) F(p_2) \rangle^0 \propto \langle \phi_4 F^{\mu\nu} F_{\mu \nu} \rangle^0  \langle \frac{\delta M_{AB}(\phi)}{\delta \phi_4}\rangle^0,
\eea
with $\langle \phi_4 F^{\mu\nu} F_{\mu \nu} \rangle^0 = - 4	\, (p_1 \! \cdot \! p_2 \epsilon_1 \! \cdot \! \epsilon_2 - p_1 \! \cdot \! \epsilon_2 p_2 \! \cdot \! \epsilon_1 )$
and $\langle \delta M_{AB}/\delta \phi_4\rangle^0 \propto \bar{v}_T$. Here we have started the practice of using
$0$,$1$ superscripts for tree level results and one loop results, in a slight overuse of notation as
$0$ also corresponds to bare parameters.

\begin{figure}[ht!]
\includegraphics[width=0.65\textwidth]{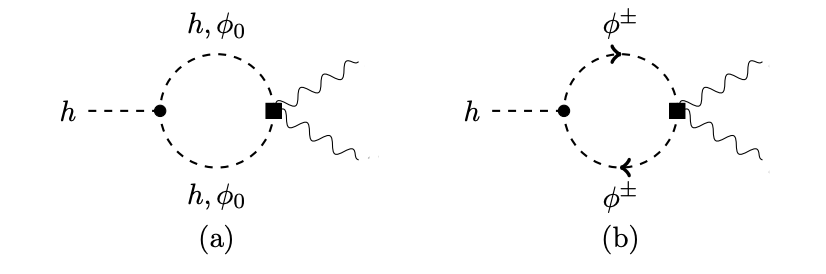}
\caption{One loop contributions to $\langle \phi_4|\, F \, F\rangle
\langle \frac{\delta M_{AB}}{\delta \phi_4} \rangle$.}
\label{fig:2a}
\end{figure}

A one-loop improvement of this result for EW corrections is given by introducing the wavefunction renormalization of the Higgs field,
introducing $\Delta v$, and calculating the one loop matrix element, which includes the diagrams
in Fig.~\ref{fig:2a}.
The sum of these terms is given by
\bea
\frac{\langle \phi_4 F(p_1) F(p_2) \rangle^1}{ \langle \phi_4 F^{\mu\nu} F_{\mu \nu} \rangle^0  \langle\frac{\delta M_{AB}(\phi)}{\delta \phi_4}\rangle^0}
&\propto& M_1
\eea
where
\bea
M_1 &\equiv
\left(\frac{\Delta R_h}{2} + \frac{\Delta v}{v}
+ \frac{(\sqrt{3} \pi - 6) \lambda}{ 16 \pi^2}
+ \frac{1}{16 \pi^2}\left(\frac{\bar{g}_1^2}{4} + \frac{3 \bar{g}_2^2}{4} + 6 \lambda \right)
\log\left[\frac{\bar{m}_h^2}{\mu^2}\right]\right), \nn
&+ \frac{1}{16 \pi^2}
\left(\frac{\bar{g}_1^2}{4} \mathcal{I}[\bar{m}_Z] + (\frac{\bar{g}_2^2}{4} + \lambda) (\mathcal{I}[\bar{m}_Z] + 2 \mathcal{I}[\bar{m}_W])\right).
\label{eq:M1defn}
\eea
Here the result is reported for $\xi =1$ for brevity of presentation. It was explicitly confirmed
that the sum of terms is gauge independent in Ref.~\cite{Hartmann:2015oia}.
For all three processes we consider, the gauge dependence in the EW corrections cancels in a common fashion
through this combination of terms.

It is important to note that all of the finite terms required for relating input parameters to one loop
in the SM, when using the BFM, and the chosen tadpole scheme, are individually gauge invariant.
As a result, different combinations of couplings that multiply the gauge invariant operators in normalization choices
do not introduce a gauge dependence to cancel against process specific matrix elements. In this sense,
the input parameter loop corrections ``factorize" in a gauge independent manner in this calculational scheme.
This is useful feature of this calculational scheme, that aids in the transparency of the results.
This fact also makes this calculational scheme have advantages for coding implementation.

Consider the case of $\Gamma(h \rightarrow \gamma \gamma)$ to illustrate this point.
The following finite terms depend on the the normalization choice of the $\mathcal{L}^{(6)}$ operators.
Expanding the three-point function $\langle \phi_4 \mathcal{A}(p_1) \mathcal{A}(p_2) \rangle$ to $\mathcal{L}^{(6)}$ one has
\begin{align}
	\label{eq:hggDim6}
\langle \phi_4|\mathcal{A} \mathcal{A} \rangle_{\mathcal{L}^{(6)}}^0 &= \frac{\langle \phi_4 \mathcal{A}^{\mu\nu} \mathcal{A}_{\mu \nu} \rangle_0}{\bar{v}_T} \left[\frac{g_2^2 \, \tilde C_{HB}^{(6)}
  + g_1^2 \, \tilde C_{HW}^{(6)} - g_1 \, g_2 \, \tilde C_{HWB}^{(6)}}{({g}^{\rm SM}_Z)^2} \right],
\end{align}
where $({g}^{\rm SM}_Z)^2 \equiv g_1^2+g_2^2$. In Ref.~\cite{Hartmann:2015oia} the choice was made to rescale
each operator by the power of gauge couplings corresponding to the field strengths present.
This normalization choice at $\mathcal{L}^{(6)}$ is indicated with an additional $N_1$ superscript, and leads to
\begin{align}
\langle \phi_4|\mathcal{A} \mathcal{A} \rangle_{\mathcal{L}^{(6)}}^{0,N_1} &= \frac{\langle \phi_4 \mathcal{A}^{\mu\nu} \mathcal{A}_{\mu \nu} \rangle_0}{\bar{v}_T} \bar{e}^2 \left[\tilde C_{HB}^{(6)}
  + \tilde C_{HW}^{(6)} - \tilde C_{HWB}^{(6)} \right].
\end{align}
With this choice, the dependence on $\Delta R_A,\Delta R_e$ cancel in the BFM,
as they dress the $S$ matrix element as
\bea
(1+ \Delta R_A) \, (1+ \Delta R_e)^2 \,\left(1 + \Delta G_F
\right)\,  \langle \phi_4|\mathcal{A}(p_1) \mathcal{A}(p_2) \rangle_{\mathcal{L}^{(6)}}^{0,N_1}.
\eea
The resulting $\Delta R_A + 2 \Delta R_e =0$ cancelation
is due to the unbroken background field symmetries, but is not required
to perform such operator rescalings for a gauge invariant result.
One can also choose to retain the normalization of Eqn.~\eqref{eq:hggDim6}. In this case, the one-loop finite terms due to mapping
to input parameters the gauge couplings multiplying each operator are given by
\bea
\langle \phi_4 | [C^{(6)}_{HB}] |\mathcal{A} \mathcal{A}\rangle^1 = \left(\Delta R_A + \Delta G_F + \frac{2 \hat{g}_1 (\Delta g_2 \hat{g}_1 - \Delta g_1 \hat{g}_2)}
{\hat{g}_2 \, (\hat{g}^{\rm SM}_Z)^2} \right)
\frac{\hat{g}_2^2 \,C^{(6)}_{HB}}{(\hat{g}^{\rm SM}_Z)^2} \frac{\hat{v}_T}{\Lambda^2}
\langle \phi_4  \mathcal{A}^{\mu\nu} \mathcal{A}_{\mu \nu} \rangle_0,
\eea
\bea
\langle \phi_4 |[C^{(6)}_{HW}] |\mathcal{A} \mathcal{A}\rangle^1 = \left(\Delta R_A + \Delta G_F
+ \frac{2 \hat{g}_2 (\Delta g_1 \hat{g}_2 - \Delta g_2 \hat{g}_1)}
{\hat{g}_1 \, (\hat{g}^{\rm SM}_Z)^2} \right)
\frac{\hat{g}_1^2 \,C^{(6)}_{HW}}{(\hat{g}^{\rm SM}_Z)^2} \frac{\hat{v}_T}{\Lambda^2}
\langle \phi_4 \mathcal{A}^{\mu\nu} \mathcal{A}_{\mu \nu} \rangle_0,
\eea
\bea
\langle \phi_4 | [C^{(6)}_{HWB}] |\mathcal{A} \mathcal{A}\rangle^1 = -\left(\Delta R_A + \Delta G_F
+ \frac{(\hat{g}_1^2-\hat{g}_2^2) (\Delta g_2 \hat{g}_1-\Delta g_1 \hat{g}_2)}{\hat{g}_1 \hat{g}_2 (\hat{g}^{\rm SM}_Z)^2}\right)
\frac{\hat{g}_1 \, \hat{g}_2 \,C^{(6)}_{HWB}}{(\hat{g}^{\rm SM}_Z)^2} \frac{\hat{v}_T}{\Lambda^2}
\langle \phi_4 \mathcal{A}^{\mu\nu} \mathcal{A}_{\mu \nu} \rangle_0. \nn
\eea
In this case, the finite terms of the calculation relating Lagrangian parameters to input measurements at one loop
are more complex in structure. Nevertheless, all of the $\Delta$
corrections at one loop in the BFM are individually gauge independent. This is the case in both the $\{\hat{\alpha}_{ew}, \hat{m}_Z, \hat{G}_F \}$
and $\{\hat{m}_{W}, \hat{m}_Z, \hat{G}_F \}$ input parameter schemes. The explicit expressions for $\Delta g_i$
are easily derived by linear perturbation theory on the input parameters defining conditions in each scheme.

Explicitly in the $\{\hat{m}_{W}, \hat{m}_Z, \hat{G}_F \}$ scheme one has
\begin{align}
\frac{\Delta g_1}{\hat{g}_1} &= \frac{\Delta G_F}{2}+ \frac{\Delta R_{m_W} \, \hat{m}_W^2- \Delta R_{m_Z} \, \hat{m}_Z^2}{\hat{m}_W^2-\hat{m}_Z^2}, \\
\frac{\Delta g_2}{\hat{g}_2} &= \frac{\Delta G_F}{2} + \Delta R_{m_W}.
\end{align}
In the $\{\hat{\alpha}_{ew}, \hat{m}_Z, \hat{G}_F \}$ scheme one has
\begin{align}
\frac{\Delta g_1}{\hat{g}_1} &= \frac{\Delta e}{\hat{e}} -\frac{\Delta c_\theta}{c_{\hat{\theta}}}, \\
\frac{\Delta g_2}{\hat{g}_2} &= \frac{\Delta e}{\hat{e}} -\frac{\Delta s_\theta}{s_{\hat{\theta}}},
\end{align}
where
\bea
\frac{\Delta s_\theta}{s_{\hat{\theta}}} = \frac{1 - s_{\hat{\theta}}^2}{2 \, (1 - 2 s_{\hat{\theta}}^2)} \left[ \frac{\Delta \alpha}{\alpha}
-  \Delta G_F -2 \, \Delta R_{M_Z} \right].
\eea
We again emphasize the benefit of using the FJ tadpole scheme in the SMEFT and the BFM,
which leads to the $\Delta$ corrections all being individually gauge independent.
Alternate tadpole schemes can be problematic when $\rm \overline{MS}$ subtraction is used, see
Ref.~\cite{Denner:2016etu} for a more detailed discussion. As the $\mathcal{L}^{(6)}$
operators of the SMEFT are renormalized in a gauge independent manner in $\rm \overline{MS}$
using background field gauge in Refs.~\cite{Jenkins:2013zja,Jenkins:2013wua,Alonso:2013hga},
in our view, it is appropriate to use the FJ tadpole scheme and the BFM
in SMEFT calculations.

\section{Three processes to $\mathcal{O}(\bar{v}_T^2/16 \pi^2 \Lambda^2)$ and $\mathcal{O}(\bar{v}_T^4/\Lambda^4)$
 consistently in geoSMEFT}\label{twopoint}
\subsection{$\sigma(\mathcal{G} \,\mathcal{G} \rightarrow h)$}\label{gghloop}

We can group the full amplitude for $\mathcal{G} \,\mathcal{G} \to h$ in the following way:
\begin{align}
\mathcal \mathcal{A}_{\mathcal{G} \mathcal{G}h} = \mathcal \mathcal{A}^{\mathcal{G} \mathcal{G}\phi_4}_{SM} + \langle \mathcal{G} \mathcal{G}|\phi_4 \rangle^0_{\mathcal O(v^2/\Lambda^2)} + \langle \mathcal{G} \mathcal{G}|\phi_4 \rangle^1_{\mathcal O(v^2/\Lambda^2)} +  \langle \mathcal{G} \mathcal{G}|\phi_4 \rangle^0_{\mathcal O(v^4/\Lambda^4)} + \cdots
\end{align}
the SM contribution, $\mathcal O(v^2_T/\Lambda^2)$ contributions at tree and loop order, and the $\mathcal O(v^4_T/\Lambda^4)$ contributions respectively.

\subsubsection{SM Result}
The leading result is already at first order in the loop expansion and is given by
\begin{align}
\mathcal{A}^{\mathcal{G} \mathcal{G}\phi_4}_{SM} = -\sum_f \frac{\alpha_s [Y]_{ff}/\sqrt 2}{16\pi \hat M_f} A_{1/2}(\tau_f)\langle \mathcal G^{\mu \nu} \mathcal G_{\mu \nu} \phi_4 \rangle_0,
\label{eq:SMggh}
\end{align}
where $ [Y]_{ff}/\sqrt{2} = [\hat{M}_f]/\bar{v}_T$ and $\tau_p = 4 m_p^2/\bar{m}_h^2$. Here $f$ sums over all coloured fermions in the mass eigenstates.
We have introduced the function~\cite{Shifman:1979eb,Bergstrom:1985hp}
\begin{align}
A_{1/2}(\tau_p)=- 2 \tau_p \left[1+ (1- \tau_p)f(\tau_p)\right],
\label{eq:A12}
\end{align}
where $f(\tau_p)$ is defined in Appendix \ref{appendix}.

\subsubsection{$\langle \mathcal{G} \mathcal{G}|\phi_4 \rangle^0_{\mathcal{L}^{(6)}}$}
The general geoSMEFT result for the $\mathcal{G} \mathcal{G}\phi_4$ three-point function at tree level is
\bea
\mathcal \mathcal{A}_{SMEFT,0}^{\mathcal{G}\mathcal{G}\phi_4} = - \frac{\sqrt{h}^{44} \sqrt \kappa^2}{4} \,
\langle \frac{\delta \kappa(\phi)}{\delta \phi_4} \rangle \, \langle \mathcal G^{\mu \nu} \mathcal G_{\mu \nu} \phi_4 \rangle _0,
\label{eq:hggvertex}
\eea
Here $\delta \kappa(\phi)/\delta \phi_4$ indicates the variation of the Higgs-gluon field space connection
with respect to $\phi_4$, and $\langle \rangle$ indicates the expectation value of the composite operator
forms for the $\mathcal{G} \,\mathcal{G}h$ amplitude. The result in Eqn.~\eqref{eq:hggvertex} contains an infinite tower of
higher order corrections in the $\bar{v}_T^2/\Lambda^2$ expansion. For $\mathcal{G} \,\mathcal{G}h$ we have
\bea
\langle \frac{\delta \kappa(\phi)}{\delta \phi_4} \rangle = -\frac{ 4}{\bar{v}_T}
\sum_{n=0}^\infty \tilde{C}_{HG}^{(6+2n)}.
\eea
Which, expanded out to $\mathcal O(v^2_T/\Lambda^2)$, gives
\begin{align}
\langle \mathcal{G} \mathcal{G}| \phi_4 \rangle^0_{\mathcal O(v^2/\Lambda^2)} =\, \frac{\tilde C^{(6)}_{HG}}{\bar{v}_T} \langle \mathcal G^{\mu \nu} \mathcal G_{\mu \nu} \phi_4 \rangle _0.
\label{eq:gghdim6tree}
\end{align}

\subsubsection{$\langle  \mathcal{G} \mathcal{G}|\phi_4 \rangle^1_{\mathcal{L}^{(6)}}$}
For the one loop matrix element, there are QCD and EW one loop corrections to the matrix element Eq.~\eqref{eq:gghdim6tree},
one loop operator mixing effects, plus $\mathcal O(v^2_T/\Lambda^2)$ corrections to the (one-loop) SM result.

We express these various terms as
\bea
\langle \mathcal{G} \mathcal{G}|\phi_4\rangle^1_{\mathcal{L}^{(6)}} =
\langle \mathcal{G} \mathcal{G} |[\tilde C^{(6)}_{HG}] |\phi_4 \rangle^1 +
\left(\frac{\tilde C_i \, f_i}{16 \, \pi^2 \, \hat v^2_T}\right) \, \hat{v}_T  \, \langle \mathcal{G}^{\mu\nu} \mathcal{G}_{\mu \nu} \phi_4  \rangle^0.
\eea
where $\langle \mathcal{G} \mathcal{G} |[\tilde C^{(6)}_{HG}] | \phi_4 \rangle^1$ contains the loop corrections to Eq.~\eqref{eq:gghdim6tree}
\begin{eqnarray}
\langle \mathcal{G} \mathcal{G} |[\tilde C^{(6)}_{HG}] | \phi_4 \rangle^1 &=& [\Delta G_F + M_1]\, \frac{\tilde C_{HG}^{(6)}}{\hat v_T} \langle \mathcal G^{\mu \nu} \mathcal G_{\mu \nu} \phi_4 \rangle_0
+  \frac{\tilde C_{HG}^{(6)}}{\hat v_T} \, \langle \mathcal G^{\mu \nu} \mathcal G_{\mu \nu} \phi_4 \rangle^1_{\alpha_s},
\label{eq:gghdim6loop}
\end{eqnarray}
and $\tilde C_i\, f_i$ contains all corrections -- from operator mixing and  $\mathcal O(\bar v^2_T/\Lambda^2)$ corrections to the SM -- that are not proportional to $\tilde C^{(6)}_{HG}$. In Eq.~\eqref{eq:gghdim6loop}, we have separated the EW effects (first term) from the QCD effects (second term, denoted with $\alpha_s$ subscript) as the latter have IR issues we will address shortly; $\Delta G_F$ and $M_1$ have been defined in Eq.~\eqref{eq:deltaGFdefn} and Eq.~\eqref{eq:M1defn} respectively.

The $\mathcal O(v^2_T/\Lambda^2)$ corrections to the SM loop come from corrections to the Yukawa coupling. Specifically, generalizing Eq.~\eqref{eq:SMggh}, to $[Y]_{ff}/\sqrt{2} \rightarrow {\rm Re}[\mathcal{Y}]_{ff}$ in
geoSMEFT  \cite{Alonso:2013hga}
\bea\label{genyukawa}
[\mathcal{Y}]_{ff} =\frac{\sqrt{h}^{44} \, [\hat{M}_f]}{\bar{v}_T} - \frac{\sqrt{h}^{44}}{\sqrt{2}} \sum_{n=0}^\infty \frac{n+1}{2^n} \tilde{C}_{\substack{\psi' H \\ ff}}^{\star, (6+2n)},
\label{eq:Yukawa}
\eea
with $\psi' = \{u,d,e\}$. Here the definition of the Yukawa coupling has convention
$\mathcal{L} = - h \bar{\psi'}_r [\mathcal{Y}]_{rs} P_L \, \psi_s+ h.c.$ where $\psi = \{q,l\}$.
The square root of the expectation of the background field Higgs field space connection
$\sqrt{h}^{44}$ is defined at all orders in the $\bar{v}_T/\Lambda$
expansion in Ref.~\cite{Helset:2020yio}. Since the Yukawa interaction enters at one loop level,
we restrict our results to $\mathcal O(\bar v^2_T/\Lambda^2)$ pieces in Eq.~\eqref{eq:Yukawa},
\bea
[\mathcal{Y}]_{\mathcal{O}(\bar{v}_T^2/\Lambda^2)}^{ff} = \frac{\hat{M}_f}{\hat{v}_T} \left(\tilde C_{H\Box}^{(6)} - \frac{1}{4}\tilde C_{HD}^{(6)} - \frac{\delta G_F^{(6)}}{\sqrt{2}} \right)
- \frac{1}{\sqrt{2}} \tilde{C}_{\substack{\psi' H \\ ff}}^{\star, (6)} + \cdots
\eea
so that
\bea
\label{eq:cifi}
f_{H\Box} &=& -\sum_f \, \alpha_s \pi \,  A_{1/2}(\tau_f), \\
f_{HD} &=& \frac{1}{4} \, \sum_f \, \alpha_s \pi \,  A_{1/2}(\tau_f), \\
f_{\delta G_F} &=& \frac{1}{\sqrt{2}} \, \sum_f \, \alpha_s \pi \, A_{1/2}(\tau_f), \\
\left[Y_{\psi'_{ff}}\right] \, f_{\substack{\psi' H \\ ff}} &=&  \sum_f \, \alpha_s \pi \,  A_{1/2}(\tau_f).
\eea
In practice, contributions from light fermions to the $f_i$ are suppressed since $A_{1/2}(\tau_f \ll 1) \sim \tau \sim \left[Y_{\psi'_{ff}}\right]^2 $, so we will only include effects from the top and bottom quarks.

There are no corrections due to the top/Higgs mass, as we consider the masses as input parameters.
In addition in principle there are a set of unknown corrections to the extraction of $\alpha_s$
in the SMEFT, that are necessarily neglected as they are unknown at this time.

The other $\tilde C_i f_i$ piece comes from chromomagnetic dipole, which mixes with $\tilde C^{(6)}_{HG}$ at one-loop. The UV divergent piece of the dipole loop is cancelled
by a SMEFT $\tilde C^{(6)}_{HG}$ counter-term \cite{Jenkins:2013zja,Jenkins:2013wua,Alonso:2013hga}, leaving a finite remnant.
As the dipoles enter at one loop, the only term which enter at $\mathcal O(v^2_T/16\pi^2\Lambda^2)$ (again retaining only the top and bottom quark terms)
are the dimension six operators $\tilde{C}_{\substack{uG\\ tt}}$ and $\tilde{C}_{\substack{dG\\ bb}}$:
\bea\label{oneloopfirstterm}
\langle \mathcal{G} \mathcal{G}|\phi_4 \rangle^1_{\mathcal O(v^2/\Lambda^2)} \supset  \frac{\sqrt{\kappa} \, \bar{g}_3}{32\pi^2 \, \bar{v}_T}
 \, \left(f_{uG}\, \tilde{C}_{\substack{uG\\ tt}}[Y]_{tt}+ f_{dG}\, \tilde{C}_{\substack{dG\\ bb}}[Y]_{bb} + h.c. \right) \langle \, \mathcal{G}_{\mu\nu} \,\mathcal{G}^{\mu\nu} \phi_4\rangle_0,
\eea
where
\bea
f_{uG}  &=&  \left[-1 +  2 \, \log \left(\frac{\Lambda^2}{\hat{m}_h^2} \right) +  \, \log \left(\frac{4}{\tau_{t}} \right)
 \right] - 2 \, \mathcal{I}_{y} [m_t^2] -  \mathcal{I}[m_t^2]. \nn
 f_{dG} & = & \left[-1 + 2 \, \log \left(\frac{\Lambda^2}{\hat{m}_h^2} \right) +  2\, \log \left(\frac{4}{\tau_{b}} \right)\right] + \frac{\tau_b}{4} \Big[\log{\Big(\frac{1 + \sqrt{1-\tau_b}}{1 -\sqrt{1-\tau_b}} \Big)}  - i\pi \Big]^2 \nn
 & &\quad\quad - 2\,\sqrt{1-\tau_b} \Big[\log{\Big(\frac{1 + \sqrt{1-\tau_b}}{1 -\sqrt{1-\tau_b}} \Big)}  - i\pi \Big]
\eea
These results are consistent with a coupling rescaling of the results in Ref.~\cite{Hartmann:2015aia}.
Results for the finite terms of this mixing (with different function and operator conventions)
have also been reported in Ref.~\cite{Deutschmann:2017qum,Grazzini:2016paz} and we agree with this
result.\footnote{Note the appearance of a rescaling by $\sqrt{\kappa}$ in Eqn.\eqref{oneloopfirstterm}, as this is a common
rescaling of the gluon field in the dipole operator in both diagrams giving this result,
but that $[Y]_{tt}$ is used, not $[\mathcal{Y}]_{tt}$.
The relationship between the top mass and the Yukawa is
redefined order by order in the SMEFT. The
Feynman diagram results do not combine in the same manner as in the SM at $\mathcal{L}^{(8)}$ and above,
as the generalization of the mass and Yukawa couplings in the geoSMEFT expansion differ
from those in the SM (due to cross terms in the SMEFT expansion).
This is as expected. The one loop anomalous dimension
for mixing of $\mathcal{L}^{(6)}$ operators does not trivially generalise to higher orders in $\bar{v}_T/\Lambda$.
As the full class of $\mathcal{L}^{(8)}$ operator effects at one loop do
not directly follow from the $\mathcal{L}^{(6)}$ operator effects at one loop, we take $\sqrt{\kappa} \rightarrow 1$
in Eqn.\eqref{oneloopfirstterm} and restrict our results to $\mathcal{O}(\bar{v}_T^2/16 \pi^2 \Lambda^2)$.} Here and below we have chosen $\mu = \Lambda$ in these expressions, implicitly renormalizing the one loop effects of the SMEFT operators at the high scale $\Lambda$.
This corresponds to operator mixing (including diagonal running terms in the RGE) being conventionally introduced
to redefine the Wilson coefficients as they run from $\Lambda$ down to the measurement scales $\mu \simeq \hat{m}_h$. For evaluations associated with the one loop improvement of input parameters and on shell renormalization conditions, we use $\mu = \hat m_h$.

We define the interference of the above $\langle \mathcal{G} \mathcal{G}|\phi_4 \rangle^1_{\mathcal O(v^2/\Lambda^2)}$ terms with $\mathcal{A}^{\mathcal{G} \mathcal{G}\phi_4}_{SM}$ as $\Delta \sigma(\mathcal{G} \,\mathcal{G} \to h)_{EW}$, the SM $\times$ one-loop $\mathcal L_6$ SMEFT contribution to the $\mathcal G \mathcal G \to h$ cross section from all loops {\it except } QCD corrections to $\tilde C^{(6)}_{HG}$:
\bea
\Delta \sigma(\mathcal{G} \,\mathcal{G} \to h)_{EW} = \delta \sigma(\mathcal{G} \,\mathcal{G} \to h)_{\mathcal O(C_{HG}\alpha_s)}
\left(\left[\Delta G_F + M_1 \right] \, \tilde C^{(6)}_{HG} + \sum_{i} \,  \frac{{\rm Re} \, \tilde C_i^{(6)} f_i^{(6)}}{16 \pi^2}
\right) \delta(1-z), \nn
\label{eq:siggghEW}
\eea
where we have introduced the energy sharing variable $z = \bar{m}^2_h/\hat s$, $i = \{H\Box, HD, \delta G_F, \psi' H,  uG\}$ and $\delta \sigma(\mathcal{G} \,\mathcal{G} \to h)_{\mathcal O(C_{HG}\alpha_s)}$ is the SM $\times$ tree level $\mathcal L_6$ cross section (in $d=4-2\epsilon$ dimensions),
\bea
\delta \sigma(\mathcal{G} \,\mathcal{G} \to h)_{\mathcal O(C_{HG}\alpha_s)} = - \frac{\alpha_s \text{Re}(A_{1/2}(\tau))\, \tilde C^{(6)}_{HG}\, \bar{m}^2_h\, \mu^{2\epsilon}}{32\, \hat s\, v^2_{T} (1- \epsilon)} \delta(1-z).
\eea

\begin{figure}[ht!]
\includegraphics[width=0.5\textwidth]{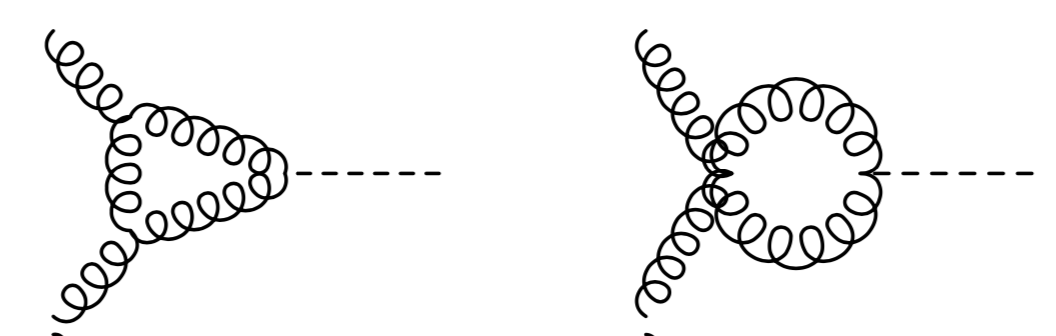}
\caption{QCD one loop contribution to $\mathcal{G} \,\mathcal{G} \rightarrow h$.}
\label{fig:2}
\end{figure}

The final ingredient is the QCD correction to Eq.~\eqref{eq:gghdim6tree}, shown in Fig.~\ref{fig:2}\,\cite{Dawson:1990zj, Spira:1995rr}:
\begin{align}
 \langle \mathcal G^{\mu \nu} \mathcal G_{\mu \nu} \phi_4 \rangle^1_{\alpha_s} = \frac{-4}{\hat{v}_T}\, \frac{\alpha_s}{2\pi}\, C_A\, \Big(\frac{\mu^2}{\bar{m}^2_h}\Big)^{\epsilon}\, c_\Gamma\, \Big(-\frac{2}{\epsilon^2}  - \frac{b_0}{C_A\, \epsilon} - \frac{2\,i\,\pi}{\epsilon} - \frac{b_0}{C_A}\log(\frac{\bar{m}^2_h}{\mu^2})+ \pi^2\Big)\,
 \langle \mathcal G^{\mu \nu} \mathcal G_{\mu \nu} \phi_4 \rangle_0,
\end{align}
after $\alpha_s \to \alpha_s(\mu)^{\overline{MS}} = \alpha_s \left(1- (\mu^2/\mu_R^2)^\epsilon \, \alpha_S \, c_\Gamma \, b_0/2 \pi \epsilon\right), b_0 = \frac 1 2 \left(11 - \frac {2\, N_F}{3} \right)$ is renormalized in $\rm \overline{MS}$. Here we consider the initial state gluons as
quantum fields here in the BFM as the gluons are convoluted with the PDF's. This choice
does not modify the expression. The same result is obtained with classical initial gluon fields
as we have explicitly verified. These results below are largely a rescaling of the results in Ref.~\cite{Djouadi:1991tka,Spira:1995rr}. The
$1/\epsilon^2, 1/\epsilon$ terms that remain are IR divergences\footnote{Here, $c_\Gamma = (4\pi)^{\epsilon}\Gamma(1+\epsilon)\Gamma(1-\epsilon)^2/\Gamma(1-2\epsilon) $}.
Taking the interference of this term with
$\mathcal{A}^{\mathcal{G} \mathcal{G}\phi_4}_{SM}$ results in an $\mathcal O(\tilde C^{(6)}_{HG} \alpha^2_s)$ contribution to $\sigma(\mathcal{G} \,\mathcal{G} \to h)$:
\begin{align}
\sigma(\mathcal{G} \,\mathcal{G} \to h)_{\mathcal O(C_{HG}\alpha^2_s)} =&\, \sigma(\mathcal{G} \,\mathcal{G} \to h)_{\mathcal O(C_{HG}\alpha_s)}\, \frac{\alpha_s}{2\pi}\, C_A\, \Big(\frac{\mu^2}{\bar{m}^2_h}\Big)^{\epsilon}\, c_\Gamma\nn
&\quad\quad \times \Big(-\frac{2}{\epsilon^2}  - \frac{2b_0}{C_A\, \epsilon} - \frac{2b_0}{C_A}\log(\frac{\bar{m}^2_h}{\mu^2})+ \pi^2\Big)\delta(1-z).
\label{eq:gghvirt}
\end{align}
To cancel these IR divergences, we need to include gluon emission, $\mathcal{G} \,\mathcal{G} \to \mathcal{G} h$.  The SM amplitude for $\mathcal{G} \,\mathcal{G} \to\mathcal{G} h$ is $\mathcal O(g^3_s)$ and the $\tilde C^{(6)}_{HG}$ amplitude is $\mathcal O(\tilde C^{(6)}_{HG}\,g_s)$, therefore the piece with the right coupling order to match the loop term above is the interference between the SM and $\tilde C^{(6)}_{HG}$ pieces.\footnote{There are other SMEFT contributions to $\mathcal{G} \,\mathcal{G} \to h \mathcal{G}$ coming from Yukawa-like and dipole terms, however they enter at one-loop (amplitude level) and therefore are only needed to fix IR issues with $\mathcal{G} \,\mathcal{G} \to h$ arising at two loops.} The SM contribution to $\mathcal{G} \,\mathcal{G} \to h \mathcal{G}$ involves loop functions of the momenta and the top mass. Extracting the IR divergences from these loops, maintaining the full mass dependence, is difficult\footnote{ For the SM alone, the full dependence is known, Ref.~\cite{Spira:1995rr}. Reference~\cite{Deutschmann:2017qum} includes the two-loop results for $\tilde C^{(6)}_{HG}$, Yukawa-like, and dipole-like operators at dimension six, retaining the full top mass dependence. This is combined with one-loop $\mathcal G \mathcal G \to \mathcal G h$ via the SMEFTNLO~\cite{Degrande_2021,durieux2019proposal} framework. However, their results are not easily adapted to the framework used here due to different operator convention choices.}, so we will expand in $m_t \to \infty$.

Taking $m_t \to \infty$, $A_{1/2}(\tau_f \gg 1) \to -\frac 4 3 + \mathcal O(1/\tau_f)$, and the SM top quark loop in Eq.~\eqref{eq:SMggh} can be combined with $\mathcal L_6$ SMEFT tree level term:
\begin{align}
\frac 1 {\hat{v}_T}\,\Big(\frac{\alpha_s}{12\pi} + \tilde C^{(6)}_{HG}\Big) \langle \mathcal G^{\mu \nu} \mathcal G_{\mu \nu} \phi_4 \rangle_0.
\label{eq:hGGcoeff}
\end{align}
Using this modified coefficient to calculate, the ${\mathcal O(C_{HG}\alpha_s)}$ contribution to $\sigma(\mathcal{G} \,\mathcal{G} \to h)_{\mathcal O(C_{HG}\alpha_s)}$ reduces to
\bea
\delta \sigma^{m_t \to \infty}(\mathcal{G} \,\mathcal{G} \to h)_{\mathcal O(C_{HG}\alpha_s)} = \frac{\alpha_s \tilde C^{(6)}_{HG}\, \bar{m}^2_h\, \mu^{2\epsilon}}{24\, \hat s\, \hat{v}^2_{T}\, (1- \epsilon)},
\eea
and we must replace the $f_i $ in Eq.~\eqref{eq:siggghEW} with their large $m_t$ expressions\footnote{In the $m_t \to \infty$ limit, $f_{uG} \to -2\left(1-\log(\Lambda^2/m^2_t) \right)$}. The SM expression can be improved by including the $\mathcal O(\alpha^2_s)$, 2-loop matching contribution~\cite{Inami:1982xt,Djouadi:1991tka,Kniehl:1995tn,Kilian:1995tra,Chetyrkin:1996ke,Chetyrkin:1996wr} and the leading $\mathcal O(\bar m^2_h/\bar m^2_t)$ correction
\begin{align}
\frac 1 {\hat{v}_T}\,\Big(\frac{\alpha_s}{12\pi} + \tilde C^{(6)}_{HG}\Big)  \to \frac 1 {\hat{v}_T}\,\Big(\frac{\alpha_s}{12\pi}\Big(1 + \frac{11\,\alpha_s}{4\pi} + \frac{7\, \bar{m}^2_h}{120\, \bar{m}^2_t}\Big) + \tilde C^{(6)}_{HG}\Big).
\label{eq:hGGcoeff2}
\end{align}
While the $\mathcal O(\alpha^2_s)$ piece is not the full two loop correction to $\mathcal{A}_{SM}$ for this process,
it is a well defined matching correction that is numerically relevant. Given that the $\mathcal O(\alpha^2_s)$ SM piece interfering with tree level $\tilde C^{(6)}_{HG}$ is formally the same order as the $\mathcal O(\alpha_s)$ SM times the $\mathcal O(\tilde C^{(6)}_{HG}\alpha_s)$ QCD loop-corrected $\mathcal L_6$ piece, we choose to include it. We emphasize that our goal is to capture and compare SMEFT effects at $\mathcal O(\bar v^2_T/16\pi^2\Lambda^2)$ and $\mathcal O(\bar v^4_T/\Lambda^4)$ rather than to push to higher orders in perturbative corrections. For SMEFT results in the $m_t \to \infty$ limit (excluding the chromomagnetic dipole operators) at NNLO, see Ref~\cite{Brooijmans:2016vro}, and Ref~\cite{Contino:2014aaa, Harlander:2016hcx, Anastasiou:2016hlm} for N$^3$LO.

Working in the $m_t \to \infty$ limit (and $d = 4-2\epsilon$ dimensions), the $\mathcal{G} \,\mathcal{G} \to \mathcal{G} h$, amplitude squared is
\begin{align}\label{eq:ggg}
|\mathcal A(\mathcal{G} \,\mathcal{G} \to h \mathcal{G})_{m_t \to \infty}|^2 \propto \alpha_s |C_{hGG}|^2 \frac{(m^8_H + s^4 + t^4 + u^4)(1-2\epsilon) + \frac 1 2 \epsilon\, (\bar{m}^2_h + s^2 + t^2 + u^2)^2}{s\, t\, u}
\end{align}
where $C_{hGG}$ is the coefficient of $\langle h\, \mathcal{G} \,\mathcal{G}\rangle^0$ and $s,t,u$ are the usual Mandelstam variables.
Inserting the Eq.~\eqref{eq:hGGcoeff} coefficient, we can extract the $\mathcal O(\tilde C^{(6)}_{HG}\alpha^2_s)$ piece.
Converting this into a cross section,  expanding the result in $1/\epsilon$ and regulating divergences as $z \to 1, \epsilon \to 0$ via
the $+$ prescription. The net result can be written as
\begin{align}
\sigma(\mathcal{G} \,\mathcal{G} \to \mathcal{G} h)_{\mathcal O(C_{HG}\alpha^2_s)}^{m_t \to \infty} =& \sigma(\mathcal{G} \,\mathcal{G} \to h)_{\mathcal O(C_{HG}\alpha_s)}^{m_t \to \infty}\times \frac{\alpha_s}{2\pi}\, C_A\, \Big(\frac{\mu^2}{\bar{m}^2_h}\Big)^{\epsilon}\, c_\Gamma\, \times \nn
& \Big[ \Big(\frac 2 {\epsilon^2} + \frac{2\,b_0}{C_A\,\epsilon} - \frac{\pi^2}{3} \Big)\delta(1-z) - \frac 2 {\epsilon} p_{\mathcal{G} \,\mathcal{G}}(z) - \frac {11} 3 \frac{(1-z)^3}{z}  \\
& - 4 \frac{(1-z)^2(1+z^2) + z^2}{z(1-z)}\log{z} + 4\frac{1 + z^4 + (1-z)^4}{z}\Big(\frac{\log{(1-z)}}{1-z}\Big)_+ \Big],\nonumber
\label{eq:gghg}
\end{align}
where $p_{\mathcal{G} \,\mathcal{G}}(z)$ is the colour-stripped Altarelli-Parisi splitting function\footnote{$p_{\mathcal{G} \,\mathcal{G}}(z) = 2\Big( \frac{z}{(1-z)_+} + \frac{1-z}{z}  + z(1-z)\Big) + \frac{b_0}{C_A} \delta (1-z)$}. The $1/\epsilon$ divergence proportional to $p_{\mathcal{G} \,\mathcal{G}}$ is absorbed into the parton distribution functions, formally by introducing a counter-term
\begin{align}
\sigma_{c.t.} = \sigma(\mathcal{G} \,\mathcal{G} \to h)_{\mathcal O(C_{HG}\alpha_s)}^{m_t \to \infty}\, 2\,\frac{\alpha_s}{2\pi}\, C_A\, \Big(\frac{\mu^2}{\mu^2_F}\Big)^{\epsilon}\, c_\Gamma\, p_{\mathcal{G} \,\mathcal{G}}(z).
\end{align}
Including this last counter-term and combining $\sigma(\mathcal{G} \,\mathcal{G} \to h)^{m_t \to \infty}_{\mathcal O(C_{HG}\alpha^2_s)}$ -- both from the $m_t \to \infty$ limit of Eq.~\eqref{eq:gghvirt} and from the interference of the $\mathcal O(\alpha^2_s)$ SM piece (Eq.~\eqref{eq:hGGcoeff2}) and the tree level $\tilde C^{(6)}_{HG}$ amplitude  -- and $\sigma(\mathcal{G} \,\mathcal{G} \to h \mathcal{G})^{m_t \to \infty}_{\mathcal O(C_{HG}\alpha^2_s)}$, we have:
\begin{align}
\Delta \sigma(\mathcal{G} \,\mathcal{G} \to h)^{m_t \to \infty}_{\mathcal O(C_{HG}\alpha_s)} &=\, \sigma(\mathcal{G} \,\mathcal{G} \to h)_{\mathcal O(C_{HG}\alpha_s)}^{m_t \to \infty}\frac{\alpha_s}{2\pi}\, C_A\, \Big(\frac{\mu^2}{\bar{m}^2_h}\Big)^{\epsilon}\, c_\Gamma\, \times \nn
&\Big[
  \Big(\frac{2\,b_0\log(\frac{\mu^2}{\bar{m}^2_h})}{C_A} + \frac{2\pi^2}{3} + \frac {11}{6}\Big)\delta(1-z) +  2\, p_{\mathcal{G} \,\mathcal{G}}(z)\, \log{(\frac{\bar{m}^2_h}{\mu^2_F})} - \frac {11} 3 \frac{(1-z)^3}{z}  \nn
& - 4 \frac{(1-z)^2(1+z^2) + z^2}{z(1-z)}\log{z} + 4\frac{1 + z^4 + (1-z)^4}{z}\Big(\frac{\log{(1-z)}}{1-z}\Big)_+\Big].
\end{align}

\subsubsection{$\langle \mathcal{G} \mathcal{G}|\phi_4 \rangle^0_{\mathcal{L}^{(8)}}$}

The $\mathcal O(v^4_T/\Lambda^4)$ terms for the $\mathcal{G} \,\mathcal{G} \to h$ amplitude comes from expanding out Eq.~\eqref{eq:hggvertex}
\begin{align}
 \langle \mathcal{G}\mathcal{G}|\phi_4 \rangle^0_{\mathcal O(v^4/\Lambda^4)} &=  \langle \sqrt{h}^{44}\rangle_{\mathcal O(v^2/\Lambda^2)} \langle \mathcal{G}\mathcal{G}|\phi_4 \rangle^0_{\mathcal O(v^2/\Lambda^2)} \\ \nonumber
&\quad\quad\quad\quad\quad\quad  + 2\,\frac{ v_T [\langle \mathcal{G}\mathcal{G}|\phi_4 \rangle^0_{\mathcal O(v^2/\Lambda^2)}]^2}{\langle \mathcal G\, \mathcal G | \phi_4 \rangle^0}  + \, (\langle \mathcal{G}\mathcal{G}|\phi_4 \rangle^0_{\mathcal O(v^2/\Lambda^2)})\Big|_{\tilde C^{(6)}_{HG} \to \tilde C^{(8)}_{HG}}
\end{align}
where $\langle \sqrt{h}^{44}\rangle_{\mathcal O(v^2/\Lambda^2)} = \tilde C^{(6)}_{H\Box} - \frac 1 4 \tilde C^{(6)}_{HD}$. A term from
the redefinition of $\bar{v}_T$ in its relation to input observables is formally present but suppressed as it cancels
when the SM amplitude is interfered with, which is $\propto 1/\bar{v}_T$.

These contributions lead to the dimension eight corrections -- in the $m_t \to \infty$ limit:
\begin{align}
\delta \sigma(\mathcal{G} \,\mathcal{G} \to h)^{m_t \to \infty}_{\mathcal O(\alpha_s \bar{v}_T^4/\Lambda^4)} = \left[\left(\langle \sqrt{h}^{44}\rangle_{\mathcal O(v^2/\Lambda^2)} + 2 \tilde C^{(6)}_{HG}\right) \tilde C^{(6)}_{HG} +  \tilde C^{(8)}_{HG}\right]\,\frac{\alpha_s \, \bar{m}^2_h\, \mu^{2\epsilon}}{24\, \hat s\, \hat{v}^2_{T}\, (1- \epsilon)}
\end{align}

Combining all of these contributions we have
\bea
\sigma(\mathcal{G} \,\mathcal{G} \to h) &=\sigma(\mathcal{G} \,\mathcal{G} \to h)^{1/m^2_t}_{SM, \mathcal O(\alpha^2_s, \alpha^3_s)} + \delta \sigma(\mathcal{G} \,\mathcal{G} \to h)^{1/m^2_t}_{\mathcal O(C_{HG}\alpha_s)}+
\delta \sigma(\mathcal{G} \,\mathcal{G} \to h)^{m_t \to \infty}_{\mathcal O(\alpha_s \bar{v}_T^4/\Lambda^4)}, \nn
&+ \Delta \sigma(\mathcal{G} \,\mathcal{G} \to h)^{m_t \to \infty}_{EW} + \Delta \sigma(\mathcal{G} \,\mathcal{G} \to h)^{m_t \to \infty}_{\mathcal O(C_{HG}\alpha^2_s)}  + \cdots
\eea
The $1/m^2_t$ superscript on the $\sigma(\mathcal{G} \,\mathcal{G} \to h)^{1/m^2_t}_{SM}$ and $\delta \sigma(\mathcal{G} \,\mathcal{G} \to h)^{1/m^2_t}_{\mathcal O(C_{HG}\alpha_s)}$ terms indicates that we include the $\mathcal O(m^2_H/m^2_t)$ corrections Eq.~\eqref{eq:hGGcoeff2}. For all other terms, we use the $\mathcal O(\alpha_s), m_t \to \infty$ SM expression\footnote{At $\mathcal O(\tilde C^{(6)}_{HG}\, \alpha^2_s\, m^2_H/m^2_t)$, there are additional contributions additional higher dimensional operators (suppressed by $1/\bar{m}^2_t$) generated which can contribute to $\mathcal{G} \,\mathcal{G} \to h\,\mathcal G$~\cite{Harlander:2013oja,Dawson:2014ora}.}.

\subsection{$\Gamma(h \rightarrow \mathcal{G} \mathcal{G})$ vs $\sigma(\mathcal{G} \mathcal{G} \rightarrow h)$}

When considering $\Gamma(h \rightarrow \mathcal{G} \mathcal{G})$ compared to $\sigma(\mathcal{G} \mathcal{G} \rightarrow h)$,
it is important to note the constraints that are present in the vertex being a three point momentum conserving interaction.
This, combined with the common matrix elements due to crossing symmetry lead to a significant overlap
in the (inclusive) results.

Nevertheless, some subtle differences are present. Using the BFM, we have checked if the one loop matrix
elements  ($\langle \mathcal{G} \mathcal{G} \phi_4\rangle^1$ compared to $\langle \phi_4 \mathcal{G} \mathcal{G}\rangle^1$)
change treating the external gluon fields as final state `classical' background fields as opposed to quantum fields to
convolute with the parton distribution functions.\footnote{Here we have used the Feynman rules package in Ref.~\cite{Corbett:2020bqv}.}
We find the one loop matrix element is identical for the two cases, as expected.
The final results are not identical even so, as we show below.

\subsubsection{SM Result, and $\langle \phi_4| \mathcal{G} \mathcal{G} \rangle^0_{\mathcal{L}^{(6,8)}}$ corrections}
In the SM, the leading order result for the decay width (still in the $m_t \to \infty$ limit) is given by \cite{Cahn:1983ip,Bergstrom:1985hp}
\begin{align}
\Gamma(h \to \mathcal{G} \,\mathcal{G})^{m_t\to \infty}_{SM} \simeq \frac{\alpha^2_s \, \hat{m}_h^3}{72 \pi^3 \hat{v}_T^2}.
\end{align}
The lowest order partial width at $\mathcal O(\tilde C^{(6)}_{HG}\, \alpha_s)$ (in $d = 4-2 \epsilon$ dimensions) is
\begin{align}
\delta \Gamma(h \to \mathcal{G} \,\mathcal{G})^{m_t\to \infty}_{\mathcal O(C_{HG} \alpha_s)} = \frac{\alpha_s\, \tilde C^{(6)}_{HG}\, \bar{m}^3_h}{3\,\pi^2\, \hat{v}^2_T}\Big(\frac{4\pi\mu^2}{\bar{m}^2_h}\Big)^{\epsilon}\frac{\Gamma(2-\epsilon)}{\Gamma(2-2\epsilon)},
\end{align}
and the $\mathcal O(v^4_T/\Lambda^4)$ contribution is a simple rescaling
\begin{align}
\delta \Gamma(h \to \mathcal{G} \,\mathcal{G})^{m_t\to \infty}_{\mathcal O(\alpha_s\bar{v}_T^4/\Lambda^4)} = \left[\left(\langle \sqrt{h}^{44}\rangle_{\mathcal O(v^2/\Lambda^2)} + 2 \tilde C^{(6)}_{HG}\right) \tilde C^{(6)}_{HG} +  \tilde C^{(8)}_{HG}\right]\,\frac{\alpha_s\, \bar{m}^3_h}{3\,\pi^2\, \hat v^2_T}\Big(\frac{4\pi\mu^2}{\bar{m}^2_h}\Big)^{\epsilon}\frac{\Gamma(2-\epsilon)}{\Gamma(2-2\epsilon)}.
\end{align}

\subsubsection{$\langle \phi_4| \mathcal{G} \mathcal{G} \rangle^1_{\mathcal{L}^{(6)}}$}
The EW correction is identical to the case of $\sigma(\mathcal{G} \mathcal{G} \rightarrow h)$,
\begin{align}
\Delta \Gamma(h \to \mathcal{G} \,\mathcal{G})^{m_t\to \infty}_{EW} = \frac{\alpha_s\, \bar{m}^3_h}{3\,\pi^2\, \hat v^2_T}\Big(\frac{4\pi\mu^2}{\bar{m}^2_h}\Big)^{\epsilon}\frac{\Gamma(2-\epsilon)}{\Gamma(2-2\epsilon)}
\times \left(\left[\Delta G_F + M_1 + \Delta R_\mathcal{G}\right] \, \tilde C^{(6)}_{HG} + \sum_{i} \,  \frac{{\rm Re} \, \tilde C_i^{(6)} f_i^{(6)}}{16 \pi^2}
\right)
\end{align}
In this expression we also include the
BFM wavefunction renormalization finite  factor of the final state gluon: ($\Delta R_\mathcal{G}$).
This result is easy to obtain in the background field method, see Ref.~\cite{Abbott:1981ke}.
As in the case of the QCD $\beta$ function, only two diagrams are required for the
gauge interactions in the BFM. As $p^2 \rightarrow 0$ for on-shell renormalization,
for $\Delta R_\mathcal{G}$ these diagrams are scaleless in dimensional regularization, and the finite terms vanish.
The contribution from the massive quarks are not scaleless and gives the finite terms
\bea
\Delta R_\mathcal{G} = \frac{1}{24 \pi^2} \sum_f \log \left(\frac{m_f^2}{\mu^2}\right)
\eea
The $\tilde C^{(6)}_{HG}$ operator was not redefined to
rescale it by $g_3^2$. Note that in the BFM this has the result
of the $\Delta R_\mathcal{G}$ contribution not canceling against a corresponding finite term for $g_3^2$ but contributing.

\begin{figure}[ht!]
\includegraphics[width=0.5\textwidth]{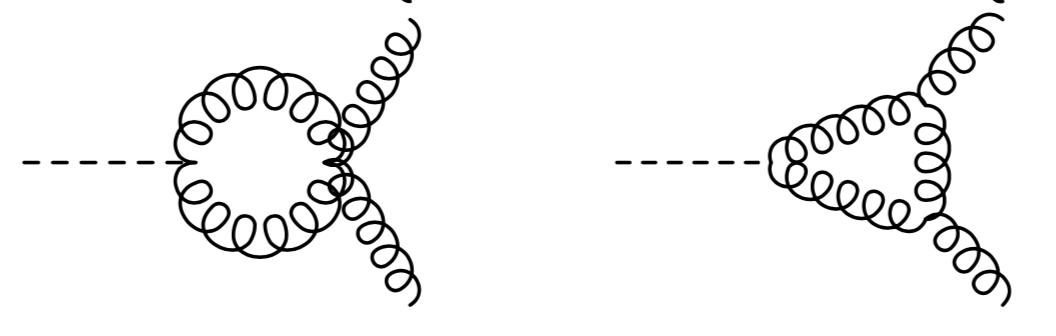}
\caption{QCD one loop contribution to $\phi_4\rightarrow \mathcal{G} \,\mathcal{G}$.}
\label{fig:3}
\end{figure}

Adding the QCD loop correction ($\langle \phi_4 \mathcal{G} \mathcal{G}\rangle^1$ shown in Fig.~\ref{fig:3}), the $\alpha_s$ counter-term, and the interference between the tree level $\tilde C^{(6)}_{HG}$ and $\mathcal O(\alpha^2_s)$ SM pieces from Eq~\eqref{eq:hGGcoeff2}, the result is the LO partial width times a factor
\begin{align}
\Delta \Gamma(h \to \mathcal{G} \,\mathcal{G})^{m_t\to \infty}_{\mathcal O(C_{HG} \alpha^2_s)} =&\, \Gamma(h \to \mathcal{G} \,\mathcal{G})^{m_t\to \infty}_{\mathcal O(C_{HG} \alpha_s)}  \frac{\alpha_s}{2\pi}\, C_A\, \Big(\frac{\mu^2}{\bar{m}^2_h}\Big)^{\epsilon}\, c_\Gamma\nn
&\times \Big( -\frac{2}{\epsilon^2} - \frac{2\,b_0}{C_A\,\epsilon} + \frac{2}{C_A}\log\Big(\frac{\mu^2}{\bar{m}^2_h}\Big) + \pi^2 + \frac {11}{6} \Big).
\end{align}
 As with $\mathcal{G} \,\mathcal{G} \to h$, there are IR divergences proportional $\mathcal O(\tilde C^{(6)}_{HG} \alpha_s)$ in the amplitude,
 or, upon interfering with the SM contribution, $\mathcal O(\tilde C^{(6)}_{HG}\alpha^2_s)$ in the partial width.
 These IR divergences are cancelled by $\Gamma(h \to \mathcal{G} \,\mathcal{G} \,\mathcal{G})$. While the matrix elements for
 $\mathcal{G} \,\mathcal{G} \to h \leftrightarrow h \to\mathcal{G} \,\mathcal{G}$, and $\mathcal{G} \,\mathcal{G} \to h \,\mathcal{G} \leftrightarrow h\to \mathcal{G} \,\mathcal{G}\,\mathcal{G}$ are identical, the phase space and kinematics differ, which leads to slightly different factors when computing the partial widths.
 It is also not sufficient to add $\Gamma(h \to \mathcal{G} \,\mathcal{G})$ and $\Gamma(h \to \mathcal{G} \,\mathcal{G}\,\mathcal{G})$ alone to cancel all IR issues; $\Gamma(h \to \mathcal{G} \bar q q)$ must be included as well.
 The two three-body processes do not interfere with each other, but both come about with the same power of $\alpha_s$ and are indistinguishable in the collinear limit.

For $\Gamma(h \to \mathcal{G} \,\mathcal{G} \,\mathcal{G})$, the matrix element (squared) in the $m_t \to \infty$ limit is Eq.~\eqref{eq:ggg}, with prefactor set by Eq.~\eqref{eq:hGGcoeff}. Integrating over three-body phase space and including the appropriate initial state factors,
\begin{align}
\Delta\Gamma(h\to \mathcal{G} \,\mathcal{G}g)^{m_t\to \infty}_{\mathcal O(C_{HG} \alpha^2_s)}  = \Gamma(h\to \mathcal{G} \,\mathcal{G})^{m_t\to \infty}_{\mathcal O(C_{HG} \alpha_s)}\, \frac{\alpha_s}{2\pi}\, C_A\, \Big(\frac{\mu^2}{\bar m^2_h}\Big)^{\epsilon}\, c_\Gamma\,\Big(\frac 2 {\epsilon^2} + \frac{11}{3\,\epsilon}  + \frac{73}{6} - \pi^2 \Big).
\end{align}

As described above, we need to include $\Gamma(h \to \mathcal G\bar q q)$ $\propto \tilde C^{(6)}_{HG}$. This comes from a $\tilde C^{(6)}_{HG}$ $h$-$\mathcal{G}$-$\mathcal{G}$ vertex with one of the gluons splitting to a $\bar q q$ pair. Interfering this term with the SM contribution -- diagrammatically identical to the $\tilde C^{(6)}_{HG}$ piece in the $m_t \to \infty$ limit -- and including phase space, yields:
\begin{align}
\Delta\Gamma(h \to g\bar q q)^{m_t\to \infty}_{\mathcal O(C_{HG} \alpha^2_s)}  = \Gamma(h\to \mathcal{G} \,\mathcal{G})^{m_t\to \infty}_{\mathcal O(C_{HG} \alpha_s)}\times \frac{\alpha_s}{2\pi}\, C_A\, \Big(\frac{\mu^2}{\bar m^2_h}\Big)^{\epsilon}\, c_\Gamma\, N_F\,\Big(-\frac{2}{9\,\epsilon} - \frac{7}{9}\Big),
\end{align}
where $N_F$ is the number of fermion flavors the Higgs can decay into.

Adding the three contributions, the IR divergences cancel and we have $\Gamma(h \to \mathcal{G} \,\mathcal{G})^{m_t \to \infty}_{\mathcal O(C_{HG})}$
\begin{align}
\Delta\Gamma(h \to \mathcal{G} \,\mathcal{G})^{m_t \to \infty}_{\mathcal O(C_{HG})}  =&\, \Gamma(h\to \mathcal{G} \,\mathcal{G})^{m_t \to \infty}_{\mathcal O(C_{HG} \alpha_s)} \Big(\frac{\alpha_s}{2\pi}\, C_A\, \Big(\frac{\mu^2}{\bar m^2_h}\Big)^{\epsilon}\, c_\Gamma
\Big(14 - \frac{7\, N_F}{9} + \frac{2}{C_A}\log\Big(\frac{\mu^2}{\bar m^2_h}\Big) \Big).
\end{align}
Given the appearance of the $p_{\mathcal{G} \,\mathcal{G}}(z)$ color-stripped Altarelli-Parisi
function in $\sigma(\mathcal{G} \,\mathcal{G} \to h)$, one may have expected fragmentation functions to appear in $h \to \mathcal G \mathcal G$. No such function is explicitly present in the inclusive result as a final state phase space integration over a normalized fragmentation function
implicitly occurs.

Combining all of these results gives the decay width
\begin{align}
\Gamma_{\rm SMEFT}(h \to \mathcal{G} \,\mathcal{G}) &=\Gamma(h \to \mathcal{G} \,\mathcal{G})^{1/m^2_t}_{SM, \mathcal O(\alpha^2_s, \alpha^3_s)}
+ \delta \Gamma(h \to \mathcal{G} \,\mathcal{G})^{1/m^2_t}_{\mathcal O(C_{HG} \alpha_s)}
+ \delta \Gamma(h \to \mathcal{G} \,\mathcal{G})^{m_t\to \infty}_{\mathcal O(\alpha_s\bar{v}_T^4/\Lambda^4)} \nn
&\quad\quad+\Delta\Gamma(h \to \mathcal{G} \,\mathcal{G})^{m_t \to \infty}_{\mathcal O(C_{HG}\alpha^2_s)}
+ \Delta \Gamma(h \to \mathcal{G} \,\mathcal{G})^{m_t\to \infty}_{EW} + \cdots
\end{align}
where the $1/m^2_t$ superscript indicates which terms include the $\mathcal O(\bar m^2_h/\bar m^2_t)$ corrections from Eq.~\eqref{eq:hGGcoeff2}.

\subsection{$\Gamma(h \rightarrow \gamma \gamma)$}

The full amplitude for $\Gamma(h \rightarrow \mathcal{A} \mathcal{A})$ that determines
the result to $\mathcal{O}(\bar{v}_T^2/16 \pi^2 \Lambda^2)$ and $\mathcal{O}(\bar{v}_T^4/\Lambda^4)$
is given by
\bea
\langle \phi_4|\mathcal{A} \mathcal{A}\rangle = \langle \phi_4|\mathcal{A} \mathcal{A}\rangle_{SM}^1
+ \langle \phi_4|\mathcal{A} \mathcal{A}\rangle^0_{\mathcal{O}(\bar{v}_T^2/\Lambda^2)} + \langle \phi_4|\mathcal{A}\mathcal{A} \rangle_{{\cal{O}}(\bar{v}_T^4/\Lambda^4)}^0 + \langle \phi_4|\mathcal{A} \mathcal{A}\rangle^1_{\mathcal{L}^{(6)}}.
\eea
We define each of these contributions to the amplitude systematically in the following sections.
\subsubsection{SM result}
In the SM, $\Gamma(h \rightarrow \gamma \gamma)$ is loop-suppressed, and the leading order result was developed in
Refs.~\cite{Ellis:1975ap,Shifman:1979eb,Bergstrom:1985hp}. We define
\begin{align}
&\langle \phi_4|\mathcal{A} \mathcal{A}\rangle_{SM}^1 = \frac{- \hat{g}_2 \, \hat{e}^2}{64 \, \pi^2 \, \hat{m}_W}
\Bigg(A_1(\tau_W)+ \sum_i \, N_c^i \,Q^2_i \, A_{1/2} (\tau_{\psi^i})\Bigg) \, \langle \phi_4 \mathcal{A}^{\mu\nu} \mathcal{A}_{\mu \nu} \rangle^0,
\end{align}
where we are using the notation that $\psi^i$ indicates a mass eigenstate fermion,
\bea
A_1(\tau_p) &=&2 + 3 \tau_p \left[1+ (2- \tau_p)f(\tau_p)\right],
\eea
and $A_{1/2}(\tau_p)$ is given in Eq.~\eqref{eq:A12}. As before, $\tau_p = 4 m_p^2/\bar{m}_h^2$ and $f (\tau_p)$ is defined in Appendix~\ref{functions}.
\subsubsection{$\langle \phi_4| \mathcal{A} \mathcal{A} \rangle^0_{\mathcal{L}^{(6)}}$}
Directly from the general definition in Eqn.~\eqref{threepointamps} one has
\begin{align}
	\label{eq:hgamgamDim6}
\langle \phi_4|\mathcal{A} \mathcal{A}\rangle^0_{\mathcal{O}(\bar{v}^2/\Lambda^2)} &= \left[\frac{\hat{g}_2^2 \,   \tilde{C}_{HB}^{(6)}
  + \hat{g}_1^2 \,  \tilde{C}_{HW}^{(6)} - \hat{g}_1 \, \hat{g}_2 \,  \tilde{C}_{HWB}^{(6)}}{(\hat{g}^{\rm SM}_Z)^2 \, \bar{v}_T^2} \right] \,\hat{v}_T \, \langle \phi_4 \mathcal{A}^{\mu\nu} \mathcal{A}_{\mu \nu} \rangle^0.
\end{align}
\subsubsection{$\langle \phi_4| \mathcal{A} \mathcal{A} \rangle^1_{\mathcal{L}^{(6)}}$}
Here we present one loop results adapting and extending Refs.~\cite{Hartmann:2015oia,Hartmann:2015aia}
collecting all SMEFT corrections at $\mathcal{O}(\bar{v}_T^2/16 \pi^2 \Lambda^2)$.
We modify these past BFM results so that they are consistent with the dimension eight results by removing
the choice of scaling operators by their corresponding gauge couplings (this choice effects one loop finite terms),
and simultaneously complete this set of results to one loop in the BFM using the results in Section \ref{oneloopBFM}.

We express the matrix element to one loop as
\bea
\langle \phi_4|\mathcal{A} \mathcal{A}\rangle^1_{\mathcal{L}^{(6)}} &=&
\sum_{i=HB,HW,HWB} \hspace{-0.75cm}\langle \phi_4 | [C^{(6)}_{i}] |\mathcal{A} \mathcal{A}\rangle^1
+ \langle \phi_4|\mathcal{A} \mathcal{A}\rangle_{SM}^1 \left[\frac{\delta \alpha}{\alpha}- \frac{\delta G_F}{\sqrt{2}} \right], \nn
&+& \left(C^{1}_{\mathcal{A} \mathcal{A}}   +  \frac{\tilde{C}_i \, f_i}{16 \, \pi^2 \,\hat{v}_T^2}\right) \, \hat{v}_T  \, \langle \phi_4 \mathcal{A}^{\mu\nu} \mathcal{A}_{\mu \nu} \rangle^0.
\eea
where we have absorbed the remaining input parameter one loop corrections into $C^{1}_{\mathcal{A} \mathcal{A}}$ and
\begin{align}
C^{1}_{\mathcal{A} \mathcal{A}} &= \left[\frac{\hat{g}_2^2 \, \tilde{C}_{HB}^{(6)}
  + \hat{g}_1^2 \, \tilde{C}_{HW}^{(6)} - \hat{g}_1 \, \hat{g}_2 \, \tilde{C}_{HWB}^{(6)}}{(\hat{g}^{\rm SM}_Z)^2 \, \bar{v}_T^2} \right] \, M_1
  +  \frac{2.1 \, \hat{e}^2}{16 \, \pi^2 \, \hat{v}_T^2} \, \frac{\delta M_W^2}{\hat{M}_W^2}.
\end{align}
The last term in this expression is an input parameter scheme dependent contribution,
only present in the $\{\alpha_{EW},\hat{m}_Z,\hat{G}_F\}$ input scheme. In the $\{\hat{m}_W,\hat{m}_Z,\hat{G}_F\}$
scheme $\delta M_W^2 = 0$ by definition. This term is directly evaluated
from the $\bar{m}_W$ dependence in the SM loop contribution.\footnote{We note that it is easier to evaluate this term as a shift
in the $W$ mass after the loop integral is calculated to the level of Feynman parameter integrations
as in Eqn.~(4.4) of Ref.~\cite{Hartmann:2015oia,Hartmann:2015aia}.}

The remaining $f_i$'s were given in Refs.~\cite{Hartmann:2015oia,Hartmann:2015aia} and are as follows.
In terms of the one loop function $\mathcal{I}_y[m_p] $
\bea
\frac{\hat{g}_1 \,\hat{g}_2}{\hat{e}^2} f_{HWB} &=&  \left(- 3 \, \hat{g}_2^2 +4 \, \lambda  \right) \,  \log \left( \frac{\bar{m}_h^2}{\Lambda^2}\right) + (4 \lambda - \hat{g}_2^2) \, \mathcal{I} [\bar{m}_W^2]
- 4 \, \hat{g}_2^2 \, \mathcal{I}_y [\bar{m}_W^2] - 2 \, \hat{g}_2^2 \left[1+ \log \left( \frac{\tau_{W}}{4}\right) \right] \nonumber \\
&+& \hat{e}^2 \, (2+ 3 \, \tau_{W})+ 6 \, \hat{e}^2 (2 - \tau_{W}) \, \, \mathcal{I}_y [\bar{m}_W^2].
\eea

In terms of the function $ \mathcal{I}_{xx}$
\bea
\frac{\hat{g}_2^2}{\hat{e}^2} f_{HW} &=& - \hat{g}_2^2 \Bigg[3 \, \tau_{W} + \left(16 - \frac{16}{\tau_{W}}- 6 \, \tau_{W}\right) \, \mathcal{I}_y [\bar{m}_W^2] \Bigg],  \\
\frac{\hat{g}_2^3}{\hat{e}^2} f_{W} &=& - 9 \, \hat{g}_2^4 \,  \log \left( \frac{\bar{m}_h^2}{\Lambda^2}\right) - 9 \, \hat{g}_2^4 \,  \mathcal{I} [\bar{m}_W^2] - 6 \, \hat{g}_2^4 \,  \mathcal{I}_y [\bar{m}_W^2]
+ 6  \, \hat{g}_2^4 \,  \mathcal{I}_{xx} [\bar{m}_W^2] \, \left(1-1/\tau_{W}\right) - 12 \, \hat{g}_2^4, \nn
\eea
For the dipole leptonic operators, we find,
\bea
\frac{\hat{g}_1}{\hat{e}^2} f_{\substack{eB \\ ss}}  &=& 2 \, Q_\ell \, [Y_\ell]_{ss}  \left[-1 +  2 \, \log \left(\frac{\Lambda^2}{\bar{m}_h^2} \right) +  \, \log \left(\frac{4}{\tau_{s}} \right) \right]
- 2 \, Q_\ell \, [Y_\ell]_{ss} \, \Big[ 2 \, \mathcal{I}_{y} [m_s^2] + \mathcal{I}[m_s^2] \Big].
\eea
and $\hat{g}_2 f_{\substack{eW \\ ss}} \rightarrow  - \hat{g}_1 \, f_{\substack{eB \\ ss}}$. Here, $s=\{1,2,3\}$
sums over the flavours of the leptons and we note that the Wilson coefficients
are summed with their Hermitian conjugates for each flavour, and the normalization is such that $f_{\substack{eB \\ ss}}$ multiplies
${\rm Re} \, C_{\substack{eB \\ ss}}$. The remaining dipole $f_i$'s for the quarks are
\bea
\frac{\hat{g}_1}{\hat{e}^2} f_{\substack{uB \\ ss}}  &=& 2 \, N_c \, Q_u \, [Y_u]_{ss}  \left[-1 +  2 \, \log \left(\frac{\Lambda^2}{\bar m_h^2} \right) +  \, \log \left(\frac{4}{\tau_{s}} \right) \right]
- 2 \, Q_u \, [Y_u]_{ss} \, \Big[ 2 \, \mathcal{I}_{y} [m_s^2] + \mathcal{I}[m_s^2] \Big], \nn
\frac{\hat{g}_1}{\hat{e}^2} f_{\substack{dB \\ ss}}  &=& 2 \, N_c \, Q_d \, [Y_d]_{ss}  \left[-1 +  2 \, \log \left(\frac{\Lambda^2}{\bar{m}_h^2} \right) +  \, \log \left(\frac{4}{\tau_{s}} \right) \right]
- 2 \, Q_d \, [Y_d]_{ss} \, \Big[ 2 \, \mathcal{I}_{y} [m_s^2] + \mathcal{I}[m_s^2] \Big]. \nonumber
\eea
In the case of up quarks $\hat{g}_2 f_{\substack{uW \\ ss}} \rightarrow  \hat{g}_1 f_{\substack{uB \\ ss}}$, while in the case
of down quarks $\hat{g}_2 f_{\substack{dW \\ ss}} \rightarrow  - \hat{g}_1 f_{\substack{dB \\ ss}}$.
The remaining contributions proportional to the SM loop functions are
\begin{align}
[Y_e]_{ss} \, f_{\substack{eH \\ ss}} &= \hat{e}^2\, \frac{Q_\ell^2}{2} A_{1/2}(\tau_s), \nonumber \\
[Y_u]_{ss} \, f_{\substack{uH \\ ss}} &= N_c \, \hat{e}^2\, \frac{Q_u^2}{2} A_{1/2}(\tau_s),  \nonumber \\
 [Y_d]_{ss} \, f_{\substack{dH \\ ss}} &= N_c \, \hat{e}^2\, \frac{Q_d^2}{2} A_{1/2}(\tau_s),  \nonumber \\
f_{H \Box} &= -\hat{e}^2 \frac{Q_\ell^2}{2} A_{1/2}(\tau_p) - N_c \, \hat{e}^2\, \frac{Q_u^2}{2} A_{1/2}(\tau_r)
 -  N_c \, \hat{e}^2 \,\frac{Q_d^2}{2} A_{1/2}(\tau_s) - \frac{1}{2} \, \hat{e}^2\, A_1(\tau_{W}),
\end{align}
and $f_{HD}  = -f_{H \Box} /4$. Here $p,r,s$ run over $1,2,3$ as flavor indices. Several of these results
have been cross checked against Ref.~\cite{Ghezzi:2015vva}.
\subsubsection{$\langle \phi_4| \mathcal{A} \mathcal{A} \rangle^0_{\mathcal{L}^{(8)}}$}
Directly from the general definition in Eqn.~\eqref{threepointamps}, the  ${\cal{O}}(v^4/\Lambda^4)$ terms in the full three-point function are \cite{Hays:2020scx}
\begin{align}
	\label{eq:hggDim8}
	\langle \phi_4|\mathcal{A}\mathcal{A} \rangle_{{\cal{O}}(v^4/\Lambda^4)}^0 &=
   \langle \sqrt{h}^{44}\rangle_{{\cal{O}}(v^2/\Lambda^2)} \, \langle \phi_4|\mathcal{A} \mathcal{A}\rangle^0_{\mathcal{O}(\bar{v}^2/\Lambda^2)}
  + 2  \, \frac{\bar{v}_T \, [\langle \phi_4|\mathcal{A}\mathcal{A}\rangle_{\mathcal{O}(\bar{v}^2/\Lambda^2)}^0]^2}{\langle \phi_4 \mathcal{A}^{\mu\nu} \mathcal{A}_{\mu \nu} \rangle^0}, \\
   &+ 2 \, \langle \phi_4|\mathcal{A}\mathcal{A}\rangle_{\mathcal{O}(\bar{v}^2/\Lambda^2)}^0|_{C_i^{(6)} \rightarrow C_i^{(8)}}.\nonumber
\end{align}

Here we have used the short-hand notation
\begin{align}
	\langle \sqrt{h}^{44}\rangle_{{\cal{O}}(v^2/\Lambda^2)}
	&= \tilde C_{H\Box}^{(6)} - \frac{1}{4}\tilde C_{HD}^{(6)}, & \quad C_{HB}^{(6)} &\rightarrow \frac{1}{2} C_{HB}^{(8)},  \\
	C_{HW}^{(6)} &\rightarrow \frac{1}{2} \left(  C_{HW}^{(8)}+C_{HW,2}^{(8)}\right), &  \quad
	C_{HWB}^{(6)} &\rightarrow \frac{1}{2} C_{HWB}^{(8)}.
\end{align}

Combining all of these results we have the desired expression for the decay width
\bea
\Gamma_{\rm SMEFT}(h \rightarrow \gamma \gamma) &=& \frac{\hat{m}_h^3}{4 \pi}
|\langle \phi_4|\mathcal{A} \mathcal{A}\rangle_{SM}^1
+ \langle \phi_4|\mathcal{A} \mathcal{A}\rangle^0_{\mathcal{O}(\bar{v}_T^2/\Lambda^2)} + \langle \phi_4|\mathcal{A}\mathcal{A} \rangle_{{\cal{O}}(\bar{v}_T^4/\Lambda^4)}^0 + \langle \phi_4|\mathcal{A} \mathcal{A}\rangle^1_{\mathcal{L}^{(6)}}|^2 \nn
\eea

\begin{center}
\begin{table}
\centering
\tabcolsep 8pt
\begin{tabular}{|c|c|c|}
\hline
Input parameters&Value&Ref.\\
\hline
$\hat m_Z$ [GeV]&$91.1876\pm0.0021$&\cite{Zyla:2020zbs}\\
$\hat m_W$ [GeV]&$80.387\pm0.016$&\cite{Aaltonen:2013iut}\\
$\hat m_h$ [GeV] &$125.10\pm0.14$&\cite{Zyla:2020zbs}\\
$\hat m_t$ [GeV] &$172.4\pm0.7$&\cite{Zyla:2020zbs}\\
$\hat{m}_b$ [GeV]&     $4.18\pm0.03$ & \cite{Olive:2016xmw}\\
$\hat{m}_c$ [GeV]&     $1.27\pm0.02$   & \cite{Olive:2016xmw}\\
$\hat{m}_\tau$ [GeV]&  $1.77686\pm0.00012$   & \cite{Olive:2016xmw}\\
$\hat{G}_F$ [GeV$^{-2}$] & 1.1663787 $\cdot 10^{-5}$&  \cite{Olive:2016xmw,Mohr:2012tt} \\
$\hat \alpha_{EW}$&1/137.03599084(21)&\cite{Zyla:2020zbs}\\
$\Delta\alpha$&$0.0590\pm0.0005$&\cite{Dubovyk:2019szj}\\
$\hat \alpha_s$&$0.1179\pm0.0010$&\cite{Zyla:2020zbs}\\
\hline
$m_W^{\hat\alpha}$&$80.36\pm0.01$&--\\
$\Delta\alpha^{\hat m_W}$&$0.0576\pm0.0008$&--\\
\hline
\end{tabular}
\caption{Input parameter values used from Ref.~\cite{Corbett:2021eux}.
$m_W^{\hat\alpha}$ is the value of $m_W$ inferred in the $\{\hat\alpha,\hat m_Z,\hat G_F\}$ scheme
using the interpolation formula of Refs.~\cite{Freitas:2014hra,Awramik:2003rn,Awramik:2006uz,Dubovyk:2019szj},
while $\Delta\alpha^{\hat m_W}$ is the shift in the
value of alpha due to hadronic effects for the $\{\hat m_W,\hat m_Z,\hat G_F\}$ scheme.  The remaining SM inputs are taken from the central values in the PDG \cite{Olive:2016xmw}.}
\label{tab:inputs}
\end{table}
\end{center}

\section{Numerics}\label{numerics}
Using the equations in the previous section, we can now present numerical replacement formulae
that can be used to generate results at one loop order and dimension eight from a leading order
simulation result using SMEFTsim.\footnote{This approach to next to leading order corrections
is essentially that laid out in Ref.~\cite{Trott:2021vqa}.} We use the numerical inputs in Table.~\ref{tab:inputs}

At one loop many gauge independent finite terms feed into the results as common numerical pre-factors in the various predictions. Here we collect these intermediate numerical results in both input parameter schemes.

Before presenting the numerics, it is important to emphasize the scope of our results. We have assumed a bottom-up SMEFT construction and calculated the effects in Higgs observables in a joint expansion of SMEFT coefficients and $\alpha_s$. The bottom-up perspective means we have not assumed anything about the sizes of the Wilson coefficients. When large hierarchies between coefficients are present, as can arise in certain UV scenarios, it is possible that terms that are higher order than what we have calculated, e.g. $\mathcal O(\bar v^2_T/(16\pi^2)^2\Lambda^2), \mathcal O(\bar v^4_T/(16\pi^2\Lambda^4)$ may be imprtant\footnote{As a specific example, in scenarios with vector-like quarks, $C^{(6)}_{HG}$ is generated at 1-loop level while $C^{(6)}_{uH}$ is generated at tree level~\cite{2013,2000}.}. As mentioned earlier, we have also assumed CP conservation.

\begin{table}[t]\centering
\renewcommand{\arraystretch}{1.2}
 \begin{tabular}{cc|cc}\hline
 $\Delta R^{\hat{m}_W}_{\mathcal{A}}$ & 0.12 & $\Delta R^{\hat{\alpha}_{ew}}_{\mathcal{A}}$ & 0.12 \\
 $\Delta G_F^{\hat{m}_W}$ & 0.024 & $\Delta G_F^{\hat{\alpha}_{ew}}$ & 0.024\\
 $\Delta R_{M^2_W}^{\hat{m}_W}$ & -0.040 & $\Delta R_{M^2_W}^{\hat{\alpha}_{ew}}$ &-0.040\\
 $\Delta R_{M^2_Z}^{\hat{m}_W}$ & -0.054  & $\Delta R_{M^2_Z}^{\hat{\alpha}_{ew}}$ &-0.054\\
 $\frac{\Delta R_{\phi_4}^{\hat{m}_W}}{2}+ \frac{\Delta v}{v}$ & -0.003 & $\frac{\Delta R_{\phi_4}^{\hat{m}_W}}{2}+ \frac{\Delta v}{v}$  & -0.003\\
 $M_1^{\hat{m}_W}$ & -0.0097 & $M_1^{\hat{\alpha}_{ew}}$ & -0.0098\\
 $\Delta g_1^{\hat{m}_W}$ & -0.014 & $\Delta g_1^{\hat{\alpha}_{ew}}$ & -0.096\\
 $\Delta g_2^{\hat{m}_W}$ & -0.0055 & $\Delta g_2^{\hat{\alpha}_{ew}}$ & 0.040 \\
 \hline
 \end{tabular}\caption{Numerical values of the one loop corrections to various Lagrangian parameters
 and matrix element corrections in both input schemes.
 We only report gauge independent combinations of parameters. We have chosen $\mu= \hat{m}_h$ in these evaluations
 for the scale dependence associated with the one loop improvement of input parameters and
 finite on shell renormalization conditions in the LSZ formula. For operator mixing effects, we
set $\mu = \Lambda$.}\label{tab:inputs2}
  \end{table}

\subsection{$\sigma(\mathcal G \mathcal G \rightarrow h)$}
For the numerics, we use NNPDF3.0 NLO parton distributions~\cite{Hartland_2013,Ball_2015} for $\alpha_s = 0.118$ and set all $\mu$ scales to $\hat m_h$ except for those associated with operator mixing, which are fixed at $\mu = \Lambda$. With these inputs, and using values from Table~\ref{tab:inputs}, the SM cross section for $\mathcal G \mathcal G \to h$ ($\sqrt s = 13\, \text{TeV}$), in the $m_t \to \infty$ limit and retaining $\mathcal O(\alpha^3_s)$ and $\mathcal O(\bar m^2_h/m^2_t)$ corrections, is
\begin{align}
\sigma^{1/m^2_t}_{SM}(\mathcal G \mathcal G \to h)_{\mathcal O(\alpha^2_s, \alpha^3_s)} = \frac{\pi}{4}|\mathcal{A}^{\mathcal{G} \mathcal{G}\phi_4}_{SM}|^2_{\mathcal O(\alpha^2_s, \alpha^3_s)}\frac{\partial \mathcal L_{gg}}{\partial \tau} =  18.15\, \text{pb}
\end{align}
Forming the ratio of the inclusive cross section at $\mathcal O(\bar v^4_T/\Lambda^4, \bar v^2_T/16\pi^2\Lambda^2)$ with the SM in the $\hat \alpha$ input scheme, we find:
\begin{align}
\frac{ \sigma^{\hat{\alpha}}_{\rm SMEFT}(\mathcal G\mathcal G \to h)}{\sigma^{\hat{\alpha}, 1/m^2_t}_{\rm SM}(\mathcal G\mathcal G \to h)_{\mathcal O(\alpha^2_s, \alpha^3_s)} }\simeq 1 & +  519\, \tilde C^{(6)}_{HG} \nn &+ 504\, \tilde C^{(6)}_{HG}\Big(\tilde C^{(6)}_{H\Box} - \frac 1 4 \tilde C^{(6)}_{HD} \Big) + 8.15\times10^4\, (\tilde C^{(6)}_{HG})^2 + 504\, \tilde C^{(8)}_{HG} \nn
& + 1.58\, \Big(\tilde C^{(6)}_{H\Box} - \frac 1 4 \tilde C^{(6)}_{HD} \Big) + 362\, \tilde C^{(6)}_{HG} -1.59\, \tilde C^{(6)}_{uH} - 12.6\, {\rm Re }\, \tilde C^{(6)}_{uG} \nn
& - 1.12\,\delta G^{(6)}_F - 7.70\, {\rm Re }\, \tilde C^{(6)}_{uG}\,\log\Big(\frac{\hat m^2_h}{\Lambda^2} \Big) - 0.19\,{\rm Re}\,\tilde C^{(6)}_{dG}\,\log\Big(\frac{\hat m^2_h}{\Lambda^2} \Big)  \nn
&-0.09\,{\rm Re}\,\tilde C^{(6)}_{dG} + 3.54\, \tilde C^{(6)}_{dH}
\label{eq:gghrationumeric}
\end{align}
The first line of Eq.~\eqref{eq:gghrationumeric} is the linear tree-level $\mathcal L_6$ correction, the second line is the (tree-level) $(\mathcal L_6)^2$ and $\mathcal L_8$ corrections, and the last two lines are the one loop $\mathcal L_6$ corrections. Extracting the $\bar v^2_T/\Lambda^2, (\bar v^4_T/\Lambda^4)$ factors out of  $\tilde C^{(6)}\, (\tilde C^{(8)})$, one can convert these expressions to the un-tilded coefficients and $\Lambda$. In this form, the terms on the second line will be suppressed relative to the others by one factor of $\bar v^2_T/\Lambda^2$.
To the accuracy we have quoted the numerical factors, there is no difference between the $\hat{\alpha}$ and $\hat m_W$ input schemes. The $\sim 3\%$ difference between the $\tilde C^{(6)}_{HG}$ coefficient on the first line and the $\tilde C^{(6)}_{HG}\,C^{(6)}_{H\Box}$ and $\tilde C^{(8)}_{HG}$ coefficients on the second line is due to the fact we've included $\mathcal O(\bar m^2_h/\bar m^2_t)$ corrections for the former.

Inspecting the terms on the right hand side the largest coefficient accompanies the $(\tilde C^{(6)}_{HG})^2$ term. The large coefficient arises as it is a tree-level term (squared), divided by the loop level SM result -- a relation we can make more explicit by rewriting the $(\tilde C^{(6)}_{HG})^2$ term as $7.19\, (\frac{4\pi}{\alpha_s})^2\, (\tilde C^{(6)}_{HG})^2$.  Introducing hierarchies among coefficients motivated by UV assumptions, such as suggested by Ref.~\cite{Craig:2019wmo}, may change which term dominates on the right hand side. Comparisons between the $\bar v^4_T/\Lambda^4$ terms with the loop level $\mathcal L_6$ terms depend strongly on the choice of $\Lambda$ (as the latter are suppressed relative to the former by $\bar v^2_T/\Lambda^2$), and also on assumptions on the relative sizes of the coefficients.  The dependence of Eq.~\eqref{eq:gghrationumeric} on these choices and the implications for error truncation studies, will be explored in more detail in a separate publication.

\subsection{$\Gamma(h\, \rightarrow \mathcal G \mathcal G)$}
The leading SM result for $\Gamma(h \to \mathcal G \mathcal G)$, working to the same coupling and $m_t$ order as in $\sigma(\mathcal G \mathcal G \to h)$ is
\begin{align}
\Gamma_{\rm SM}(h \to \mathcal G \mathcal G)_{\mathcal O(\alpha^2_s, \alpha^3_s)} = \frac{2\, \hat m^3_h}{\pi}|\mathcal{A}^{\mathcal{G} \mathcal{G}\phi_4}_{SM}|^2_{\mathcal O(\alpha^2_s, \alpha^3_s)} = 2.55\times 10^{-4}\,\text{GeV}
\end{align}
Dividing the $\mathcal O(\bar v^4_T/\Lambda^4, \bar v^2_T/16\pi^2\Lambda^2)$ by this result and working in the $\hat{\alpha}$ scheme:
\begin{align}
\frac{\Gamma^{\hat{\alpha}}_{SMEFT}(h \to \mathcal G \mathcal G)}{\Gamma^{\hat{\alpha}}_{SM}(h \to \mathcal G \mathcal G)} \simeq 1 & +  519\, \tilde C^{(6)}_{HG} \nn
&+ 504\, \tilde C^{(6)}_{HG}\Big(\tilde C^{(6)}_{H\Box} - \frac 1 4 \tilde C^{(6)}_{HD} \Big) + 8.15\times10^4\, (\tilde C^{(6)}_{HG})^2 + 504\, \tilde C^{(8)}_{HG} \nn
& + 1.58\, \Big(\tilde C^{(6)}_{H\Box} - \frac 1 4 \tilde C^{(6)}_{HD} \Big) + 296\, \tilde C^{(6)}_{HG} -1.59\, \tilde C^{(6)}_{uH} - 12.6\, {\rm Re }\, \tilde C^{(6)}_{uG} \nn
& - 1.12\,\delta G^{(6)}_F - 7.70\,{\rm Re }\,  \tilde C^{(6)}_{uG}\,\log\Big(\frac{\hat m^2_h}{\Lambda^2} \Big)- 0.19\,{\rm Re}\,\tilde C^{(6)}_{dG}\,\log\Big(\frac{\hat m^2_h}{\Lambda^2} \Big)  \nn
&-0.09\,{\rm Re}\,\tilde C^{(6)}_{dG} + 3.54\, \tilde C^{(6)}_{dH},
\end{align}
with the linear tree-level $\mathcal L_6$ correction on the first line, the (tree-level) $(\mathcal L_6)^2$ and $\mathcal L_8$ corrections on the second line, and one loop $\mathcal L_6$ corrections on the third and fourth lines. As with $\sigma(\mathcal G\mathcal G \to h)$, there is no difference between the $\hat{\alpha}$ and $\hat m_W$ input schemes.

Comparing the expressions here with Eq.~\eqref{eq:gghrationumeric}, the results are identical except for the $\tilde C^{(6)}_{HG}$ terms. To the order we have worked, the parton distribution functions factor out of $\sigma_{\rm SM}(\mathcal G \mathcal G \to h)$ and cancel out in the ratio for all SMEFT pieces proportional to $\delta(1-z)$, so these pieces should match in the ratios for $\sigma(\mathcal G \mathcal G \to h)$ and $\Gamma(h \to \mathcal G \mathcal G)$. The only exception is $\tilde C^{(6)}_{HG}$, which is sensitive to how the IR divergences in the QCD loop and extra emission cancel, and the kinematic differences in the two processes lead to slightly different coefficients. Additionally, there are  $\tilde C^{(6)}_{HG}$ pieces of $\sigma_{\rm SMEFT}(\mathcal G \mathcal G \to h)$ with different parton distribution function dependency that do not cancel neatly in the ratio with the SM.

\subsection{$\Gamma(h \rightarrow \gamma \gamma)$}
Using the input parameters in Table~\ref{tab:inputs}, we find the following SM leading-order
$h \rightarrow \gamma \gamma$ partial width in the $\hat{m}_W$ scheme:
\bea
\Gamma^{ \hat{m}_W}_{\rm SM}(h \rightarrow \gamma \gamma) &=& \frac{\hat{m}_h^3}{4 \pi} \, \bigg|\mathcal{A}_{\rm SM}^{h\gamma\gamma}  \bigg|^2 \nn
&=& 1.00 \times 10^{-5}\, {\rm GeV}.
\eea

The full result to $\mathcal{O}(\bar{v}_T^4/\Lambda^4)$ was reported in Ref.~\cite{Hays:2020scx}.
Adding the $\mathcal{O}(\bar{v}_T^2/ 16 \pi^2\Lambda^2)$ loop corrections with consistent
theoretical scheme conventions,  in the $m_W$ input parameter scheme, we have the result
for $h \rightarrow \gamma \gamma$
\begin{align}
\frac{\Gamma^{\hat{m}_{W}}_{SMEFT}}{\Gamma^{\hat{m}_{W}}_{\rm SM}}
&\simeq 1 -  788 f^{\hat{m}_W}_1, \\
 &+ 394^2 \, (f^{\hat{m}_W}_1)^2
- 351 \, (\tilde{C}_{HW}^{(6)} - \tilde{C}_{HB}^{(6)})\, f^{\hat{m}_W}_3 + 2228 \, \delta G_F^{(6)} \, f^{\hat{m}_W}_1, \nonumber \\
&+  979 \, \tilde{C}_{HD}^{(6)}(\tilde{C}_{HB}^{(6)} +0.80\, \, \tilde{C}_{HW}^{(6)} -1.02  \, \tilde{C}_{HWB}^{(6)})
-788 \left[ \left(\tilde C_{H\Box}^{(6)} - \frac{\tilde C_{HD}^{(6)}}{4}\right)  \, f^{\hat{m}_W}_1+ f^{\hat{m}_W}_2\right], \nonumber \\
&+2283 \, \tilde{C}_{HWB}^{(6)}(\tilde{C}_{HB}^{(6)} +0.66 \, \, \tilde{C}_{HW}^{(6)} -0.88  \, \tilde{C}_{HWB}^{(6)})
- 1224 \, (f^{\hat{m}_W}_1)^2, \nn
&- 117 \, \tilde{C}_{HB}^{(6)} - 23 \, \tilde{C}_{HW}^{(6)} + \left[51 + 2 \log \left(\frac{\hat{m}_h^2}{\Lambda^2}\right)\right] \, \tilde{C}_{HWB}^{(6)}
+ \left[-0.55 + 3.6 \log \left(\frac{\hat{m}_h^2}{\Lambda^2}\right)\right]\, \tilde{C}_{W}^{(6)}, \nn
&+ \left[27 - 28 \log \left(\frac{\hat{m}_h^2}{\Lambda^2}\right)\right]\, {\rm{Re} \, \tilde{C}_{\substack{uB \\ 33}}^{(6)}}
+ \left[14 - 15\log \left(\frac{\hat{m}_h^2}{\Lambda^2}\right)\right]\, {\rm{Re} \, \tilde{C}_{\substack{uW \\ 33}}^{(6)}}
+ 0.56\, {\rm{Re} \, {\tilde{C}_{\substack{uH \\ 33}}^{(6)}}}, \nn
&- 0.31\, {\rm Re} \,{\tilde{C}_{\substack{dH \\ 33}}^{(6)}} + 2 \, \tilde{C}_{H \Box}^{(6)} - \frac{\tilde{C}_{HD}^{(6)}}{2} +2.0 \, \tilde{C}_{HD}^{(6)} -7.5 \, \tilde{C}_{HWB}^{(6)} - 3 \,\sqrt{2} \,\delta G_F^{(6)}. \nonumber
\end{align}
The results have been presented in a manner to make clear the origin of the various contributions.
First the corrections are up to $\mathcal{O}(\bar{v}_T^4/\Lambda^4)$ terms in the operator expansion.
Next the one loop contributions involving novel one loop diagrams and operator mixing in the SMEFT
are given. The contributions from rescaling the SM amplitude for a series of corrections are then reported.
Finally, the last line is due to input parameter corrections to the SM amplitude.
Several numerically small corrections that follow from the formulae given are neglected here as the contributions are
negligible compared to the retained terms. These neglected corrections are generally further suppressed by small
Yukawa couplings. Here $f^{\hat{m}_W}_i \simeq f^{\hat{\alpha}_{ew}}_i$ for $i=1,2,3$ and
\begin{align}
 \delta G_F^{(6)} &= \frac{1}{\sqrt2} \left(\tilde C^{(3)}_{\substack{Hl \\ee}}+\tilde C^{(3)}_{\substack{Hl \\ \mu \mu}} - \frac{1}{2}(\tilde C'_{\substack{ll \\ \mu ee \mu}}+\tilde C'_{\substack{ll \\ e \mu \mu e}})\right),\\
f^{\hat{m}_W}_1 &=  \left[\tilde{C}_{HB}^{(6)} +0.29 \, \, \tilde{C}_{HW}^{(6)} -0.54  \, \tilde{C}_{HWB}^{(6)}\right],\\
f^{\hat{m}_W}_2 &=   \left[\tilde{C}_{HB}^{(8)} +0.29 \, \, (\tilde{C}_{HW}^{(8)}+ \tilde{C}_{HW,2}^{(8)}) -0.54  \, \tilde{C}_{HWB}^{(8)}\right],\\
f^{\hat{m}_W}_3 &= \left[\tilde{C}_{HW}^{(6)} - \tilde{C}_{HB}^{(6)} -0.66  \, \tilde{C}_{HWB}^{(6)}\right].
\end{align}
From Table~\ref{tab:inputs}, we also find the following SM leading-order
$h \rightarrow \gamma \gamma$ partial width in the $\hat{\alpha}$ scheme:
\bea
\Gamma^{ \hat{\alpha}}_{\rm SM}(h \rightarrow \gamma \gamma) &=& \frac{\hat{m}_h^3}{4 \pi} \, \bigg|\mathcal{A}_{\rm SM}^{h\gamma\gamma}  \bigg|^2 \nn
&=& 1.06 \times 10^{-5}\, {\rm GeV}.
\eea
and the corrections to the decay width
\begin{align}
\frac{\Gamma_{SMEFT}^{\hat{\alpha}_{ew}}}{ \Gamma^{\hat{\alpha}_{ew}}_{\rm SM}} &\simeq
1 -  758 f^{\hat{\alpha}_{ew}}_1, \nn
&+379^2 \, (f^{\hat{\alpha}_{ew}}_1)^2
- 350 \, (\tilde{C}_{HW}^{(6)} - \tilde{C}_{HB}^{(6)})^2 - 1159 \, (f^{\hat{\alpha}_{ew}}_1)^2 \nn
&- 61\,\tilde{C}_{HWB}^{(6)} \, \left(\tilde{C}_{HB}^{(6)} +7.2 \tilde{C}_{HW}^{(6)} -9.2 \tilde{C}_{HWB}^{(6)} \right)
- 13.5\,\tilde{C}_{HD}^{(6)} \, \left(\tilde{C}_{HB}^{(6)} +16 \tilde{C}_{HW}^{(6)} -15 \tilde{C}_{HWB}^{(6)} \right) \nn
&+ 1383\, \delta G_F^{(6)} \, \left(\tilde{C}_{HB}^{(6)}-0.13 \tilde{C}_{HW}^{(6)} -0.15 \tilde{C}_{HWB}^{(6)} \right)
-  758 \left[\left(\tilde C_{H\Box}^{(6)} -\frac{\tilde C_{HD}^{(6)}}{4}\right) f^{\hat{\alpha}_{ew}}_1
+ f^{\hat{\alpha}_{ew}}_2\right],\nn
&- 218 \, \tilde{C}_{HB}^{(6)} +22 \, \tilde{C}_{HW}^{(6)} + \left[-17 + 2.0 \log \left(\frac{\hat{m}_h^2}{\Lambda^2}\right)\right] \, \tilde{C}_{HWB}^{(6)}
+ \left[-0.60 + 3.6 \log \left(\frac{\hat{m}_h^2}{\Lambda^2}\right)\right]\, \tilde{C}_{W}^{(6)}, \nn
&+ \left[26 -27 \log \left(\frac{\hat{m}_h^2}{\Lambda^2}\right)\right]\, {\rm{Re} \, \tilde{C}_{\substack{uB \\ 33}}^{(6)}}
+ \left[14 - 15\log \left(\frac{\hat{m}_h^2}{\Lambda^2}\right)\right]\, {\rm{Re} \, \tilde{C}_{\substack{uW \\ 33}}^{(6)}} + 0.56\, {\rm{Re} \, {\tilde{C}_{\substack{uH \\ 33}}^{(6)}}}, \nn
&- 0.31\, {\rm Re} \,{\tilde{C}_{\substack{dH \\ 33}}^{(6)}} + 2 \, \tilde{C}_{H \Box}^{(6)} - \frac{\tilde{C}_{HD}^{(6)}}{2} 
- \sqrt{2} \,\delta G_F^{(6)}. \nonumber
\end{align}
Note that the numerical values in the operator expansion, and the value of the
total decay width in this scheme have slightly shifted compared to Ref.~\cite{Hays:2020scx}
due to an update of the $\alpha$ input parameter numerical value. The corrections are reported in the same
order as in the $m_W$ scheme. In this scheme the degeneracy of
the combination of Wilson coefficients $\tilde{C}_{H \Box} - \tilde{C}_{HD}/4$ is not broken
at one loop in this observable. This is because the individual dependence on $\tilde{C}_{HD}$ due to
the shift in $\alpha$ that breaks the degeneracy in the $m_W$ scheme is absent.

Past results in the literature include the dimension eight results in Ref.~\cite{Hays:2020scx}.
One loop results were previously reported in Refs.~\cite{Hartmann:2015oia,Hartmann:2015aia} calculated in the BFM, and
also results calculated using $R_\xi$ gauge fixing outside the BFM in Refs.~\cite{Ghezzi:2015vva,Dedes:2018seb,Dawson:2018liq,Dawson:2018pyl}.
Some theoretical aspects of the loop results are common between calculation schemes. For example,
the breaking of the $\tilde{C}_{H \Box} - \tilde{C}_{HD}/4$ degeneracy in the $m_W$  scheme as in Ref.~\cite{Dawson:2018pyl}
is confirmed in this work, as it is a feature of input parameter dependence.
Other aspects of calculation scheme dependence are distinct, and should not be interpreted
as an error when comparing between schemes. For example, in the BFM each of the input parameter loop corrections
is individually gauge invariant. While in a on-shell scheme as employed in
\cite{Ghezzi:2015vva,Dedes:2018seb,Dawson:2018liq,Dawson:2018pyl} $\Delta G_F$ carries gauge dependence.
There is not a contradiction {\it per se}, although arguably the BFM offers several advantages, such as the gauge independence
of $\Delta G_F$, and also affords cross checks, while aiding in the transparency of results. We have stressed some of these
cross checks in this paper. We strongly caution against over-interpreting scheme dependence
differences in calculations of this form when comparing results as a calculation error
and again emphasize that scheme choices and operator normalizations effect both the loop expansion and
the operator expansion in the SMEFT demanding consistent conventions be defined and adhered to when combining
results.

\section{Discussion and conclusion}\label{discussion}
In this paper we have reported formulae for three processes
$\sigma(\mathcal{G} \,\mathcal{G}\rightarrow h)$, $\Gamma(h \rightarrow \mathcal{G} \,\mathcal{G})$
and $\Gamma(h \rightarrow \gamma \gamma)$. We have including a complete set of
$\mathcal{O}(\bar{v}_T^2/16 \pi^2 \Lambda^2)$ and $\mathcal{O}(\bar{v}_T^4/\Lambda^4)$
corrections at the amplitude level for these processes, with common calculation
scheme dependence across the results. We have stressed the point that the operator expansion and the
loop expansion are not independent at sub-leading order, but are correlated by calculational scheme dependence.
The formulas reported include a consistent
set of SMEFT corrections not only to the processes, and also to the input parameter measurements
defining Lagrangian parameters in both expansions. To our knowledge these are the first complete results of this form
in the literature. These calculations were enabled due to the fundamental interplay of the
Background Field Method approach to gauge fixing with the geoSMEFT (i.e. scalar background field) formulation
of the SMEFT. In a real sense, this completes at sub-leading order the theoretical program
of studying the central production and decay processes of the Higgs at LHC in a model independent EFT, as initiated
in Ref.~\cite{Manohar:2006gz} by Manohar and Wise.

The semi-analytic formulae for these inclusive processes are of interest
as they can be used to generate a consistent set of higher order terms in each of
the expansions present, from the LO simulation results using SMEFTsim \cite{Brivio:2017btx,Brivio:2019myy}.
This is possible as the three-point vertices in inclusive observables have common kinematics (populations of phase space) at LO
and also at $\mathcal{O}(\bar{v}_T^2/16 \pi^2 \Lambda^2)$ and $\mathcal{O}(\bar{v}_T^4/\Lambda^4)$.
The methodology of how to use these formulae was recently laid out in Ref.~\cite{Trott:2021vqa};
the LO dependence on the SMEFT perturbation can be used to define a simple replacement/swap formulae
to generate a set of
$\mathcal{O}(\bar{v}_T^2/16 \pi^2 \Lambda^2)$ and $\mathcal{O}(\bar{v}_T^4/\Lambda^4)$ corrections,
in these inclusive observables. Importantly, this can be done {\it post-facto} using LO simulation results
bypassing the need for costly, and redundant, Monte Carlo simulation on all of the parameters in the sub-leading terms
in both expansions.

Knowing the dependence to sub-leading order
on the SMEFT parameters in the operator and loop expansions in these measurable processes
is primarily of interest to inform a theoretical
error estimate, when such sub-leading terms are neglected in studies of LHC data. The formulae are also of intrinsic interest
as they demonstrate the convergence (or lack there of) of the SMEFT perturbations to the SM.
The large size of the sub-leading (``quadratic") terms is due to the interference with the
loop suppressed SM amplitude at LO in each case. A SMEFT perturbation
that is directly self-squared at sub-leading order avoids the corresponding loop suppression in the
SM amplitudes, leading to a very large numerical enhancement, overwhelming the naive suppression by
$\bar{v}_T^2/\Lambda^2$ compared to the LO SMEFT perturbation. Even so, at sub-leading order
in the operator expansion, a set of new
parameters enter into the predictions, so that a consistent use of quadratic terms
when studying experimental data should include this larger set of SMEFT parameters we have reported
or a well defined theoretical error if these extra terms are neglected.
We leave a detailed numerical study on the implications of the results reported here to a future publication.

These results are reported in a transparent semi-analytic fashion to aid cross checking of the results
by external theory groups.

\acknowledgments
We thank Chris Hays and Andreas Helset for insightful discussions.
M.T. acknowledges support from the Villum Fund, project number 00010102.
We thank Peter Stoffer for useful discussions on the LEFT results used.
T.C. acknowledges funding from European Union’s Horizon 2020 research and innovation programme under the Marie Sklodowska-Curie grant agreement No. 890787.
The work of A.M. was supported in part by the National Science Foundation under Grant Number PHY-1820860.

\appendix
\section{Appendix}\label{appendix}
Here we give explicit results for the corrections to $m_{W,Z}^2$ at one loop in the BFM.
These expressions are reported with $\xi =1$ for brevity of presentation.
We have confirmed that the combination of these corrections with $2 \Delta v/v$ is gauge independent, as expected.
\begin{equation}
\begin{array}{rl}
-\frac{\Delta M_Z^2}{\bar m_Z^2} &=
\frac{\bar g_1^2+\bar g_2^2}{576\pi^2}\frac{288\bar m_W^6-3\bar m_h^4\bar m_Z^2+408\bar m_W^4\bar m_Z^2+18\bar m_h^2\bar m_Z^4-88\bar m_W^2\bar m_Z^4-79\bar m_Z^6}{\bar m_Z^6}\\
&-\frac{\bar g_1^2+\bar g_2^2}{96\pi^2}\frac{7\bar m_Z^4-42\bar m_W^4+10\bar m_W^2\bar m_Z^2}{\bar m_Z^4}\log\left(\frac{\mu^2}{\bar m_h^2}\right)-\frac{\bar g_1^2+\bar g_2^2}{192\pi^2}\frac{\bar m_Z^4-84\bar m_W^4+20\bar m_W^2\bar m_Z^2}{\bar m_Z^4}\log\left(\frac{\bar m_h^2}{\bar m_W^2}\right)\\
&+\frac{\bar g_1^2+\bar g_2^2}{384\pi^2}\frac{\bar m_h^6-6\bar m_h^4\bar m_Z^2+18\bar m_h^2\bar m_Z^4-26\bar m_Z^6}{\bar m_Z^6}\log\left(\frac{\bar m_h^2}{\bar m_Z^2}\right)\\
&-\frac{\bar g_1^2+\bar g_2^2}{192\pi^2}\frac{\bar m_h^4-4\bar m_h^2\bar m_Z^2+12\bar m_Z^4}{\bar m_Z^4}{\rm Disc}\left[\bar m_Z,\bar m_Z,\bar m_h\right]\\
&+\frac{\bar g_1^2+\bar g_2^2}{192\pi^2}\frac{48\bar m_W^6+68\bar m_W^4\bar m_Z^2-16\bar m_W^2\bar m_Z^4-\bar m_Z^6}{\bar m_Z^6}{\rm Disc}\left[\bar m_Z,\bar m_W,\bar m_W\right]\\
&-\sum_\nu\frac{\bar g_1^2+\bar g_2^2}{288\pi^2}\left[5+3\log\left(-\frac{\mu^2}{\bar m_Z^2}\right)\right]-\sum_l\frac{\bar g_1^2+\bar g_2^2}{288\pi^2}\frac{5\bar m_Z^2(8\bar m_W^4-12\bar m_W^2\bar m_Z^2+5\bar m_Z^4)+6\bar m_l^2(16\bar m_W^4-24\bar m_W^2\bar m_Z^2+7\bar m_Z^4)}{\bar m_Z^6}\\
&-\sum_l\frac{\bar g_1^2+\bar g_2^2}{96\pi^2}\frac{8\bar m_W^4-3(\bar m_l^2+4\bar m_W^2)\bar m_Z^2+5\bar m_Z^4}{\bar m_Z^4}\log\left(\frac{\mu^2}{\bar m_l^2}\right)\\
&-\sum_l\frac{\bar g_1^2+\bar g_2^2}{96\pi^2}\frac{8\bar m_W^4\bar m_Z^2-12\bar m_W^2\bar m_Z^4+5\bar m_Z^6+\bar m_l^2(16\bar m_W^4-24\bar m_W^2\bar m_Z^2+7\bar m_Z^4)}{\bar m_Z^6}{\rm Disc}\left[\bar m_Z,\bar m_l,\bar m_l\right]\\
&-\sum_u\frac{(\bar g_1^2+\bar g_2^2)N_c}{2592\pi^2}\frac{6\bar m_u^2(64\bar m_W^4-80\bar m_W^2\bar m_Z^2+7\bar m_Z^4)+5\bar m_Z^2(32\bar m_W^4-40\bar m_W^2\bar m_Z^2+17\bar m_Z^4)}{\bar m_Z^6}\\
&-\sum_u\frac{(\bar g_1^2+\bar g_2^2)N_c}{864\pi^2}\frac{32\bar m_W^4-(27\bar m_u^2+40\bar m_W^2)\bar m_Z^2+17\bar m_Z^4}{\bar m_Z^4}\log\left(\frac{\mu^2}{\bar m_u^2}\right)\\
&-\sum_u\frac{(\bar g_1^2+\bar g_2^2)N_c}{864\pi^2}\frac{32\bar m_W^4\bar m_Z^2-40\bar m_W^2\bar m_Z^4+17\bar m_Z^6+\bar m_u^2(64\bar m_W^4-80\bar m_W^2\bar m_Z^2+7\bar m_Z^4)}{\bar m_Z^6}{\rm Disc}\left[\bar m_Z,\bar m_u,\bar m_u\right]\\
&-\sum_d\frac{(\bar g_1^2+\bar g_2^2)N_c}{2592\pi^2}\frac{6\bar m_d^2(16\bar m_W^4-8\bar m_W^2\bar m_Z^2-17\bar m_Z^4)+5(8\bar m_W^4\bar m_Z^2-4\bar m_W^2\bar m_Z^4+5\bar M_Z^6)}{\bar m_Z^6}\\
&-\sum_d\frac{(\bar g_1^2+\bar g_2^2)N_c}{864\pi^2}\frac{8\bar m_W^4-(27\bar m_d^2+4\bar m_W^2)\bar m_Z^2+5\bar m_Z^4}{\bar m_Z^4}\log\left(\frac{\mu^2}{\bar m_d^2}\right)\\
&-\sum_d\frac{(\bar g_1^2+\bar g_2^2)N_c}{864\pi^2}\frac{8\bar m_W^4\bar m_Z^2-4\bar m_W^2\bar m_Z^4+5\bar m_Z^6+\bar m_d^2(16\bar m_W^4-8\bar m_W^2\bar m_Z^2-17\bar m_Z^4)}{\bar m_Z^6}{\rm Disc}\left[\bar m_Z,\bar m_d,\bar m_d\right],
\end{array}
\end{equation}
\begin{equation}
\begin{array}{rl}
-\frac{\Delta M_W^2}{\bar m_W^2} &=
\frac{\bar g_2^2}{576\pi^2}\frac{610\bar m_W^4-3\bar m_Z^4-3\bar m_h^4+18\bar m_h^2\bar m_W^2-78\bar m_W^2\bar m_Z^2}{\bar m_W^4}+\frac{\bar g_2^2}{96\pi^2}\frac{31\bar m_W^2-6\bar m_Z^2}{\bar m_W^2}\log\left(\frac{\mu^2}{\bar m_W^2}\right)\\
&+\frac{\bar g_2^2}{384\pi^2}\frac{\bar m_h^6-6\bar m_h^4\bar m_W^2+18\bar m_h^2\bar m_W^4}{\bar m_W^6}\log\left(\frac{\bar m_h^2}{\bar m_W^2}\right)+\frac{\bar g_2^2}{384\pi^2}\frac{78\bar m_W^4\bar m_Z^2-14\bar m_W^2\bar m_Z^4-\bar m_Z^6}{\bar m_W^6}\log\left(\frac{\bar m_W^2}{\bar m_Z^2}\right)\\
&-\frac{\bar g_2^2}{192\pi^2}\frac{\bar m_h^4-4\bar m_h^2\bar m_W^2+12\bar m_W^4}{\bar m_W^4}{\rm Disc}\left[\bar m_W,\bar m_W,\bar m_h\right]\\
&+\frac{\bar g_2^2}{192\pi^2}\frac{48\bar m_W^6+68\bar m_W^4\bar m_Z^2-16\bar m_W^2\bar m_Z^4-\bar m_Z^6}{\bar m_W^4\bar m_Z^2}{\rm Disc}\left[\bar m_W,\bar m_W,\bar m_Z\right]\\
&+\sum_{l\nu}\frac{\bar g_2^2}{288\pi^2}\frac{3\bar m_l^4+6\bar m_l^2\bar m_W^2-10\bar m_W^4}{\bar m_W^4}+\sum_{l\nu}\frac{\bar g_2^2}{288\pi^2}\frac{9\bar m_l^2\bar m_W^4-6\bar m_W^6}{\bar m_W^6}\log\left(\frac{\mu^2}{\bar m_l^2}\right)\\
&-\sum_{l\nu}\frac{\bar g_2^2}{96\pi^2}\frac{\bar m_l^6-3\bar m_l^2\bar m_W^4+2\bar m_W^6}{\bar m_W^6}\log\left(\frac{\bar m_l^2}{\bar m_l^2-\bar m_W^2}\right)+\sum_{u_id_j}\frac{\bar g_2^2N_cV_{ij}V^*_{ji}}{288\pi^2}\frac{3(\bar m_{d_j}^2-\bar m_{u_i}^2)^2\bar m_W^2+6(\bar m_{d_j}^2+\bar m_{u_i} ^2)\bar m_W^4-10\bar m_W^6}{\bar m_W^6}\\
&+\sum_{u_id_j}\frac{\bar g_2^2N_c V_{ij}V^*_{ji}}{16\pi^2}\left[\frac{\bar m_{d_j}^2}{\bar m_W^2}\log\left(\frac{\mu^2}{\bar m_d^2}\right)-\frac{3\bar m_{d_j}^2-3\bar m_{u_i}^2+2\bar m_W^2}{6\bar m_W^2}\log\left(\frac{\mu^2}{\bar m_{u_i}^2}\right)\right]\\
&-\sum_{u_id_j}\frac{\bar g_2^2N_cV_{ij}V^*_{ji}}{128\pi^2}\frac{(\bar m_{d_j}^2-\bar m_{u_i}^2)^3+3(\bar m_{u_i}^2-3\bar m_{d_j}^2)\bar m_W^4-2\bar m_W^6}{\bar m_W^6}\log\left(\frac{\bar m_{d_j}^2}{\bar m_{u_i}^2}\right)\\
&+\sum_{u_id_j}\frac{\bar g_2^2N_cV_{ij}V^*_{ji}}{96\pi^2}\frac{(\bar m_{d_j}^2-\bar m_{u_i}^2)^2+(\bar m_{d_j}^2+\bar m_{u_i}^2)\bar m_W^2-2\bar m_W^4}{\bar m_W^4}{\rm Disc}\left[\bar m_W,\bar m_{d_j},\bar m_{u_i}\right].
\end{array}
\end{equation}
\subsection{One Loop Functions}\label{functions}

Here, we have used the notation
\bea
{\rm Disc[a,b,c]} = \frac{\sqrt{a^4-2 a^2 b^2-2 a^2 c^2+b^4-2 b^2 c^2+c^4} \log \left(\frac{-a^2+b^2+c^2+\sqrt{a^4-2 a^2 b^2-2 a^2 c^2+b^4-2 b^2
   c^2+c^4}}{2 b c}\right)}{a^2}.
   \eea
Further
\bea
f (\tau_p) =
\begin{cases}
   \arcsin^2 \sqrt{1/\tau_p} ,&  \tau_p \ge 1\\
    - \frac{1}{4} \left[\ln \frac{1 + \sqrt{1- \tau_p}}{1 - \sqrt{1- \tau_p}} - i \pi \right]^2,              & \tau_p <1,
\end{cases}
\eea
and we also define
\begin{align}
\mathcal{I}[m^2] & \equiv   \int_0^1 dx \,\log \left( \frac{m^2-\bar{m}_h^2\, x \, (1-x)}{\bar{m}_h^2} \right)
& \quad
\mathcal{J}_{x}[m^2] & \equiv  \int_0^1 dx \, \frac{x \, m^2}{m^2 - \bar{m}_h^2\, x\, (1-x)}, \\
\mathcal{I}_y[m^2] & \equiv \int_0^{1-x} dy \int_0^1 dx \, \frac{m^2}{m^2 - m_h^2\, x\, (1-x-y)}.
\end{align}
Here the expressions for $\mathcal{I}, \mathcal{I}_y, \mathcal{I}_{xx} $ for $\tau \ge 1$ (while restricting our results to top loops) are
\bea
\mathcal{I}[m_p] & \equiv& \log(\frac{\tau_p}{4}) + 2 \sqrt{\tau_p -1} \, \arctan \left(\frac{1}{\sqrt{\tau_p -1}} \right) - 2, \\
\mathcal{I}_y [m_p] &\equiv& \frac{\tau_p}{2} \arcsin^2 (1/\sqrt{\tau_p}), \\
\mathcal{I}_{xx}[m_p] & \equiv& \frac{\tau_p}{\sqrt{\tau_{p} - 1}} \,  \arctan \left(\frac{1}{\sqrt{\tau_{p} - 1}}\right).
\eea

\bibliographystyle{JHEP}
\bibliography{bibliography1.bib}

\providecommand{\href}[2]{#2}\begingroup\raggedright\begin{thebibliography}{10}

\bibitem{Brivio:2017vri}
I.~Brivio and M.~Trott, \emph{{The Standard Model as an Effective Field
  Theory}}, \href{https://doi.org/10.1016/j.physrep.2018.11.002}{\emph{Phys.
  Rept.} {\bfseries 793} (2019) 1}
  [\href{https://arxiv.org/abs/1706.08945}{{\ttfamily 1706.08945}}].

\bibitem{Glashow:1961tr}
S.~L. Glashow, \emph{{Partial Symmetries of Weak Interactions}},
  \href{https://doi.org/10.1016/0029-5582(61)90469-2}{\emph{Nucl. Phys.}
  {\bfseries 22} (1961) 579}.

\bibitem{Weinberg:1967tq}
S.~Weinberg, \emph{{A Model of Leptons}},
  \href{https://doi.org/10.1103/PhysRevLett.19.1264}{\emph{Phys. Rev. Lett.}
  {\bfseries 19} (1967) 1264}.

\bibitem{Salam:1968rm}
A.~Salam, \emph{{Weak and Electromagnetic Interactions}}, {\emph{Conf. Proc.}
  {\bfseries C680519} (1968) 367}.

\bibitem{Grzadkowski:2010es}
B.~Grzadkowski, M.~Iskrzynski, M.~Misiak and J.~Rosiek, \emph{{Dimension-Six
  Terms in the Standard Model Lagrangian}},
  \href{https://doi.org/10.1007/JHEP10(2010)085}{\emph{JHEP} {\bfseries 1010}
  (2010) 085} [\href{https://arxiv.org/abs/1008.4884}{{\ttfamily 1008.4884}}].

\bibitem{Alonso:2013hga}
R.~Alonso, E.~E. Jenkins, A.~V. Manohar and M.~Trott, \emph{{Renormalization
  Group Evolution of the Standard Model Dimension Six Operators III: Gauge
  Coupling Dependence and Phenomenology}},
  \href{https://doi.org/10.1007/JHEP04(2014)159}{\emph{JHEP} {\bfseries 1404}
  (2014) 159} [\href{https://arxiv.org/abs/1312.2014}{{\ttfamily 1312.2014}}].

\bibitem{Brivio:2017btx}
I.~Brivio, Y.~Jiang and M.~Trott, \emph{{The SMEFTsim package, theory and
  tools}}, \href{https://doi.org/10.1007/JHEP12(2017)070}{\emph{JHEP}
  {\bfseries 12} (2017) 070}
  [\href{https://arxiv.org/abs/1709.06492}{{\ttfamily 1709.06492}}].

\bibitem{Helset:2020yio}
A.~Helset, A.~Martin and M.~Trott, \emph{{The Geometric Standard Model
  Effective Field Theory}},
  \href{https://doi.org/10.1007/JHEP03(2020)163}{\emph{JHEP} {\bfseries 03}
  (2020) 163} [\href{https://arxiv.org/abs/2001.01453}{{\ttfamily
  2001.01453}}].

\bibitem{Brivio:2020onw}
I.~Brivio, \emph{{SMEFTsim 3.0 \textemdash{} a practical guide}},
  \href{https://doi.org/10.1007/JHEP04(2021)073}{\emph{JHEP} {\bfseries 04}
  (2021) 073} [\href{https://arxiv.org/abs/2012.11343}{{\ttfamily
  2012.11343}}].

\bibitem{Cirigliano_2016}
V.~Cirigliano, W.~Dekens, J.~de~Vries and E.~Mereghetti, \emph{Is there room
  forcpviolation in the top-higgs sector?},
  \href{https://doi.org/10.1103/physrevd.94.016002}{\emph{Physical Review D}
  {\bfseries 94} (2016) }.

\bibitem{Corbett:2019cwl}
T.~Corbett, A.~Helset and M.~Trott, \emph{{Ward Identities for the Standard
  Model Effective Field Theory}},
  \href{https://doi.org/10.1103/PhysRevD.101.013005}{\emph{Phys. Rev. D}
  {\bfseries 101} (2020) 013005}
  [\href{https://arxiv.org/abs/1909.08470}{{\ttfamily 1909.08470}}].

\bibitem{Hays:2020scx}
C.~Hays, A.~Helset, A.~Martin and M.~Trott, \emph{{Exact SMEFT formulation and
  expansion to $\mathcal{O}(v^4/\Lambda^4)$}},
  \href{https://doi.org/10.1007/JHEP11(2020)087}{\emph{JHEP} {\bfseries 11}
  (2020) 087} [\href{https://arxiv.org/abs/2007.00565}{{\ttfamily
  2007.00565}}].

\bibitem{Helset:2018fgq}
A.~Helset, M.~Paraskevas and M.~Trott, \emph{{Gauge fixing the Standard Model
  Effective Field Theory}},
  \href{https://doi.org/10.1103/PhysRevLett.120.251801}{\emph{Phys. Rev. Lett.}
  {\bfseries 120} (2018) 251801}
  [\href{https://arxiv.org/abs/1803.08001}{{\ttfamily 1803.08001}}].

\bibitem{Brivio:2019myy}
I.~Brivio, T.~Corbett and M.~Trott, \emph{{The Higgs width in the SMEFT}},
  \href{https://doi.org/10.1007/JHEP10(2019)056}{\emph{JHEP} {\bfseries 10}
  (2019) 056} [\href{https://arxiv.org/abs/1906.06949}{{\ttfamily
  1906.06949}}].

\bibitem{DeWitt:1967ub}
B.~S. DeWitt, \emph{{Quantum Theory of Gravity. 2. The Manifestly Covariant
  Theory}}, \href{https://doi.org/10.1103/PhysRev.162.1195}{\emph{Phys. Rev.}
  {\bfseries 162} (1967) 1195}.

\bibitem{tHooft:1973bhk}
G.~'t~Hooft, \emph{{An algorithm for the poles at dimension four in the
  dimensional regularization procedure}},
  \href{https://doi.org/10.1016/0550-3213(73)90263-0}{\emph{Nucl. Phys. B}
  {\bfseries 62} (1973) 444}.

\bibitem{Abbott:1981ke}
L.~F. Abbott, \emph{{Introduction to the Background Field Method}}, {\emph{Acta
  Phys. Polon.} {\bfseries B13} (1982) 33}.

\bibitem{Hartmann:2015oia}
C.~Hartmann and M.~Trott, \emph{{On one-loop corrections in the standard model
  effective field theory; the $\Gamma(h \rightarrow \gamma \, \gamma)$ case}},
  \href{https://doi.org/10.1007/JHEP07(2015)151}{\emph{JHEP} {\bfseries 07}
  (2015) 151} [\href{https://arxiv.org/abs/1505.02646}{{\ttfamily
  1505.02646}}].

\bibitem{Ghezzi:2015vva}
M.~Ghezzi, R.~Gomez-Ambrosio, G.~Passarino and S.~Uccirati, \emph{{NLO Higgs
  effective field theory and $\kappa$-framework}},
  \href{https://doi.org/10.1007/JHEP07(2015)175}{\emph{JHEP} {\bfseries 07}
  (2015) 175} [\href{https://arxiv.org/abs/1505.03706}{{\ttfamily
  1505.03706}}].

\bibitem{Dedes:2017zog}
A.~Dedes, W.~Materkowska, M.~Paraskevas, J.~Rosiek and K.~Suxho, \emph{{Feynman
  rules for the Standard Model Effective Field Theory in R$_{\xi}$ -gauges}},
  \href{https://doi.org/10.1007/JHEP06(2017)143}{\emph{JHEP} {\bfseries 06}
  (2017) 143} [\href{https://arxiv.org/abs/1704.03888}{{\ttfamily
  1704.03888}}].

\bibitem{Misiak:2018gvl}
M.~Misiak, M.~Paraskevas, J.~Rosiek, K.~Suxho and B.~Zglinicki,
  \emph{{Effective Field Theories in R$_\xi$ gauges}},
  \href{https://doi.org/10.1007/JHEP02(2019)051}{\emph{JHEP} {\bfseries 02}
  (2019) 051} [\href{https://arxiv.org/abs/1812.11513}{{\ttfamily
  1812.11513}}].

\bibitem{Lehmann:1954rq}
H.~Lehmann, K.~Symanzik and W.~Zimmermann, \emph{{On the formulation of
  quantized field theories}},
  \href{https://doi.org/10.1007/BF02731765}{\emph{Nuovo Cim.} {\bfseries 1}
  (1955) 205}.

\bibitem{'tHooft:1975vy}
G.~'t~Hooft, \emph{{The Background Field Method in Gauge Field Theories}},  in
  \emph{{Functional and Probabilistic Methods in Quantum Field Theory. 1.
  Proceedings, 12th Winter School of Theoretical Physics, Karpacz, Feb 17-March
  2, 1975}}, pp.~345--369, 1975.

\bibitem{Corbett:2020bqv}
T.~Corbett, \emph{{The Feynman rules for the SMEFT in the background field
  gauge}}, \href{https://doi.org/10.1007/JHEP03(2021)001}{\emph{JHEP}
  {\bfseries 03} (2021) 001}
  [\href{https://arxiv.org/abs/2010.15852}{{\ttfamily 2010.15852}}].

\bibitem{Shore:1981mj}
G.~M. Shore, \emph{{Symmetry Restoration and the Background Field Method in
  Gauge Theories}},
  \href{https://doi.org/10.1016/0003-4916(81)90198-6}{\emph{Annals Phys.}
  {\bfseries 137} (1981) 262}.

\bibitem{Einhorn:1988tc}
M.~B. Einhorn and J.~Wudka, \emph{{Screening of Heavy Higgs Radiative
  Effects}}, \href{https://doi.org/10.1103/PhysRevD.39.2758}{\emph{Phys. Rev.}
  {\bfseries D39} (1989) 2758}.

\bibitem{Denner:1994xt}
A.~Denner, G.~Weiglein and S.~Dittmaier, \emph{{Application of the background
  field method to the electroweak standard model}},
  \href{https://doi.org/10.1016/0550-3213(95)00037-S}{\emph{Nucl. Phys.}
  {\bfseries B440} (1995) 95}
  [\href{https://arxiv.org/abs/hep-ph/9410338}{{\ttfamily hep-ph/9410338}}].

\bibitem{Hartmann:2015aia}
C.~Hartmann and M.~Trott, \emph{{Higgs Decay to Two Photons at One Loop in the
  Standard Model Effective Field Theory}},
  \href{https://doi.org/10.1103/PhysRevLett.115.191801}{\emph{Phys. Rev. Lett.}
  {\bfseries 115} (2015) 191801}
  [\href{https://arxiv.org/abs/1507.03568}{{\ttfamily 1507.03568}}].

\bibitem{Hartmann:2016pil}
C.~Hartmann, W.~Shepherd and M.~Trott, \emph{{The $Z$ decay width in the SMEFT:
  $y_t$ and $\lambda$ corrections at one loop}},
  \href{https://doi.org/10.1007/JHEP03(2017)060}{\emph{JHEP} {\bfseries 03}
  (2017) 060} [\href{https://arxiv.org/abs/1611.09879}{{\ttfamily
  1611.09879}}].

\bibitem{Jenkins:2017jig}
E.~E. Jenkins, A.~V. Manohar and P.~Stoffer, \emph{{Low-Energy Effective Field
  Theory below the Electroweak Scale: Operators and Matching}},
  \href{https://doi.org/10.1007/JHEP03(2018)016}{\emph{JHEP} {\bfseries 03}
  (2018) 016} [\href{https://arxiv.org/abs/1709.04486}{{\ttfamily
  1709.04486}}].

\bibitem{Jenkins:2017dyc}
E.~E. Jenkins, A.~V. Manohar and P.~Stoffer, \emph{{Low-Energy Effective Field
  Theory below the Electroweak Scale: Anomalous Dimensions}},
  \href{https://doi.org/10.1007/JHEP01(2018)084}{\emph{JHEP} {\bfseries 01}
  (2018) 084} [\href{https://arxiv.org/abs/1711.05270}{{\ttfamily
  1711.05270}}].

\bibitem{Dekens:2019ept}
W.~Dekens and P.~Stoffer, \emph{{Low-energy effective field theory below the
  electroweak scale: matching at one loop}},
  \href{https://doi.org/10.1007/JHEP10(2019)197}{\emph{JHEP} {\bfseries 10}
  (2019) 197} [\href{https://arxiv.org/abs/1908.05295}{{\ttfamily
  1908.05295}}].

\bibitem{Manohar:2000dt}
A.~V. Manohar and M.~B. Wise, \emph{{Heavy quark physics}}, {\emph{Camb.
  Monogr. Part. Phys. Nucl. Phys. Cosmol.} {\bfseries 10} (2000) 1}.

\bibitem{Fleischer:1980ub}
J.~Fleischer and F.~Jegerlehner, \emph{{Radiative Corrections to Higgs Decays
  in the Extended Weinberg-Salam Model}},
  \href{https://doi.org/10.1103/PhysRevD.23.2001}{\emph{Phys. Rev.} {\bfseries
  D23} (1981) 2001}.

\bibitem{Corbett:2020ymv}
T.~Corbett and M.~Trott, \emph{{One loop verification of SMEFT Ward
  Identities}},  \href{https://arxiv.org/abs/2010.08451}{{\ttfamily
  2010.08451}}.

\bibitem{Sirlin:1980nh}
A.~Sirlin, \emph{{Radiative Corrections in the SU(2)-L x U(1) Theory: A Simple
  Renormalization Framework}},
  \href{https://doi.org/10.1103/PhysRevD.22.971}{\emph{Phys. Rev. D} {\bfseries
  22} (1980) 971}.

\bibitem{Wells:2005vk}
J.~D. Wells, \emph{{TASI lecture notes: Introduction to precision electroweak
  analysis}},  in \emph{{Physics in D >= 4. Proceedings, Theoretical Advanced
  Study Institute in elementary particle physics, TASI 2004, Boulder, USA, June
  6-July 2, 2004}}, pp.~41--64, 2005,
  \href{https://arxiv.org/abs/hep-ph/0512342}{{\ttfamily hep-ph/0512342}}.

\bibitem{Agashe:2014kda}
{\scshape Particle Data Group} collaboration, \emph{{Review of Particle
  Physics}},
  \href{https://doi.org/10.1088/1674-1137/38/9/090001}{\emph{Chin.Phys.}
  {\bfseries C38} (2014) 090001}.

\bibitem{Baikov:2012zm}
P.~A. Baikov, K.~G. Chetyrkin, J.~H. Kuhn and J.~Rittinger, \emph{{Vector
  Correlator in Massless QCD at Order O($\alpha_s^4$) and the QED beta-function
  at Five Loop}}, \href{https://doi.org/10.1007/JHEP07(2012)017}{\emph{JHEP}
  {\bfseries 07} (2012) 017} [\href{https://arxiv.org/abs/1206.1284}{{\ttfamily
  1206.1284}}].

\bibitem{PhysRevD.22.971}
A.~Sirlin, \emph{Radiative corrections in the su(2)l x u(1) theory: A simple
  renormalization framework},
  \href{https://doi.org/10.1103/PhysRevD.22.971}{\emph{Phys. Rev. D} {\bfseries
  22} (1980) 971}.

\bibitem{Denner:2018opp}
A.~Denner, S.~Dittmaier and J.-N. Lang, \emph{{Renormalization of mixing
  angles}}, \href{https://doi.org/10.1007/JHEP11(2018)104}{\emph{JHEP}
  {\bfseries 11} (2018) 104}
  [\href{https://arxiv.org/abs/1808.03466}{{\ttfamily 1808.03466}}].

\bibitem{Denner:2019vbn}
A.~Denner and S.~Dittmaier, \emph{{Electroweak Radiative Corrections for
  Collider Physics}},
  \href{https://doi.org/10.1016/j.physrep.2020.04.001}{\emph{Phys. Rept.}
  {\bfseries 864} (2020) 1} [\href{https://arxiv.org/abs/1912.06823}{{\ttfamily
  1912.06823}}].

\bibitem{Corbett:2021jox}
T.~Corbett, \emph{{The one-loop tadpole in the geoSMEFT}},
  \href{https://arxiv.org/abs/2106.10284}{{\ttfamily 2106.10284}}.

\bibitem{Kallen:1968gaa}
G.~Kallen, \emph{{Radiative corrections in elementary particle physics}},
  \href{https://doi.org/10.1007/BFb0045559}{\emph{Springer Tracts Mod. Phys.}
  {\bfseries 46} (1968) 67}.

\bibitem{Green:1980bd}
M.~Green and M.~J.~G. Veltman, \emph{{Weak and Electromagnetic Radiative
  Corrections to Low-Energy Processes}},
  \href{https://doi.org/10.1016/0550-3213(80)90257-6}{\emph{Nucl. Phys. B}
  {\bfseries 169} (1980) 137}.

\bibitem{Denner:2016etu}
A.~Denner, L.~Jenniches, J.-N. Lang and C.~Sturm, \emph{{Gauge-independent
  $\overline{MS}$ renormalization in the 2HDM}},
  \href{https://doi.org/10.1007/JHEP09(2016)115}{\emph{JHEP} {\bfseries 09}
  (2016) 115} [\href{https://arxiv.org/abs/1607.07352}{{\ttfamily
  1607.07352}}].

\bibitem{Jenkins:2013zja}
E.~E. Jenkins, A.~V. Manohar and M.~Trott, \emph{{Renormalization Group
  Evolution of the Standard Model Dimension Six Operators I: Formalism and
  lambda Dependence}},
  \href{https://doi.org/10.1007/JHEP10(2013)087}{\emph{JHEP} {\bfseries 1310}
  (2013) 087} [\href{https://arxiv.org/abs/1308.2627}{{\ttfamily 1308.2627}}].

\bibitem{Jenkins:2013wua}
E.~E. Jenkins, A.~V. Manohar and M.~Trott, \emph{{Renormalization Group
  Evolution of the Standard Model Dimension Six Operators II: Yukawa
  Dependence}}, \href{https://doi.org/10.1007/JHEP01(2014)035}{\emph{JHEP}
  {\bfseries 1401} (2014) 035}
  [\href{https://arxiv.org/abs/1310.4838}{{\ttfamily 1310.4838}}].

\bibitem{Shifman:1979eb}
M.~A. Shifman, A.~I. Vainshtein, M.~B. Voloshin and V.~I. Zakharov,
  \emph{{Low-Energy Theorems for Higgs Boson Couplings to Photons}},
  {\emph{Sov. J. Nucl. Phys.} {\bfseries 30} (1979) 711}.

\bibitem{Bergstrom:1985hp}
L.~Bergstrom and G.~Hulth, \emph{{Induced Higgs Couplings to Neutral Bosons in
  $e^+ e^-$ Collisions}}, \href{https://doi.org/10.1016/0550-3213(86)90074-X,
  10.1016/0550-3213(85)90302-5}{\emph{Nucl. Phys.} {\bfseries B259} (1985)
  137}.

\bibitem{Deutschmann:2017qum}
N.~Deutschmann, C.~Duhr, F.~Maltoni and E.~Vryonidou, \emph{{Gluon-fusion Higgs
  production in the Standard Model Effective Field Theory}},
  \href{https://doi.org/10.1007/JHEP12(2017)063}{\emph{JHEP} {\bfseries 12}
  (2017) 063} [\href{https://arxiv.org/abs/1708.00460}{{\ttfamily
  1708.00460}}].

\bibitem{Grazzini:2016paz}
M.~Grazzini, A.~Ilnicka, M.~Spira and M.~Wiesemann, \emph{{Modeling BSM effects
  on the Higgs transverse-momentum spectrum in an EFT approach}},
  \href{https://doi.org/10.1007/JHEP03(2017)115}{\emph{JHEP} {\bfseries 03}
  (2017) 115} [\href{https://arxiv.org/abs/1612.00283}{{\ttfamily
  1612.00283}}].

\bibitem{Dawson:1990zj}
S.~Dawson, \emph{{Radiative corrections to Higgs boson production}},
  \href{https://doi.org/10.1016/0550-3213(91)90061-2}{\emph{Nucl. Phys.}
  {\bfseries B359} (1991) 283}.

\bibitem{Spira:1995rr}
M.~Spira, A.~Djouadi, D.~Graudenz and P.~M. Zerwas, \emph{{Higgs boson
  production at the LHC}},
  \href{https://doi.org/10.1016/0550-3213(95)00379-7}{\emph{Nucl. Phys. B}
  {\bfseries 453} (1995) 17}
  [\href{https://arxiv.org/abs/hep-ph/9504378}{{\ttfamily hep-ph/9504378}}].

\bibitem{Djouadi:1991tka}
A.~Djouadi, M.~Spira and P.~M. Zerwas, \emph{{Production of Higgs bosons in
  proton colliders: QCD corrections}},
  \href{https://doi.org/10.1016/0370-2693(91)90375-Z}{\emph{Phys. Lett. B}
  {\bfseries 264} (1991) 440}.

\bibitem{Degrande_2021}
C.~Degrande, G.~Durieux, F.~Maltoni, K.~Mimasu, E.~Vryonidou and C.~Zhang,
  \emph{Automated one-loop computations in the standard model effective field
  theory}, \href{https://doi.org/10.1103/physrevd.103.096024}{\emph{Physical
  Review D} {\bfseries 103} (2021) }.

\bibitem{durieux2019proposal}
G.~Durieux, I.~Brivio, F.~Maltoni, M.~Trott, S.~Alioli, A.~Buckley et~al.,
  \emph{Proposal for the validation of monte carlo implementations of the
  standard model effective field theory},  2019.

\bibitem{Inami:1982xt}
T.~Inami, T.~Kubota and Y.~Okada, \emph{{Effective Gauge Theory and the Effect
  of Heavy Quarks in Higgs Boson Decays}},
  \href{https://doi.org/10.1007/BF01571710}{\emph{Z. Phys. C} {\bfseries 18}
  (1983) 69}.

\bibitem{Kniehl:1995tn}
B.~A. Kniehl and M.~Spira, \emph{{Low-energy theorems in Higgs physics}},
  \href{https://doi.org/10.1007/s002880050007}{\emph{Z. Phys. C} {\bfseries 69}
  (1995) 77} [\href{https://arxiv.org/abs/hep-ph/9505225}{{\ttfamily
  hep-ph/9505225}}].

\bibitem{Kilian:1995tra}
W.~Kilian, \emph{{Renormalized soft Higgs theorems}},
  \href{https://doi.org/10.1007/s002880050008}{\emph{Z. Phys. C} {\bfseries 69}
  (1995) 89} [\href{https://arxiv.org/abs/hep-ph/9505309}{{\ttfamily
  hep-ph/9505309}}].

\bibitem{Chetyrkin:1996ke}
K.~G. Chetyrkin, B.~A. Kniehl and M.~Steinhauser, \emph{{Three loop O
  (alpha-s**2 G(F) M(t)**2) corrections to hadronic Higgs decays}},
  \href{https://doi.org/10.1016/S0550-3213(97)00051-5}{\emph{Nucl. Phys. B}
  {\bfseries 490} (1997) 19}
  [\href{https://arxiv.org/abs/hep-ph/9701277}{{\ttfamily hep-ph/9701277}}].

\bibitem{Chetyrkin:1996wr}
K.~G. Chetyrkin, B.~A. Kniehl and M.~Steinhauser, \emph{{Virtual top quark
  effects on the H ---\ensuremath{>} b anti-b decay at next-to-leading order in
  QCD}}, \href{https://doi.org/10.1103/PhysRevLett.78.594}{\emph{Phys. Rev.
  Lett.} {\bfseries 78} (1997) 594}
  [\href{https://arxiv.org/abs/hep-ph/9610456}{{\ttfamily hep-ph/9610456}}].

\bibitem{Contino:2014aaa}
R.~Contino, M.~Ghezzi, C.~Grojean, M.~Mühlleitner and M.~Spira,
  \emph{{eHDECAY: an Implementation of the Higgs Effective Lagrangian into
  HDECAY}}, \href{https://doi.org/10.1016/j.cpc.2014.06.028}{\emph{Comput.
  Phys. Commun.} {\bfseries 185} (2014) 3412}
  [\href{https://arxiv.org/abs/1403.3381}{{\ttfamily 1403.3381}}].

\bibitem{Harlander:2013oja}
R.~V. Harlander and T.~Neumann, \emph{{Probing the nature of the Higgs-gluon
  coupling}}, \href{https://doi.org/10.1103/PhysRevD.88.074015}{\emph{Phys.
  Rev. D} {\bfseries 88} (2013) 074015}
  [\href{https://arxiv.org/abs/1308.2225}{{\ttfamily 1308.2225}}].

\bibitem{Dawson:2014ora}
S.~Dawson, I.~M. Lewis and M.~Zeng, \emph{{Effective field theory for Higgs
  boson plus jet production}},
  \href{https://doi.org/10.1103/PhysRevD.90.093007}{\emph{Phys. Rev. D}
  {\bfseries 90} (2014) 093007}
  [\href{https://arxiv.org/abs/1409.6299}{{\ttfamily 1409.6299}}].

\bibitem{Cahn:1983ip}
R.~N. Cahn and S.~Dawson, \emph{{Production of Very Massive Higgs Bosons}},
  \href{https://doi.org/10.1016/0370-2693(84)91180-8}{\emph{Phys. Lett. B}
  {\bfseries 136} (1984) 196}.

\bibitem{Ellis:1975ap}
J.~R. Ellis, M.~K. Gaillard and D.~V. Nanopoulos, \emph{{A Phenomenological
  Profile of the Higgs Boson}},
  \href{https://doi.org/10.1016/0550-3213(76)90382-5}{\emph{Nucl. Phys.}
  {\bfseries B106} (1976) 292}.

\bibitem{Zyla:2020zbs}
{\scshape Particle Data Group} collaboration, \emph{{Review of Particle
  Physics}}, \href{https://doi.org/10.1093/ptep/ptaa104}{\emph{PTEP} {\bfseries
  2020} (2020) 083C01}.

\bibitem{Aaltonen:2013iut}
{\scshape CDF, D0} collaboration, \emph{{Combination of CDF and D0 $W$-Boson
  Mass Measurements}},
  \href{https://doi.org/10.1103/PhysRevD.88.052018}{\emph{Phys. Rev.}
  {\bfseries D88} (2013) 052018}
  [\href{https://arxiv.org/abs/1307.7627}{{\ttfamily 1307.7627}}].

\bibitem{Olive:2016xmw}
{\scshape Particle Data Group} collaboration, \emph{{Review of Particle
  Physics}}, \href{https://doi.org/10.1088/1674-1137/40/10/100001}{\emph{Chin.
  Phys.} {\bfseries C40} (2016) 100001}.

\bibitem{Mohr:2012tt}
P.~J. Mohr, B.~N. Taylor and D.~B. Newell, \emph{{CODATA Recommended Values of
  the Fundamental Physical Constants: 2010}},
  \href{https://doi.org/10.1103/RevModPhys.84.1527}{\emph{Rev. Mod. Phys.}
  {\bfseries 84} (2012) 1527}
  [\href{https://arxiv.org/abs/1203.5425}{{\ttfamily 1203.5425}}].

\bibitem{Dubovyk:2019szj}
I.~Dubovyk, A.~Freitas, J.~Gluza, T.~Riemann and J.~Usovitsch,
  \emph{{Electroweak pseudo-observables and Z-boson form factors at two-loop
  accuracy}}, \href{https://doi.org/10.1007/JHEP08(2019)113}{\emph{JHEP}
  {\bfseries 08} (2019) 113}
  [\href{https://arxiv.org/abs/1906.08815}{{\ttfamily 1906.08815}}].

\bibitem{Corbett:2021eux}
T.~Corbett, A.~Helset, A.~Martin and M.~Trott, \emph{{EWPD in the SMEFT to
  dimension eight}}, \href{https://doi.org/10.1007/JHEP06(2021)076}{\emph{JHEP}
  {\bfseries 06} (2021) 076}
  [\href{https://arxiv.org/abs/2102.02819}{{\ttfamily 2102.02819}}].

\bibitem{Freitas:2014hra}
A.~Freitas, \emph{{Higher-order electroweak corrections to the partial widths
  and branching ratios of the Z boson}},
  \href{https://doi.org/10.1007/JHEP04(2014)070}{\emph{JHEP} {\bfseries 1404}
  (2014) 070} [\href{https://arxiv.org/abs/1401.2447}{{\ttfamily 1401.2447}}].

\bibitem{Awramik:2003rn}
M.~Awramik, M.~Czakon, A.~Freitas and G.~Weiglein, \emph{{Precise prediction
  for the W boson mass in the standard model}},
  \href{https://doi.org/10.1103/PhysRevD.69.053006}{\emph{Phys.Rev.} {\bfseries
  D69} (2004) 053006} [\href{https://arxiv.org/abs/hep-ph/0311148}{{\ttfamily
  hep-ph/0311148}}].

\bibitem{Awramik:2006uz}
M.~Awramik, M.~Czakon and A.~Freitas, \emph{{Electroweak two-loop corrections
  to the effective weak mixing angle}},
  \href{https://doi.org/10.1088/1126-6708/2006/11/048}{\emph{JHEP} {\bfseries
  11} (2006) 048} [\href{https://arxiv.org/abs/hep-ph/0608099}{{\ttfamily
  hep-ph/0608099}}].

\bibitem{Trott:2021vqa}
M.~Trott, \emph{{A methodology for theory uncertainties in the SMEFT}},
  \href{https://arxiv.org/abs/2106.13794}{{\ttfamily 2106.13794}}.

\bibitem{2013}
J.~A. Aguilar-Saavedra, R.~Benbrik, S.~Heinemeyer and M.~Pérez-Victoria,
  \emph{Handbook of vectorlike quarks: Mixing and single production},
  \href{https://doi.org/10.1103/physrevd.88.094010}{\emph{Physical Review D}
  {\bfseries 88} (2013) }.

\bibitem{2000}
F.~d. Aguila, J.~Santiago and M.~Pérez-Victoria, \emph{Observable
  contributions of new exotic quarks to quark mixing},
  \href{https://doi.org/10.1088/1126-6708/2000/09/011}{\emph{Journal of High
  Energy Physics} {\bfseries 2000} (2000) 011–011}.

\bibitem{Hartland_2013}
N.~P. Hartland and E.~R. Nocera, \emph{A mathematica interface to nnpdfs},
  \href{https://doi.org/10.1016/j.nuclphysbps.2012.11.013}{\emph{Nuclear
  Physics B - Proceedings Supplements} {\bfseries 234} (2013) 54–57}.

\bibitem{Ball_2015}
R.~D. Ball, V.~Bertone, S.~Carrazza, C.~S. Deans, L.~Del~Debbio, S.~Forte
  et~al., \emph{Parton distributions for the lhc run ii},
  \href{https://doi.org/10.1007/jhep04(2015)040}{\emph{Journal of High Energy
  Physics} {\bfseries 2015} (2015) }.

\bibitem{Craig:2019wmo}
N.~Craig, M.~Jiang, Y.-Y. Li and D.~Sutherland, \emph{{Loops and Trees in
  Generic EFTs}}, \href{https://doi.org/10.1007/JHEP08(2020)086}{\emph{JHEP}
  {\bfseries 08} (2020) 086}
  [\href{https://arxiv.org/abs/2001.00017}{{\ttfamily 2001.00017}}].

\bibitem{Dedes:2018seb}
A.~Dedes, M.~Paraskevas, J.~Rosiek, K.~Suxho and L.~Trifyllis, \emph{{The decay
  $h\to \gamma\gamma$ in the Standard-Model Effective Field Theory}},
  \href{https://doi.org/10.1007/JHEP08(2018)103}{\emph{JHEP} {\bfseries 08}
  (2018) 103} [\href{https://arxiv.org/abs/1805.00302}{{\ttfamily
  1805.00302}}].

\bibitem{Dawson:2018liq}
S.~Dawson and P.~P. Giardino, \emph{{Electroweak corrections to Higgs boson
  decays to $\gamma\gamma$ and $W^+W^-$ in standard model EFT}},
  \href{https://doi.org/10.1103/PhysRevD.98.095005}{\emph{Phys. Rev. D}
  {\bfseries 98} (2018) 095005}
  [\href{https://arxiv.org/abs/1807.11504}{{\ttfamily 1807.11504}}].

\bibitem{Dawson:2018pyl}
S.~Dawson and P.~P. Giardino, \emph{{Higgs decays to $ZZ$ and $Z\gamma$ in the
  standard model effective field theory: An NLO analysis}},
  \href{https://doi.org/10.1103/PhysRevD.97.093003}{\emph{Phys. Rev. D}
  {\bfseries 97} (2018) 093003}
  [\href{https://arxiv.org/abs/1801.01136}{{\ttfamily 1801.01136}}].

\bibitem{Manohar:2006gz}
A.~V. Manohar and M.~B. Wise, \emph{{Modifications to the properties of the
  Higgs boson}},
  \href{https://doi.org/10.1016/j.physletb.2006.03.030}{\emph{Phys. Lett.}
  {\bfseries B636} (2006) 107}
  [\href{https://arxiv.org/abs/hep-ph/0601212}{{\ttfamily hep-ph/0601212}}].

\end{thebibliography}\endgroup

\end{document}